\documentclass[a4paper,twocolumn,11pt,accepted=2026-09-02]{quantumarticle}
\pdfoutput=1
\usepackage[utf8]{inputenc}
\usepackage[english]{babel}
\usepackage[T1]{fontenc}
\usepackage{amsmath}
\usepackage{hyperref}

\usepackage{tikz}
\usepackage{lipsum}

\usepackage{cite}
\usepackage{amsmath,amssymb,amsfonts}

\usepackage{graphicx}
\usepackage{textcomp}
\usepackage{xcolor}

\usepackage{algorithm}
\usepackage[noend]{algpseudocode}
\algnewcommand{\IfThenElse}[3]{
	\State \algorithmicif\ #1\ \algorithmicthen\ #2\ \algorithmicelse\ #3}

\usepackage{braket,amsthm,enumitem}
\usepackage[caption=false,labelformat=simple]{subfig}		

\usetikzlibrary{quantikz}


\newtheorem{definitionenv}{Definition}
\newtheorem{lemmaenv}[definitionenv]{Lemma}
\newtheorem{theoremenv}[definitionenv]{Theorem}
\newtheorem{corollaryenv}[definitionenv]{Corollary}
\newtheorem{propositionenv}[definitionenv]{Proposition}
\newtheorem{remarkenv}[definitionenv]{Remark}
\newtheorem{conjectureenv}[definitionenv]{Conjecture}
\newtheorem{exampleenv}{Example}
\newtheorem{app-lemmaenv}[section]{Lemma}

\newenvironment{definition}{\begin{definitionenv}\rm}{\end{definitionenv}}
\newenvironment{lemma}{\begin{lemmaenv}\rm}{\end{lemmaenv}}

\newenvironment{corollary}{\begin{corollaryenv}\rm}{\end{corollaryenv}}
\newenvironment{remark}{\begin{remarkenv}\rm}{\end{remarkenv}}
\newenvironment{example}{\begin{exampleenv}\rm}{\end{exampleenv}}
\newenvironment{proposition}{\begin{propositionenv}\rm}{\end{propositionenv}}

\newenvironment{app-lemma}{\begin{app-lemmaenv}\rm}{\end{app-lemmaenv}}

\DeclareMathOperator*{\argmin}{arg\,min}
\DeclareMathOperator{\wt}{wt}

\newcommand{\cM}{{\cal M}}
\newcommand{\cN}{{\cal N}}
\newcommand{\cS}{{\cal S}}

\newcommand{\cE}{{\cal E}}
\newcommand{\cF}{{\cal F}}

\newcommand{\cH}{{\cal H}}
\newcommand{\cR}{{\cal R}}

\newcommand{\mR}{{\mathbb R}}

\newcommand{\fM}{{\mathfrak M}}

\usepackage{accents} 		

\makeatletter 
\renewcommand*\env@matrix[1][*\c@MaxMatrixCols c]{%
	\hskip -\arraycolsep
	\let\@ifnextchar\new@ifnextchar
	\array{#1}}
\makeatother

\usepackage{xpatch}
\xpretocmd{\eqref}{Eq.~}{}{}

\newcommand{\rmark}[1]{{#1}}
\newcommand{\bmark}[1]{{#1}}

\begin{document}

\title{Streaming Belief Propagation on Mixed-Alphabet Tanner Graphs for Practical Quantum Memory}

\author{Kao-Yueh Kuo}
\affiliation{Department of Electrical Engineering, Yuan Ze University, Taoyuan City, Taiwan}
\orcid{0000-0002-1390-5197}
\email{kywukuo@saturn.yzu.edu.tw}
\thanks{
	K.-Y. Kuo was supported by the National Science and Technology Council (NSTC), Taiwan, under Grant 115-2222-E-155-002.
}

\author{Ching-Yi Lai}
\affiliation{Institute of Communications Engineering, National Yang Ming Chiao Tung University, Hsinchu City, Taiwan}
\affiliation{National Center for Excellence in Quantum Information Science and Engineering}
\orcid{0000-0003-1970-8167}
\email{cylai@nycu.edu.tw}

\thanks{
	C.-Y. Lai was supported by the NSTC under Grants 113-2221-E-A49-114-MY3 and 115-2119-M-A49-002, and by the Ministry of Education, Taiwan, through the Taiwan Centers of Excellence (TCE) Program.
}
\maketitle

\begin{abstract}
Reliable quantum memory under circuit-level noise requires decoders that can process syndrome information continuously and at a rate comparable to its generation. In practical quantum error correction (QEC), repeated syndrome measurements cause the number of potential error locations to grow rapidly with both code size and time. In this paper, we propose a streaming mixed-alphabet belief propagation (SM-BP) decoder.
We construct a space-time Tanner graph across multiple rounds of syndrome extraction with mixed-alphabet error variables, preserving correlations arising from multi-qubit faults. Additionally, we propose an adaptive sliding window procedure that captures long error events across window boundaries and adjusts the decoding in real time. To enhance SM-BP, we introduce a technique of probabilistic error consolidation to mitigate degeneracy effects and short cycles.  
Our simulations demonstrate high error thresholds of 0.4\%–0.87\% and strong error-floor performance for topological code families, including rotated toric, toric color, and twisted XZZX toric codes.
These results show that SM-BP provides a practical decoding framework for continuous QEC under circuit-level noise.
\end{abstract}

	\section{Introduction}

	In the pursuit of scalable quantum computing, protecting quantum information through quantum error correction (QEC) is crucial, as quantum systems are inherently vulnerable to noise~\cite{Shor95,Steane96,CS96,KL97,GotPhD}. To achieve fault-tolerant quantum computation (FTQC)~\cite{DS96,Got98,AB99,BK05}, it is essential to develop reliable quantum memory that supports the long-term storage of quantum information. Among various QEC schemes, topological codes, such as surface codes~\cite{Kit03} and color codes~\cite{BM06,BM07}, have emerged as leading candidates due to their advantageous properties, including local stabilizer measurements and high thresholds for fault tolerance~\cite{DKLP02,RH07,Ste14c,CKYZ20}. On the other hand, quantum low-density parity-check (QLDPC) codes offer a distinct advantage by providing higher code rates and reduced overhead for quantum memory, assuming higher qubit connectivity~\cite{MMM04,TZ14,KP13,Got14c,PK19,LP24}. Additionally, QLDPC codes may also exhibit high thresholds for fault tolerance~\cite{BCG+24,scruby2024high,old2024lift}, making them a promising alternative in FTQC.

	In reality, every component in a quantum circuit is potentially faulty, and errors can propagate throughout the system. This presents a challenge for noise-resilient error correction when using imperfect components. In QEC, error syndromes are generated rapidly, often within the execution time of a few quantum gates, especially for codes with low-weight checks. Active QEC must be implemented quickly enough to prevent logical errors due to error accumulation~ \cite{Ter15}. 
	When quantum memory is protected by a quantum code with  minimum distance $d$, a typical QEC cycle involves $O(d)$ rounds of syndrome measurements \cite{DKLP02}, resulting in  $O(d^3)$ potential error locations for a code of length $n=O(d^2)$.

	Traditional decoders like minimum-weight perfect matching (MWPM)~\cite{Edm65,DKLP02,RH07,RHG07,HG25}, union-find (UF)~\cite{DN21,HNB20,griffiths2024union,LL25}, and belief propagation (BP)~\cite{Gal63,Tan81,Pea88,KFL01,MMM04,PC08,KL20,KL22} combined with ordered statistic decoding (OSD)~\cite{PK19,RWBC20,KKL23} are commonly used for circuit-level decoding.        
	Both MWPM and UF are primarily tailored to specific topological codes. In contrast, BP-OSD is a more general decoder for QLDPC codes~\cite{BCG+24,scruby2024high,old2024lift}, combining BP for error likelihood estimation with OSD for better correction. However, 
	\rmark{its performance gain comes at the cost of high computational overhead for OSD-0 and exponentially higher for higher-order re-processing), which limits its scalability for real-time applications.}	
	Since large-distance quantum codes are essential for scalable quantum systems, decoders with near-linear complexity in the number of error variables are highly desirable, especially for extending quantum memory lifetimes.

	In this paper, we propose a BP-based solution for circuit-level decoding. BP has been adapted to decode quantum codes in the code capacity model  with perfect error syndromes \cite{MMM04,PC08,Wan+12,Bab+15,KL20,KL21,KL22}, quantum data-syndrome codes \cite{KCL21,KL24}, and the phenomenological noise model \cite{KL24}. \rmark{We introduce a streaming mixed-alphabet belief propagation (SM-BP) decoding algorithm incorporating an adaptive sliding window procedure. This framework is applicable to general QLDPC codes, including both CSS and non-CSS families, under the circuit-level noise model.} 
	Our approach consists of several key components, which will be explained as follows.

	\rmark{A quantum stabilizer code can be viewed as an additive quaternary code generated by a binary matrix~\cite{CRSS98}, describing single-qubit Pauli errors $X$, $Y$, and $Z$. In the circuit-level noise model, a single-qubit location may introduce $X$, $Y$, or $Z$ errors, while a two-qubit gate can produce $4^2=16$ Pauli operators (15 nontrivial errors plus the identity), and measurement errors are binary (flip or non-flip).} This leads naturally to a mixed-alphabet representation with binary, quaternary, and 16-ary symbols. 
	By introducing a mixed-alphabet syndrome representation, we construct a generalized check matrix for multi-round syndrome extraction.
	 This generalized check matrix defines the linear relationship between the actual errors and the observed binary error syndromes. 
Consequently, we propose an SM-BP decoding algorithm for this mixed-alphabet decoding problem, where all transmitted messages are scalars, extending our previous decoders in~\cite{KL20,KL22}.

	\bmark{
	\begin{table*}[ht]
	\scriptsize
		\centering
		\begin{tabular}{|l|c|c|c|l|} 
			\hline
			Decoder & Complexity & Applicability & Decoding Radius & Notes  \\
			\hline
			MWPM & $O(N^3)$ & Graph-like codes (Column weight $\le 2$) & $\lfloor \frac{d-1}{2} \rfloor$ & Empirically $O(N^{1.5})$ \cite{HG25} \\
			UF & $O(N )$ & Graph-like codes (Column weight $\le 2$) & $\lfloor \frac{d-1}{2} \rfloor$ &Linear complexity \cite{griffiths2024union} \\
			BP-OSD-0 & $O(N^3)$ & General & Heuristic & OSD step dominates latency$^\ddag$ \\
			SM-BP & $O(N T_{\max})$ $^\S$ & General (Non-CSS) & Heuristic & 
			empirically $O(N\log N)$  (see Appendix~\ref{sec:efficiency_paralellism})\\
			\hline
		\end{tabular}
		\begin{flushleft}
			\small
			$^\ddag$ Potential speedup via replacement schemes: BP-LSD (localized statistics decoding)~\cite{Hil+25}, BP-AC (ambiguity clustering)~\cite{WB24ambiguity}, BP-ADOSD (approximate degenerate OSD)~\cite{KKL24}\\
			$^\S$ The scaling parameter $\alpha$ is selected from a candidate set $\mathcal{A}$ where $|\mathcal{A}| = 100$. Since instances for different $\alpha$ are independent, they can be fully parallelized, maintaining $O(N)$ online latency.
		\end{flushleft}
		\caption{Comparison of quantum decoders. $N$ denotes the total number of error variables; $T_{\max}$ is the maximum BP iterations.}\label{tb:comparison}
	\end{table*}
}
	
	\bmark{
		The generalized-check matrix formalism is closely related to the detector-error-model (DEM) matrices~\cite{Pry20, derks2025designing} generated by simulators such as Stim~\cite{Gid21}. 
		While DEM-based approaches typically reduce the decoding problem to a binary task, our mixed-alphabet framework explicitly represents Pauli errors and their correlations,
		including those between single- and two-qubit errors.  
		For instance, a CNOT gate can produce 16 distinct Pauli outcomes, corresponding to four binary degrees of freedom with strong correlations, all of which are directly represented in our formulation. By preserving these correlations, our method retains more information about the underlying errors, which can enable more accurate decoding, particularly for non-CSS codes or asymmetric noise.}

	\rmark{For QEC schemes based on repeated syndrome measurements,  the generalized check matrix can be transformed into a sparse  lower block bidiagonal Toeplitz form using appropriate row operations.}
 This ensures that any localized error affects only a limited number of syndrome rounds, enabling the construction of a space-time Tanner graph that captures error evolution over time. 
	\bmark{This procedure is analogous to taking differences of syndromes between consecutive rounds in surface-code decoding using MWPM.} Consequently, for an $n$-qubit code with distance $d=O(\sqrt{n})$, $O(1)$-weight stabilizers, and a decoding window of $O(d)$ rounds, the complexity of SM-BP scales as 
	\bmark{$O(nd\log n)=O(d^3\log d)$,
	where $O(\log n)$ corresponds to the number of BP iterations. This cost is incurred once every $O(d)$ rounds, which on average is $O(d^2 \log d)=O(n\log n)$ per round.}
Table~\ref{tb:comparison} summarizes the basic decoder for quantum codes.


However, the presence of multiple two-qubit gates introduces many degenerate errors that produce the same syndrome and have the same residual effect on the data qubits, leading to numerous short cycles in the Tanner graph. A direct application of SM-BP on the generalized check matrix with standard error initialization results in poor decoding performance (see Fig.~\ref{fig:FT_schemes}). 
To address this issue, we propose \textit{probabilistic error consolidation} so that degenerate errors are merged into a single representative with aggregated probability, reducing the problem size.
\rmark{
More specifically, degenerate error variables with the same alphabet size are merged first. If a larger-alphabet variable (e.g., 16-ary) shares partial degeneracy with smaller ones (e.g., quaternary or binary), it is decomposed into multiple smaller variables. These are then merged accordingly, approximating the original distribution by the product of the resulting lower-dimensional distributions. For example, a 16-ary error can be decomposed into two quaternary variables or four binary variables; if the quaternary variables are degenerate with others, they are merged. The original 16-ary distribution is then approximated by the product of the two quaternary distributions.
}
This reduces decoding complexity and mitigates short cycles in the Tanner graph.

Finally, for practical memory-lifetime decoding, sliding-window decoding is necessary to maintain a reasonable computational complexity~\cite{DKLP02,TZC+23,SBB+23,GCR24}.
In this approach, the decoder processes a fixed number of syndrome extraction rounds before advancing by a specified number of rounds, referred to as the \textit{window offset}. 
We observe that long error events spanning across window boundaries can degrade the performance of sliding-window decoding with a fixed offset. To address this issue, we introduce a simple adaptive sliding-window procedure. By identifying potential boundary-crossing events, the window offset can be dynamically adjusted, leading to improved decoding performance.

	Together, these techniques form a comprehensive framework for SM-BP.
	To validate our approach, we conduct memory lifetime simulations on the $[[d^2,2,d]]$ rotated toric codes (with even $d$) \cite{Kit03,BM07}, $[[\frac{9}{8}d^2,\, 4,\, d]]$ rotated 6.6.6 toric color codes (with $d$ a multiple of 4) \cite{BM06,BM07}, and $[[\frac{d^2+1}{2},\, 1,\, d]]$ twisted XZZX toric codes (with odd $d$) \cite{KDP11,SY24,KL22isit}. In our experience, decoding problems induced by topological codes in the code capacity and phenomenological noise models 
	tend to be more difficult compared to those from quantum  codes with nonlocal stabilizers, such as bicycle codes~\cite{KL22,KL24}, which is why we focus on topological codes under the circuit-level noise model. However, our methods are applicable to general QLDPC codes.

	In our simulations, the use of adaptive sliding window decoding and probabilistic error consolidation techniques significantly enhances the SM-BP decoding performance. Additionally, a window size of $d$ to $2d$ rounds is generally sufficient for these topological codes with minimum distance $d$, as performance saturates beyond $2d$ rounds. Based on these techniques, the SM-BP decoding threshold for rotated toric codes is estimated to be $0.75\%$ using the finite-size scaling ansatz~\cite{WHP03,Har04}, which is comparable to the MWPM decoding threshold for toric and surface codes~\cite{DKLP02,RH07,WFH11,Ste14s},
	estimated to be roughly 0.75\%--1\%.
	Similarly, we observe a threshold of approximately $0.4\%$--$0.59\%$ for the rotated 6.6.6 color codes, which surpasses previous thresholds for color-type codes using modified MWPM algorithms. Furthermore, we provide the first threshold estimation of around $0.75\%$--$0.87\%$ for the twisted XZZX toric codes.

	By utilizing adaptive sliding windows and probabilistic error consolidation techniques, SM-BP significantly reduces the overhead associated with decoding while maintaining high performance. Its nearly-linear scaling makes it particularly suited for practical quantum memory applications. Moreover, it is applicable to general QLDPC codes, paving the way for the construction of improved quantum codes.

	The paper is organized as follows.
	In Sec.~\ref{sec:basics}, we introduce quantum stabilizer codes and the circuit-level noise model.
	In Sec.~\ref{sec:FT}, we construct a generalized check matrix for a syndrome extraction circuit, define the circuit-level decoding problem, and propose the SM-BP decoding algorithm.  We then derive a sparse generalized check matrix that induces a space-time Tanner graph in Sec.~\ref{sec:reduction}.   Additionally, we   discuss several problem reduction techniques, 	including error merging and probabilistic error consolidation. In Sec.~\ref{sec:memory}, 	we discuss the lifetime of  quantum memory and propose an adaptive sliding window decoding  approach.  	 Finally, we conduct several simulations in Sec.~\ref{sec:sim} to demonstrate the effectiveness of these techniques, followed by our conclusion.

	\section{ Quantum stabilizer codes} \label{sec:basics}

	The Pauli matrices are $I=[{1\atop 0}{0\atop 1}], ~ X=[{0\atop 1}{1\atop 0}], ~ Y=[{0\atop i}{-i\atop 0}], \text{ and } Z=[{1\atop 0}{0\atop -1}]$.
	For convenience, an $n$-fold Pauli operator $ E=c E_1 \otimes  \cdots \otimes E_n$ for $c\in\{\pm 1,\pm i\}$ and $E_j\in\{I,X,Y,Z\}$  may  occasionally be represented as a row vector $E=(E_1, \dots,E_n) \in\{I,X,Y,Z\}^{1\times n}$, since the phase does not impact our examination of Pauli error decoding problems.
	An identity operator $I^{\otimes n}$ for any $n$ will be simply denoted by $I$.
	The weight of $E$, denoted  $\wt(E)$, is  the number of its nonidentity Pauli components. 
	We use $X_j$ or $Z_j$ to represent a nontrivial Pauli matrix applied to qubit~$j$,
	while operating trivially  on the other qubits.
	For $u=(u_1,\dots,u_n),v=(v_1,\dots,v_n)\in\{0,1\}^n$, we define 
	\begin{align}
		Z^u X^v= \prod_{i=1}^n Z_i^{u_i}\prod_{j=1}^n X_i^{v_i}=(Z_1^{u_1}X_1^{v_1}, ~ \dots, ~ Z_n^{u_n}X_n^{v_n} ). \label{eq:bin_repr}
	\end{align}
	Thus, an $n$-fold  Pauli operator $Z^u X^v$ can be represented by a binary string $(u,v)\in \{0,1\}^{2n}$ when the global phase is not important.
	For example, $X\otimes I\otimes Z=(X,I,Z)= X_1 Z_3$ 
	\bmark{and can be represented by $u=(001)$ and $v=(100)$.}

	Any two Pauli operators either commute or anticommute. Thus, we define a bilinear form $*$ on two Pauli operators $E$ and $F$  by
	\begin{align}
		E*F=	
		\begin{cases} 
			0, &\text{if $E$ and $F$ commute};\\
			1, &\text{if $E$ and $F$ anticommute}.
		\end{cases} \label{eq:bilinear_form1}
	\end{align}
	
	\subsection{Code capacity model}
	
	Let $\cS$ be an abelian group of $n$-fold Pauli operators such that $-I \notin \cS$. The joint $(+1)$-eigenspace of $\cS$ is called a stabilizer code, which includes $n$-qubit states that are fixed by $\cS$. Suppose that $\cS$ is generated by $n-k$ independent generators; then the code is denoted as an $[[n, k]]$ stabilizer code. The group $\cS$ is called the stabilizer group, and its elements are called stabilizers for the code.

	A Pauli error $E$ acting on a codespace state can be detected if it anticommutes with any of the stabilizers.  
	Assume that $m\ge n-k$ stabilizers $g_1,g_2,\dots,g_m\in\{I,X,Y,Z\}^n$ are perfectly measured  with binary outcomes $s=(s_1,\dots,s_m)\in\{0,1\}^m$, where
	\begin{align}
		s_i = E*g_i. \label{eq:stabilizer_syndrome}
	\end{align}
	The binary vector $s$ is called the error syndrome of $E$. 
	If a Pauli error is a stabilizer, it does not affect the codespace. A stabilizer code has minimum distance $d$ if the weight of any non-stabilizer, undetectable Pauli error is at least $d$. Such a code is referred to as an $[[n, k, d]]$ stabilizer code.

	If two Pauli errors have the same effect on the codespace and cannot be distinguished by the measured error syndrome, they are referred to as \textit{degenerate} errors. 

\begin{figure}[t!]	   
		\centering
		\includegraphics[width=\columnwidth]{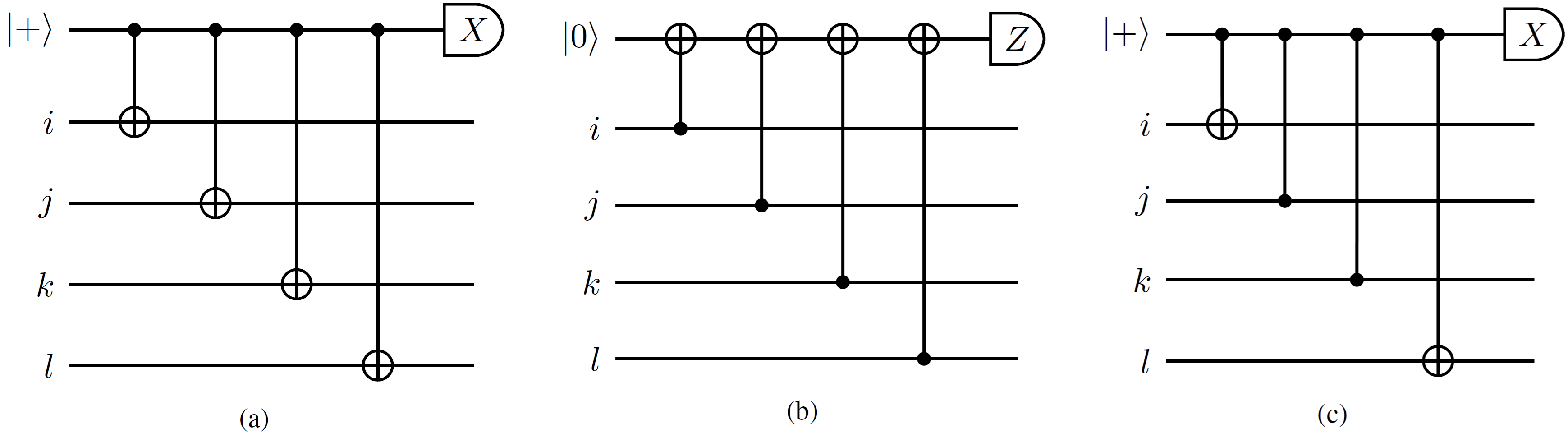}
		\caption[Roots associated to the Cartan matrix]{	 
			 \rmark{Raw syndrome extraction circuits} for (a)~$X_iX_jX_kX_l$, (b)~$Z_iZ_jZ_kZ_l$, and (c)~$X_iZ_jZ_kX_l$, applied to the corresponding data qubits $i$, $j$, $k$, $l$ in sequence. 
		} \label{fig:meas}
	\end{figure}

	Figure~\ref{fig:meas} illustrates the raw syndrome extraction process for stabilizers of weight~4. \rmark{In the code-capacity model, Pauli errors are assumed to occur on the data qubits prior to syndrome extraction, and syndrome measurements are considered perfect.} This model is commonly used to assess the theoretical limits of quantum error-correcting codes.

	\subsection{Circuit-level noise model}	

	In the circuit-level noise model~\cite{DKLP02,RH07,WFSH10}, each location in a quantum circuit, including quantum gates, measurements, state preparation, and idle qubits, is susceptible to faults. Additionally, errors can propagate through controlled-phase (CZ) and  controlled-NOT (CNOT) gates within the quantum circuit. As a result, \rmark{noise-resilient} syndrome extraction procedures are crucial to obtaining reliable measurement outcomes that can effectively prevent error accumulation~\cite{NC00}.

	For topological codes with low-weight stabilizer generators, repeatedly applying the raw syndrome extraction method is sufficient to achieve \rmark{noise-resilient} syndrome extraction~\cite{DKLP02}. 
	In this work, we follow this procedure for simplicity, with the understanding that it can be directly extended to general quantum error correction protocols. 
	Specifically, we focus on topological quantum codes with local, low-weight stabilizers~\cite{Kit03}, particularly those defined on a torus. These include the $ [[d^2,2,d]]$   rotated toric codes (with even $d$) \cite{Kit03,BM07}, $[[\frac{9}{8}d^2,\, 4,\, d]]$
	rotated 6.6.6 toric  color codes (with $d$ a multiple of 4) \cite{BM06,BM07}, and  $[[(d^2+1)/2,\, 1,\, d]]$ twisted XZZX toric codes (with odd $d$) \cite{KDP11,SY24,KL22isit}.

\begin{figure}[t] 
		\centering
		\subfloat[\label{fig:toric_2x2}]{\includegraphics[width=0.18\textwidth]{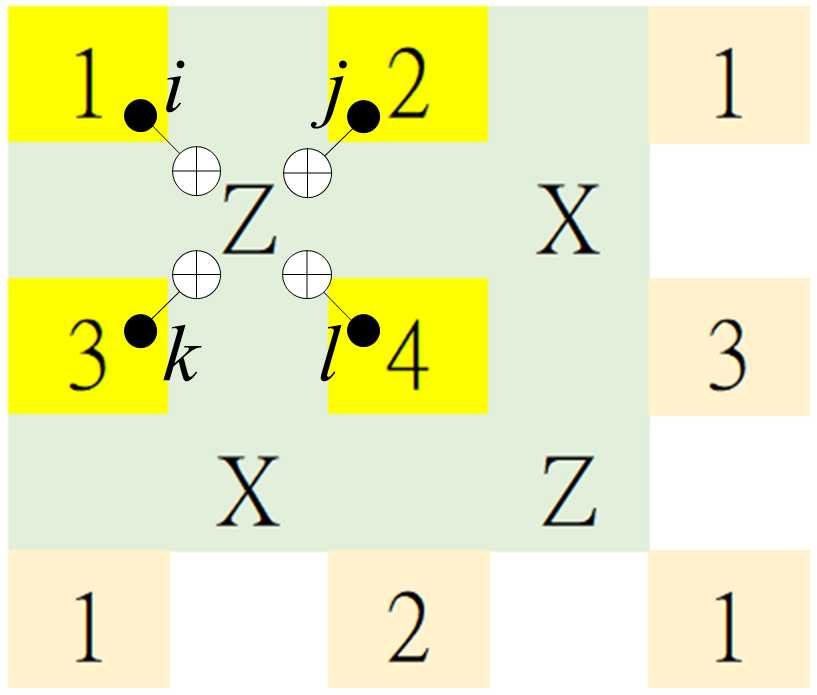}} ~~~~~~~~ 
		\subfloat[\label{fig:toric_4x4}]{\includegraphics[width=0.18\textwidth]{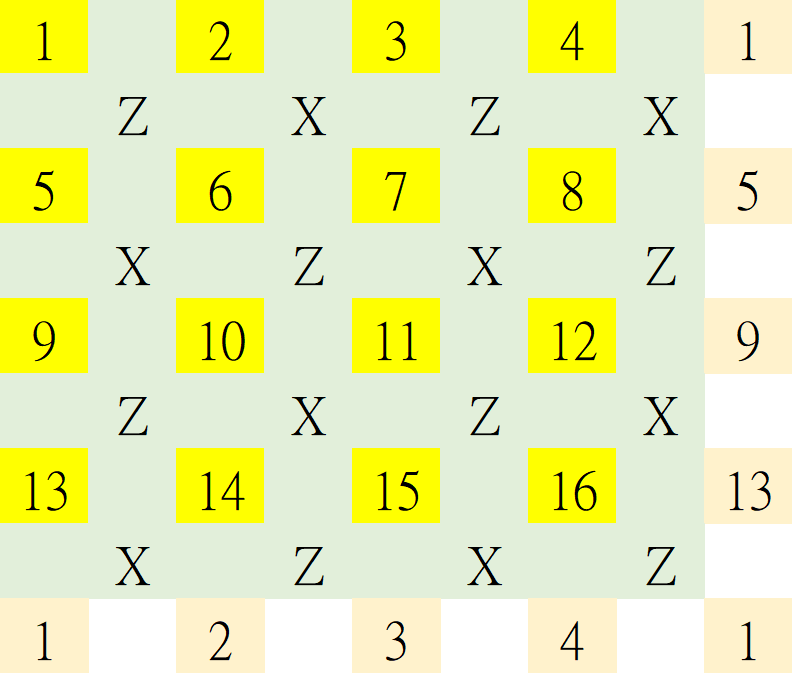}}
		\caption{
			The lattice representation of the family of $[[d^2, 2, d]]$ toric codes  is illustrated   for (a)~${d=2}$ and (b)~${d=4}$.
			Each data qubit is represented by a yellow box labeled with a number ranging from 1 to $d^2$.
			The orange boxes on the boundary correspond to the yellow boxes with the same numbers, indicating the connectivity of qubits since the lattice wraps around a torus.
			Green boxes labeled with $X$ or $Z$, surrounded by four data qubits labeled as $(i,j,k,l)$, represent stabilizers of the form $X_i X_j X_k X_l$ or $Z_i Z_j Z_k Z_l$.
			The order $(i,j,k,l)$ also signifies the order in which measurements are performed.
			For instance, in (a), the upper-left label $Z$ corresponds to the measurement order \rmark{$(1,2,3,4)$}, indicating that a stabilizer $Z_1 Z_4 Z_3 Z_2$ is measured in this specific order.
			All stabilizers in this code family follow this measurement order. This representation visually demonstrates the structure of the toric code and the sequence in which stabilizers are measured on the lattice.
		} \label{fig:toric_lat}		
	\end{figure}

\begin{figure}[t]
		\centering \includegraphics[width=0.33\textwidth]{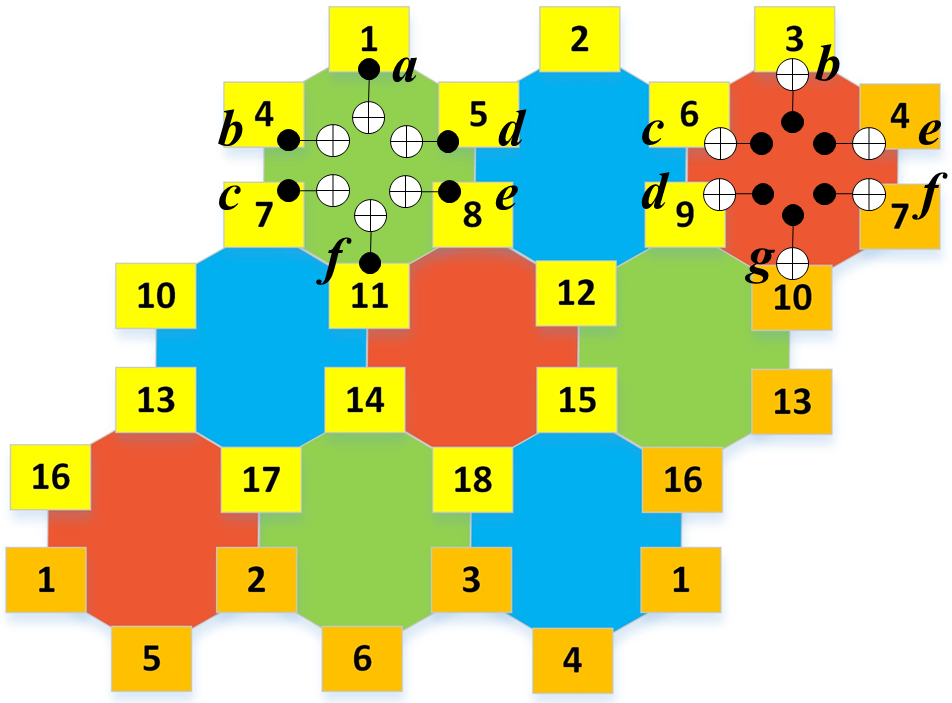}
		\caption{	
			The lattice  of  the $[[\frac{9}{8}d^2,\, 4,\, d]]$ toric color code with $d=4$.
			The qubits are  labeled similarly to those in Fig.~\ref{fig:toric_lat}.
			Every plaquette represents both an $X$ and a $Z$ stabilizer, each operating on the six data qubits surrounding the plaquette.
			For full parallelism, the $X$ and $Z$ stabilizers are measured \rmark{in similar order but with a one-clock offset \cite{BKS21}}. \rmark{Seven clocks are required: the $Z$ stabilizer starts at the first clock and follows the measurement order $(a,b,c,d,e,f)$ as shown in a green plaquette, while
			the $X$ stabilizer starts at the second clock and follows $(b,c,d,e,f,g)$ as shown in a red plaquette.}
		} \label{fig:color_torus} 
	\end{figure}

\begin{figure}[t!] 
		\centering \centering \includegraphics[width=0.33\textwidth]{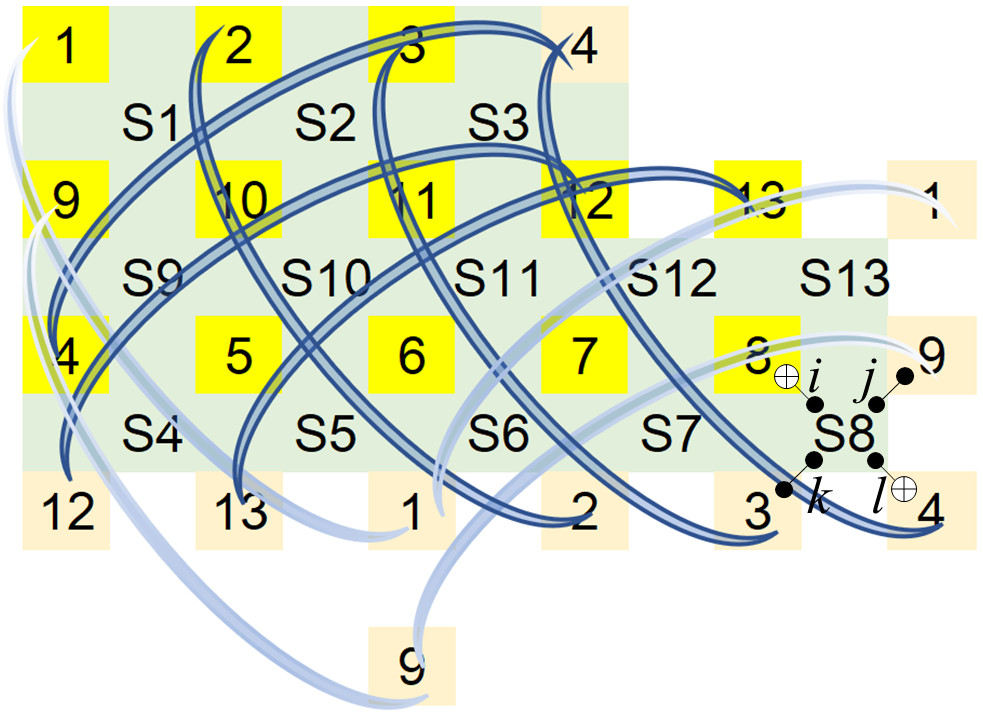}
		\caption{
			The lattice of the $[[L^2+(L-1)^2,\, 1,\, 2L-1]] = [[(d^2+1)/2,\, 1,\, d]]$ twisted XZZX toric code with  $L=3$ ($d=5$), corresponding to a $[[13, 1, 5]]$ code.  
			Its stabilizers  are cyclicly generated by \rmark{$X_1Z_2Z_9X_{10}$}.
			All the stabilizers can be measured in parallel, following a similar order. For example,  stabilizer S8  is \rmark{$X_8Z_9Z_3X_4$, as indicated by $(i,j,k,l) =(8,9,3,4)$} shown in the figure.
			The   labels are  defined similarly to those in Fig.~\ref{fig:toric_lat}.
		} \label{fig:XZZX_torus}		
	\end{figure}

	For clarity, we illustrate the code lattices in Figs.~\ref{fig:toric_lat},~\ref{fig:color_torus}, and~\ref{fig:XZZX_torus}, respectively, and define the corresponding measurement orders for the stabilizers. 
	\bmark{Note that, to maintain full parallelism and ensure a properly functioning syndrome extraction, the measurement gates are applied in a flash-like order; see \cite{Fow+12} for toric and surface codes and \cite{BKS21} for color codes. In particular, although the color code has stabilizer weight~6, it requires 7 depths (clocks). This is because $X$ and $Z$ stabilizers overlap on every color plaquette and thus cannot start simultaneously; see Fig.~\ref{fig:color_torus} and \cite{BKS21} for details. 
		For more codes, see \cite{Kan+25}.}

	\bmark{Given a code and a measurement order, one obtains a syndrome extraction sequence. We illustrate this with a simple example.}
	Figure~\ref{fig:toric_2x2} shows a lattice of a rotated $2\times 2$ toric code with  four stabilizers given by
	$
	\left[\begin{smallmatrix} Z&Z&Z&Z\\ X&X&X&X\\ Z&Z&Z&Z\\X&X&X&X \end{smallmatrix}\right],
	$
	where two of the rows are redundant.
	All four syndrome measurements can be performed in parallel across four depths. Based on the measurement order shown in
	Fig.~\ref{fig:toric_2x2}, the syndrome extraction sequence is as follows: 
	\begin{equation} \label{eq:SE_seq}
	\begin{aligned}
		\begin{matrix} 
			&\text{depth 1} &\text{depth 2} &\text{depth 3} &\text{depth 4}\\
			\text{ancilla 1} &Z_1 &Z_2 &Z_3 &Z_4\\ 
			\text{ancilla 2} &X_2 &X_1 &X_4 &X_3\\ 
			\text{ancilla 3} &Z_3 &Z_4 &Z_1 &Z_2\\
			\text{ancilla 4} &X_4 &X_3 &X_2 &X_1 
		\end{matrix}
	\end{aligned}
	\end{equation}
	This sequence defines a round of syndrome extraction.
	Similar syndrome extraction sequence can be defined for the toric color codes and twisted XZZX toric codes.

	\section{ Circuit-level noise  decoding problem} \label{sec:FT}

	In this section, we explore quantum memory protected by an $n$-qubit topological code, where quantum information is encoded in the codespace. A continuous syndrome measurement procedure is constantly applied to the encoded quantum memory, followed by error correction based on the measurement outcomes.

	We assume a circuit-level noise model in the quantum memory~\cite{DKLP02,RH07,WFSH10}. 	Each potential error source in the syndrome extraction circuit is referred to as a \textit{location}. For ancillary state preparation, qubits are initialized in either the $\ket{0}$ or $\ket{+}$ state. Since the $\ket{0}$ state is immune to $Z$ errors and the $\ket{+}$ state is immune to $X$ errors, we assume that qubits are perfectly initialized and then may subsequently experience bit-flip or phase-flip errors. During syndrome extraction, only qubit measurements in the $X$ and $Z$ bases are needed, and we assume that the binary measurement outcomes are subject to flip errors. Additionally, idle qubits are subject to Pauli errors due to imperfections in the quantum memory, referred to as idle qubit error. Furthermore, we assume that every gate operation, qubit initialization, and qubit measurement takes the same unit of time. Each operation is modeled as ideal, followed by its corresponding error.

	\subsection{Generalized check matrix} \label{sec:FT_chk}
\begin{figure*}[t!]
		\centering 
		\includegraphics[width=\linewidth]{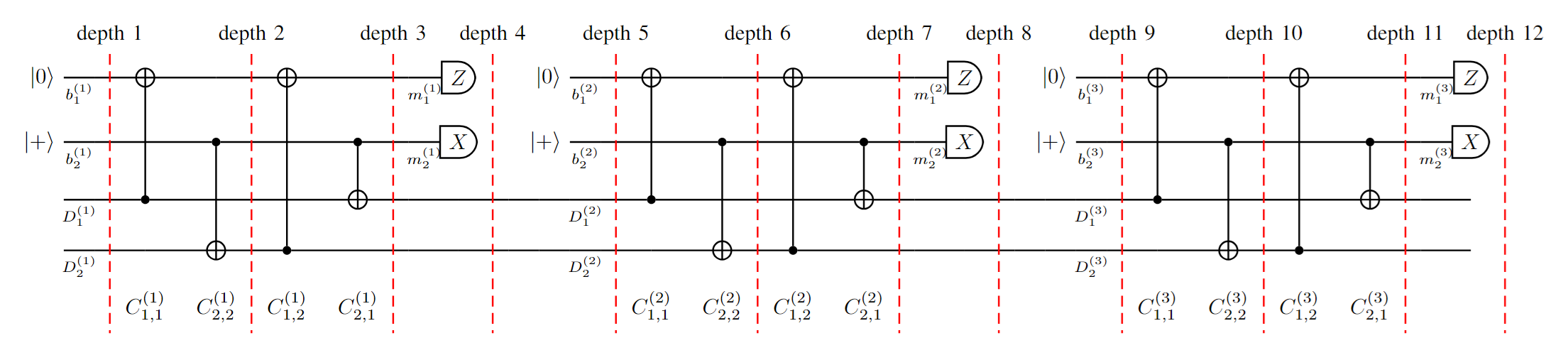}
		\caption[Roots associated to the Cartan matrix]{	
			Circuit for three rounds of raw syndrome extraction, \rmark{focusing on the top-left submatrix $\left[\begin{smallmatrix}Z&Z\\X&X\end{smallmatrix}\right]$ of \eqref{eq:SE_seq}.
			The four gates corresponding to} $Z_1Z_2$ and $X_2X_1$ are measured in two depths, with $Z_1$ and $X_2$ measured simultaneously first, followed by $Z_2$ and $X_1$.
			\rmark{Note that this parallelism arises because we treat the circuit as a subblock of the 4-qubit code in \eqref{eq:SE_seq}, focusing on its top-left portion, rather than directly using it as a 2-qubit code, which would introduce errors.}
			
			%
		} \label{fig:202_circ} \vspace*{\floatsep} \vspace*{\floatsep}
	\end{figure*}

	A syndrome extraction procedure is specified by a circuit description, which details the sequence of gates, measurements, ancillary state preparations, and the qubits involved at each circuit depth. For instance, Fig.~\ref{fig:202_circ} illustrates three rounds of raw syndrome extractions, \rmark{focusing on the top-left submatrix $\left[\begin{smallmatrix}Z&Z\\X&X\end{smallmatrix}\right]$ of \eqref{eq:SE_seq}.}

	In the context of \textit{fault-tolerant error correction}, our objective is to control errors in such a way that the remaining errors after correction can be addressed in the subsequent error correction cycle. 
	Essentially, we only need to ensure that the residual errors   can be  corrected using a perfect syndrome extraction circuit.

	We will define a corresponding decoding problem within the circuit-level noise model. To apply  BP to this circuit-level decoding problem, the first step involves constructing a check matrix that defines the linear relationship between potential errors and their corresponding error syndromes. 
	This extends the concept of the generalized data-syndrome check matrix from~\cite{ALB16,ALB20,KL24}.

	Let $\fM$ be a syndrome extraction circuit of $N$ locations and $M$ syndrome bits. Each location in $\fM$ is assigned an error variable $\cE_k$ based on its type.  In this context, we have the following types of error variables:
	\begin{enumerate}
		\item $b_i\in\{0,1\}$: Ancillary qubit preparation error at ancilla~$i$, corresponding to either $\{I,X\}$ or $\{I,Z\}$.
		\item $C_{ij}\in\{I,X,Y,Z\}^2$: CZ or CNOT gate error at ancilla $i$ and data qubit $j$.
		\item $D_i\in\{I,X,Y,Z\}$: Idle qubit error at data qubit $i$.
		\item $m_i\in\{0,1\}$: Measurement error at ancilla $i$.
	\end{enumerate}
	For illustration, the error variables at each location in Fig.~\ref{fig:202_circ} are indicated,
	where the superscript $\ell$ represents the $\ell$-th round of syndrome extraction.
	\rmark{The index $\ell$ may be omitted to simplify notation when no ambiguity arises.}

	Let $\cE=(\cE_1,\dots,\cE_N)$ be an $N$-dimensional error vector    over a mixed alphabet
	of symbols from $\{I,X,Y,Z\}$, $\{I,X,Y,Z\}^2$, and $\{0,1\}$,  with components  $b_i^{(\ell)}$, $C^{(\ell)}_{ij}$, $D_i^{(\ell)}$, and $m^{(\ell)}_i$ in the syndrome extraction circuit $\fM$.
	Thus, $\cE$ can be understood as a collection of error variables.

	
	We model the syndrome extraction circuit $\fM$ as a function that takes an error vector $\cE$ as input and outputs $M$ syndrome bits $s = \fM(\cE) \in \{0,1\}^{M\times 1}$. 
	Specifically, $\fM$ is a linear function characterized by a matrix $\cH$, such that
	\begin{align}
		s = \fM(\cE)   = \cH \star \cE, ~\label{eq:generalized_syndrome_relation}
	\end{align}
	where $\cH$, with dimensions $M \times N$, is referred to as a \textit{generalized check matrix}.
	The operator $\star$  is  a bilinear form   on  two error vectors,  $\cE=(\cE_1,\dots,\cE_N)$ and $\cF=(\cF_1,\dots,\cF_N)$, over  a mixed alphabet
	of symbols from $\{I,X,Y,Z\}$, $\{I,X,Y,Z\}^2$, and $\{0,1\}$, defined by
	\begin{align}
		\cE\star\cF = \sum_{k: \cE_k=D_i \text{ or } C_{ij}} \cE_k* \cF_k + \sum_{k: \cE_k= b_i \text{ or }m_i}  \cE_k\cdot \cF_k \mod 2. \label{eq:bilinear_form2}
	\end{align}
	This bilinear form extends the commutation bilinear form defined in~Eq.~(\ref{eq:bilinear_form1}).
	We say that $\cE$ and $\cF$ are \textit{orthogonal} if $\cE\star\cF=0$.
	For example, two Pauli operators are orthogonal if they commute.
	Equation~(\ref{eq:generalized_syndrome_relation}) extends the error syndrome calculation for stabilizer codes in the code capacity model~(\ref{eq:stabilizer_syndrome}).

	For simplicity, we will explicitly specify only the nontrivial error components in $\cE$. For instance, if $\cE_k=m_i$, then $\cE = \{ \cE_k = 1 \}$ indicates that a measurement error occurs at  location~$k$, while all the other locations remain error-free.
	\begin{proposition} \label{prop:generalized_check}
		The generalized check matrix $\cH$ can be derived as follows. 
		
		\begin{enumerate}
			\item If  location $k$  corresponds to an ancillary  preparation error $b_i$ or a measurement error $m_i$,  then the $k$-th column of $\cH$ is in $\{0,1\}^M$ and is set to be 
			$\fM(\{\cE_k=1\})$.
			
			\item  If location $k$ corresponds to an idle qubit error $D_i$, then the $k$-th column of $\cH$ is in $\{I,X,Y,Z\}^M$, as it must account for both $X$ and $Z$ syndrome measurements. Therefore, it is set to $Z^u X^v \in \{I,X,Y,Z\}^M$, where $u = \fM(\{\cE_k = X\})$ and $v = \fM(\{\cE_k = Z\})$.
			
			\item  If location $k$ represents a CZ or CNOT gate error $C_{ij}$, there are 16 potential two-qubit Pauli errors and the $k$-th column of $\cH$ is in $\{I,X,Y,Z\}^{M\times 2}$.
			Consequently, we have to establish the syndromes of the error bases $X_i$, $Z_i$, $X_j$ and $Z_j$, which are given as $u=\fM(\{\cE_k=X\otimes I\})$,  $v=\fM(\{\cE_k=Z\otimes I\})$,  $u'=\fM(\{\cE_k=I\otimes X\})$, and $v'=\fM(\{\cE_k=I\otimes Z\})$, respectively. Thus  the $k$-th column of $\cH$ is set to
			\[
			\begin{pmatrix}
				Z^{u_1}X^{v_1} & Z^{u'_1}X^{v'_1}\\
				\vdots& \vdots \\  Z^{u_M}X^{v_M} & Z^{u'_M}X^{v'_M} 
			\end{pmatrix}, 
			\]
			which  should be  interpreted as a vector of length $M$ over $\{I,X,Y,Z\}^2$.
		\end{enumerate}
	\end{proposition}

\begin{figure*}[t!]
		\includegraphics[width=\linewidth]{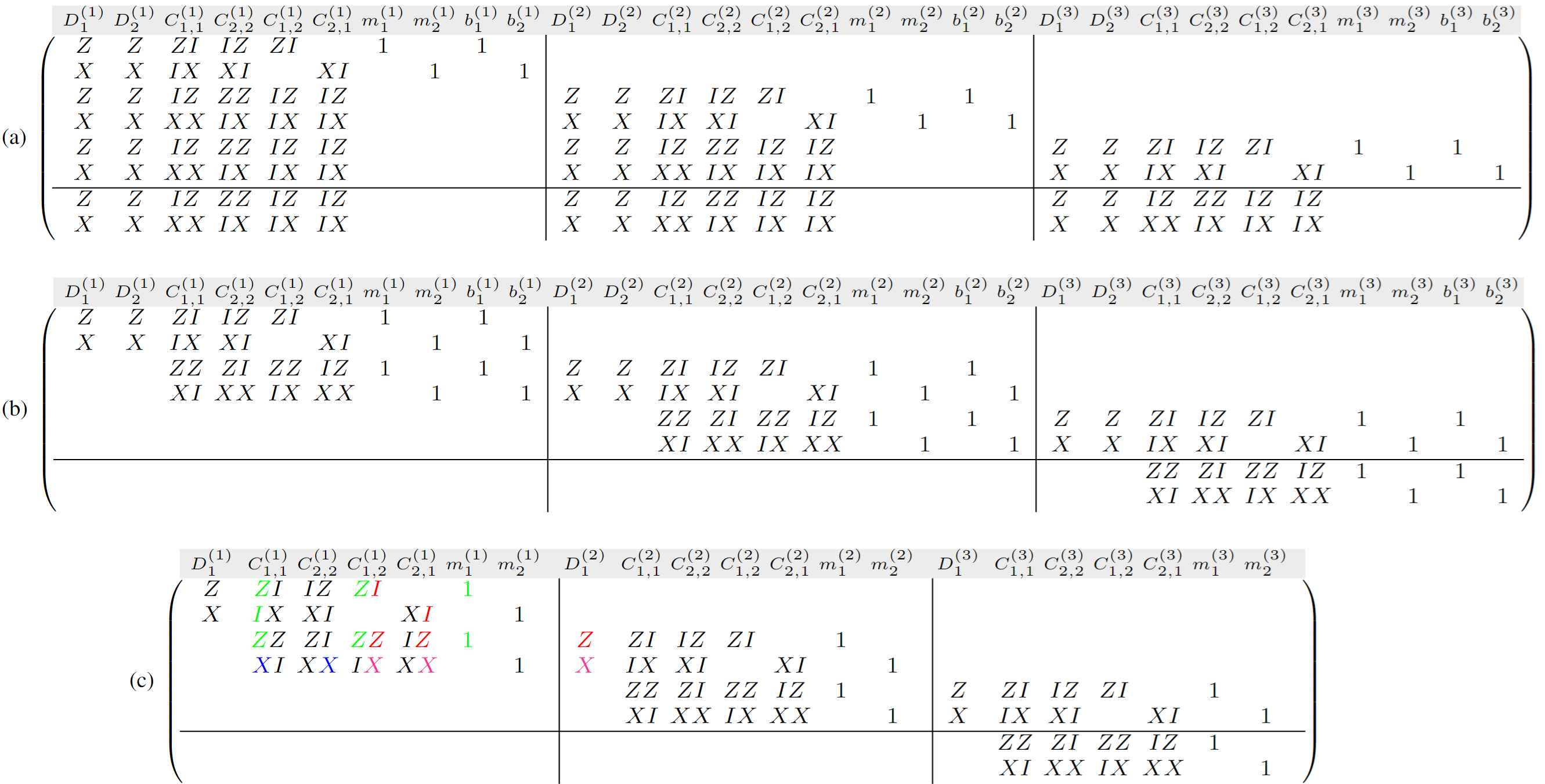}
		\caption{ 
			Matrix (a)  represents the generalized check matrix  corresponding to the syndrome extraction circuit in Fig.~\ref{fig:202_circ} by Proposition~\ref{prop:generalized_check}.
			The columns are indexed by the location variables. 
			An empty entry in the matrix  represents a trivial element, either $I$ or $0$.
			The last two rows of a matrix represent a round of perfect syndrome extraction, which is used for virtual decoding to verify whether the residual error corresponds to an \bmark{uncorrectable} logical error.
			Matrix (b) is a sparse matrix derived from Matrix (a) using the procedure outlined in Lemma~\ref{lemma:row_operation}.
			Matrix (c) is obtained by merging degenerate locations and deleting the corresponding columns from Matrix (b) by Lemma~\ref{lemma:merging_degenerate}. Several examples of degenerate error consolidation are highlighted in different colors.
		} \label{fig:FT_mtx}
	\end{figure*}

	Following the procedure outlined above,  we can construct a generalized check matrix so that  Eq.~(\ref{eq:generalized_syndrome_relation}) holds.
	Note that an entry corresponding to a two-qubit Pauli error  can be interpreted as two correlated single-qubit Pauli errors.
	Thus this generalized check matrix further extends the concept of quantum data-syndrome codes~\cite{ALB20,KL24}.
	For illustration, the syndrome extraction circuit depicted in Fig.~\ref{fig:202_circ}   corresponds to the generalized check matrix $\cH$  shown as Matrix~(a) in Fig.~\ref{fig:FT_mtx}.

	Let $\cR(\fM,\cE)\in\{I,X,Y,Z\}^n$ denote  the residual Pauli error on the data qubits of the $n$-qubit code after the execution of circuit $\fM$.
	This error can be determined by tracing the evolution of $\cE$ within $\fM$.

	\begin{definition}{\bf (Circuit-level decoding problem)}
		Given  a syndrome extraction circuit $\fM$ for an $n$-qubit  stabilizer code defined by a stabilizer group $\cS$
		and  $M$ syndrome bits  $s=\fM(\cE)\in\{0,1\}^M$   for a set of errors $\cE$, 
		output an estimate of the error set $\hat{\cE}$  such that $\fM(\hat{\cE})=s$
		and the \rmark{residual miscorrection} $\cR(\fM,\hat{\cE})\cR(\fM,\cE)$ can be corrected by an ideal quantum error correction procedure.
		\qed     \label{def:cldp}
	\end{definition}

	This concept of degenerate errors in the  code capacity model can be extended to the circuit-level noise model as follows.

	\begin{definition}  \label{def:degeneracy}
		Two errors $\cE$ and $\cF$  are said to be \textit{degenerate} with respect to   a syndrome extraction circuit~$\fM$ 
		if they produce the same error syndrome $\fM(\cE)=\fM(\cF)$ and  result in the same residual errors on the data qubits, $\cR(\fM,\cE)=\cR(\fM,\cF)$, up to stabilizers.
		\qed
	\end{definition}

\bmark{
For identifying degenerate errors, we associate each binary error variable with a residual error pattern. Specifically, we use an array to record the residual errors corresponding to each binary variable composing all location error variables. 
This array serves as a reference for determining degeneracy among error variables. 
Such an array can be generated during the construction of the generalized check matrix.
}

	\subsection{Belief propagation decoding} \label{sec:dec}

	The generalized check matrix $\cH$ and the error syndrome relation in~(\ref{eq:generalized_syndrome_relation})
	induce a Tanner graph with various types of variable nodes and check nodes 
	which allow us to develop a BP decoding algorithm.

	In general, we are given an $M\times N$ generalized check matrix $\cH$ and syndrome bits $s\in \{0,1\}^M$ for decoding.
	Herein, we generalize the  data-syndrome BP algorithms in~\cite{KCL21,KL24} and propose a decoding algorithm based on MBP in \cite{KL22} to solve the
	circuit-level decoding problem.

	We consider  error vector $\cE = (\cE_1, \dots, \cE_N)$, where $\cE_k$ is a variable of type $D_i, C_{i,j}$ or $m_i$ (or $b_i$) defined over $\{I, X, Y, Z\}$, $\{I, X, Y, Z\}^2$, or $\{0,1\}$, respectively.  These variables are generated according to the following distributions.
	
	\begin{enumerate}
		\item  Each single-qubit Pauli variable $D_j$ is independently generated with   depolarizing rate $\epsilon \in [0, 3/4)$, following the distribution $(p_j^I, p_j^X, p_j^Y, p_j^Z) = ({1-\epsilon},\, \epsilon/3,\, \epsilon/3,\, \epsilon/3)$.
		
		\item Each two-qubit Pauli variable $C_{ij}$ is independently generated with   depolarizing rate $\epsilon \in [0, 3/4)$ so that $C_{ij}=I$ with probability $1-\epsilon$ and $C_{ij}$ is a non-identity two-qubit Pauli with probability $\epsilon/15$.
		
		\item  Each syndrome bit error $m_j$  or ancillary preparation error $b_j$ is an independent bit-flip or phase-flip error with rate $\epsilon_\text{b} \in [0, 1/2)$, following the probability distribution $(q_j^{(0)}, q_j^{(1)}) = {(1-\epsilon_\text{b}, \epsilon_\text{b})}.$
		

	\end{enumerate}
	Consequently, we have initial log-likelihood ratio (LLR) vectors $\Lambda_1,\dots,\Lambda_{N}$ for each variable  as follows.
	\begin{enumerate}
		\item If $\cE_k$ is  syndrome bit error   or ancillary preparation error, then  $\Lambda_k$ is a scalar given by
		\begin{align}
			\Lambda_{k} =&   \ln \frac{1-\epsilon_\text{b}}{\epsilon_\text{b}}.
		\end{align}
		
		
		\item If $\cE_k$ is a single-qubit Pauli variable $D_j$,  then $\Lambda_k=(\Lambda_k^X, \Lambda_k^Y, \Lambda_k^Z)\in\mR^3$, where
		\begin{align}
			\Lambda_k^W = \ln \frac{3(1-\epsilon)}{\epsilon} ~\text{for } W\in\{X,Y,Z\}.
		\end{align}

		\item If $\cE_k$ is a two-qubit Pauli variable $C_{ij}$,  then $\Lambda_k=(\Lambda_k^{XI}, \Lambda_k^{YI}, \Lambda_k^{ZI},\dots, \Lambda_k^{XZ}, \Lambda_k^{YZ}, \Lambda_k^{ZZ})\in\mR^{15}$, where
		\begin{align}
			\Lambda_k^W =  \ln \frac{15(1-\epsilon) }{\epsilon} ~\text{for } W\in\{I,X,Y,Z\}^2\setminus\{I\}.
		\end{align}
	\end{enumerate}

	The Tanner graph consists of $N$ variable nodes corresponding to $\cE_1,\dots,\cE_N$ and $M$ check nodes corresponding to the rows of $\cH$.
	Define the set of neighboring nodes for a check node $i$ by
	$
	\cN(i) = \{j: \cH_{ij}\neq I \text{ or } \cH_{ij}\neq 0\},
	$
	and similarly for a variable node $j$ by
	$
	\cM(j) = \{i:  \cH_{ij}\neq I \text{ or } \cH_{ij}\neq 0\}.
	$
	BP  performs iterative message passing on the Tanner graph to generate LLR vectors $\Gamma_1, \dots, \Gamma_N$ for the error estimate based on a given error syndrome. We propose our BP algorithm, called SM-BP, for solving this circuit-level noise decoding problem, as detailed in Algorithm~\ref{alg:SM-BP}.  The SM-BP algorithm performs binary, quaternary, and 16-ary message computations, resulting in a slightly higher complexity.

	\begin{algorithm}[H]
	\small
		\begin{flushleft}
			
			\caption{: SM-BP} \label{alg:SM-BP}
			\textbf{Input}: 
			An $M\times N$ generalized check matrix $\cH$ over  a mixed alphabet
			of symbols from $\{I,X,Y,Z\}$, $\{I,X,Y,Z\}^2$, and $\{0,1\}$,
			an error syndrome $s\in\{0,1\}^M$, 
			an integer $T_{\max}>0$, 
			a real scalar $\alpha>0$, 
			and initial LLRs $\Lambda_{1},\dots,\Lambda_N$. \\
			{\bf Initialization:} 
			\begin{algorithmic}
				\For{$j\in\{1,2,\dots,N\}$ and $i\in\cM(j)$}
				\If{ $\cE_j$ is a Pauli error,} {} let 
				$\Gamma_{j\to i} = \lambda_{\cH_{ij}} (\{\Lambda_j^{V}\})$.
				\Else   {} let $\Gamma_{j\to i} = \Lambda_j$.
				\EndIf
				\EndFor
			\end{algorithmic}

			{\bf Steps:}
			\begin{itemize}
				\item 
				
				{\bf Horizontal Step (Update C-to-V Messages):}  
				\begin{algorithmic}
					\For{$i\in\{1,2\dots,M\}$ and $j\in\cN(i)$}
					\begin{align}
						\Delta_{i\to j} = (-1)^{s_i}\underset{j'\in\cN(i)\setminus \{j\}}{\boxplus} \Gamma_{j'\to i}.   \label{eq:delta_mn_BP4}
					\end{align}
					\EndFor
				\end{algorithmic}

				\item 	{\bf Vertical Step (Marginal Distribution Part):}
				\begin{algorithmic}
					\For{ $j\in\{1,\dots,N\}$}
					\If{ $\cE_j$ is a Pauli error,}
					\State  \begin{equation}			\displaystyle 
					\Gamma_{j}^W = \Lambda_j^W + \frac{1}{\alpha} \sum_{i\in\cM(j) \atop W\star \cH_{ij} =1} \Delta_{i\to j}, \label{eq:alpha1}
							\end{equation} 
					\text{\quad} for $W\neq I$.
					\Else   {\quad}  
					\begin{equation} \displaystyle 
					\Gamma_{j} = \Lambda_j + \frac{1}{\alpha} \sum_{i\in\cM(j)} \Delta_{i\to j}. \label{eq:alpha2}
					\end{equation} 
					\EndIf
					\EndFor
				\end{algorithmic}

				\item {\bf Hard Decision (and Update V-to-C Messages)}:
				\begin{algorithmic}  
					\If{$\cE_j$ is a Pauli error,}
					\If{$\Gamma_{j}^W > 0$ for all $W\neq I$,} {} $\hat \cE_j \gets I$.
					\Else {} $\hat \cE_j\gets \argmin\limits_{W\neq I} \Gamma_{j}^W$.
					\EndIf
					\Else {}
					\If{$\Gamma_{j} > 0$,}	 {} $\hat \cE_j = 0$.
					\Else  {} $\hat \cE_j = 1 $.
					\EndIf
					\EndIf 	
					
					\State 		 Let $\hat \cE = (\hat \cE_1,\dots,\hat \cE_{N})$.
					\If{$\cH\star \hat{\cE}= s$,} {} \Return ``CONVERGE'';
					\ElsIf{the iteration count reaches $T_{\max}$,} {} halt and \Return ``FAIL'';
					\Else {} Update V-to-C Messages:
					\For{ $j\in\{1,2,\dots,N\}$ and $i\in\cM(j)$,}
					\If{ $\cE_j$ is a Pauli error,}{}  
					\begin{align} 
						\Gamma_{j\to i}^W &= \Gamma_j^W -  (W\star \cH_{ij})  \Delta_{i\to j}, \text{ for }~ W\neq I, \label{eq:gammaW_ji}\\
						\Gamma_{j\to i} &= \lambda_{\cH_{ij}} (\Gamma_{j\to i}^X, \Gamma_{j\to i}^Y, \Gamma_{j\to i}^Z).	\label{eq:gamma_ji}
					\end{align}	
					\Else   {\quad}  	$\Gamma_{j\to i} = \Gamma_j - \Delta_{i\to j}.$
					\EndIf
					\EndFor
					\State Repeat from the horizontal step.
					\EndIf
					
				\end{algorithmic}
				
			\end{itemize}
			
		\end{flushleft}
	\end{algorithm}

\bmark{
In Eqs.~(11) and~(12), all messages are scaled by a factor of $\frac{1}{\alpha}$. The parameter $\alpha$ therefore controls the effective step size in message passing, similar to the scaling used in MBP~\cite{KL22}. 

When $\alpha > 1$, the messages are attenuated, which stabilizes the updates and increases the relative influence of prior iterations. When $\alpha < 1$, the messages are amplified, leading to more aggressive updates that can help the decoder escape local trapping effects.
}

	At each iteration, the variable-to-check (V-to-C) messages $\Gamma_{j\to i}$ and check-to-variable (C-to-V) messages $\Delta_{i\to j}$ are computed as described in the algorithm. A hard decision on the error estimate is made at every iteration to check if the syndrome matches.   If the syndrome does not match, the process may continue until the maximum of $T_{\max}$ iterations is reached, at which point a failure is declared.

	Given that the error syndromes are binary, Eq.~(\ref{eq:generalized_syndrome_relation}) and the LLR distributions collectively determine whether $\cE_k$ and an entry of $\cH$ are more likely to be orthogonal. This likelihood can be quantified by a single LLR value.
	Consequently, all exchanged messages are scalars.
	
	For the calculation of messages in~Algorithm~\ref{alg:SM-BP},  a real-valued function $\lambda_W$ for $W\in\{X,Y,Z\}$  is defined by
	\begin{align}
		\lambda_W(\{\gamma^V:V\in\{X,Y,Z\}\}) = \ln \tfrac{1+ e^{-\gamma^{W}}}{ e^{-\gamma^{X}}+e^{-\gamma^{Y}}+e^{-\gamma^{Z}}-e^{-\gamma^{W}} } \label{eq:la}
	\end{align} 
	for $\{\gamma^V:V\in\{X,Y,Z\}\}=\{\gamma^X,\gamma^Y,\gamma^Z\}$ \bmark{as a vector in $\mR^3$}.
	It can also be extended for $W\in\{I,X,Y,Z\}^2\setminus\{I\}$ by
	\begin{align} 
		\lambda_W( \{\gamma^{V}\}) =& \ln\left(  \displaystyle 1+ \sum_{V\in\{I,X,Y,Z\}^2\setminus\{I\}: V\star W=0} e^{-\gamma^{V}}\right)\notag \\
		&-\ln \left( \displaystyle \sum_{ V\in\{I,X,Y,Z\}^2: V\star W=1} e^{-\gamma^{V}} \right) \label{eq:la2}
	\end{align} 
	for $\{\gamma^{V}\}=\{\gamma^{XI},\gamma^{YI},\gamma^{ZI},\dots,\gamma^{ZZ} \}$ \bmark{as a vector in $\mR^{15}$}.
	We will use  $\{\gamma^{V}\}$ for both   $V\in \{X,Y,Z\}$ and $V\in \{I,X,Y,Z\}^2\setminus \{I\}$
	without specifying the domain explicitly when it can be inferred from the context.
	
	The check-node computation is performed using the operator $\boxplus$, defined for a set of $\ell$ real scalars $a_1, a_2, \dots, a_\ell \in \mR$, as follows:
	\begin{equation} \label{eq:bsum} 
		\overset{\ell}{\underset{j=1}{\boxplus}} \, a_j = 2\tanh^{-1} \textstyle \left( \prod_{j=1}^\ell \displaystyle \tanh\frac{a_j}{2} \right).
	\end{equation}

	\begin{remark}
		In case a two-qubit gate error location $C_{i,j}$ can be represented by two independent single-qubit error variables
		$C_{i,j}^{a}$ and $C_{i,j}^{d}$, where the superscripts $a$ and $d$ represent the ancilla and data qubits, respectively,
		the generalized check matrix can be treated as a matrix over $\{I,X,Y,Z\}$ and $\{0,1\}$ and the decoding problem reduces to a generalized data-syndrome decoding problem~\cite{KL24}, which can be handled by SM-BP without 16-ary variables and 16-ary calculations.
	\end{remark}

	\section{Problem reduction} \label{sec:reduction}

	In the circuit-level decoding problem, multiple error locations during the syndrome extraction process typically lead to a large generalized check matrix, which directly affects the instantaneous decoding complexity. 
	For the SM-BP algorithm, this complexity generally scales as $O(N \gamma T_{\max})$, where $N$ is the number of variables, $\gamma$ is the average column weight of the generalized check matrix, and $T_{\max}$ is the maximum number of iterations~\cite{KL21, KL22, KL24}. 
	In the following, we illustrate how the SM-BP algorithm can be applied to simplify the circuit-level decoding problem while keeping the computational cost manageable.

	\subsection{BP complexity for the circuit-level decoding problem}
	
		Consider an $n$-qubit stabilizer code with $m$ stabilizers being continuously measured. The measurement outcomes are collected over a specific duration, termed a \textit{(decoding) window}. A smaller window size is desirable, as it yields a smaller generalized check matrix. 

	If each  stabilizer measured has a weight $w$, each round of  raw syndrome extraction requires  $mw$  CNOT or CZ gates, $m$~ancilla preparations, and $m$ measurements.
	If the decoding procedure is applied to every window of $r$ rounds of syndrome extraction,   a corresponding generalized check matrix $\cH$ is of size $M \times N$, where $M = rm$ and $N = r(n + mw + 2m)$, with columns corresponding to quaternary, 16-ary, and binary entries.

	For topological codes with minimum distance $d = O(\sqrt{n})$ and constant-weight stabilizer measurements, such as toric or XZZX codes ($w=4$) or 6.6.6 toric color codes ($w=6$), we have $m = O(n)$ stabilizers measured over a window of $r = O(d)$ rounds, giving a total number of error locations $N = O(d^3) = O(n^{1.5})$. Typically, choosing $T_{\max} = O(\log N)$ suffices \bmark{to effectively propagate the beliefs across the graph}.
	
	In these codes, each data qubit is checked by a constant number of stabilizers per round. However, Pauli errors accumulate on the data qubits and trigger syndrome measurements in subsequent rounds.
Consequently, the column weights corresponding to these Pauli errors depend on the window size, as shown by the columns corresponding to locations of types $D_j$ and $C_{i,j}$ in Matrix (a) in Fig.~\ref{fig:FT_mtx}. 
	As a result, the generalized check matrix has a high mean column weight of $O(d)$ and it is not sparse. The following lemma shows how these issues can be managed.

	\begin{lemma}  \label{lemma:row_operation}
		The complexity of SM-BP for an $n$-qubit quantum code with $O(1)$ stabilizer weights and a window of $O(\sqrt{n})$ rounds of syndrome extraction is $O(n^{1.5}\log n)$, achieved using a sparse generalized check matrix.
		Additionally, each variable node connects to at most two rounds of check nodes in the corresponding Tanner graph.
	\end{lemma}

	\begin{proof}

		We begin by dividing the rows of the given generalized check matrix $\cH$, as defined in Proposition~\ref{prop:generalized_check}, into $r$ blocks corresponding to $r$ rounds of syndrome extraction. Let $R_{i,j}$ denote the row operation that adds the $i$-th block to the $j$-th block.
		(The addition of two Pauli entries should be interpreted as their multiplication within the Pauli group.)
		
		First, consider the data errors corresponding to $D_j^{(1)}$, which will accumulate on the data qubits. As a result, the entries in the first block corresponding to $D_j^{(1)}$ will appear in the other blocks as well. To eliminate these entries from the second to the $r$-th blocks, we apply the row operations $R_{1,2}, R_{1,3}, \dots, R_{1,r}$.

		Next, consider the gate errors corresponding to $C_{i,j}^{(1)}$. These errors will be detected in the second to the $r$-th rounds of syndrome extraction. However, the entries in these blocks may differ from those in the first block, as not all gate errors are detected by the first round of syndrome measurements. Thus,  we apply the row operations $R_{2,3}, R_{2,4}, \dots, R_{2,r}$ to eliminate these entries from the third to the $r$-th blocks, which also simultaneously eliminates the entries corresponding to $D_j^{(2)}$ in the third to the $r$-th blocks.
		
		By continuing this process and applying $R_{i,j}$ for $j > i$, we ensure that the columns corresponding to $D_j^{(\ell)}$ remain nontrivial in only one block, while the columns corresponding to $C_{i,j}^{(\ell)}$ remain nontrivial in only two blocks. Since each stabilizer checks a constant number of data qubits, the weight of these columns remains constant.
		
		Finally, the ancilla preparation errors and measurement errors do not propagate to other rounds of syndrome extraction. However, due to the row operations, their effects will be copied to one additional block. Consequently, the column weights of $b_j^{(\ell)}$ or $m_j^{(\ell)}$ will be two.

		By combining these results, we obtain a row operation $R$ that sequentially applies  $R_{1,2},\dots,R_{1,r},~ R_{2,3},\dots,R_{2,r},~ \dotsb,~ R_{r-1,r}$ 
		(or equivalently $R_{r-1,r},\, R_{r-2,r-1},\, \dots,\, R_{3,2}, R_{1,2}$) to transform the given generalized check matrix $\cH$ into a sparse matrix $R(\cH)$ with constant column weights. Therefore, Eq.~(\ref{eq:generalized_syndrome_relation}) becomes:
		\begin{align} 
			R(s) = R(\cH) \star \cE, ~\label{eq:generalized_syndrome_relation2}
		\end{align}
		
		This transformation allows the SM-BP algorithm to achieve a complexity of $O(N \log N) = O(d^3 \log d) = O(n^{1.5} \log n)$ using the new generalized check matrix $R(\cH)$.
	\end{proof}

	For illustration, in Fig.~\ref{fig:FT_mtx}, we apply a row operation to Matrix (a) to derive the sparse Matrix (b) using Lemma~\ref{lemma:row_operation}. Matrix (b) takes the form of a lower block bidiagonal Toeplitz matrix, with the same block repeated along each bidiagonal line. This structure allows the matrix to be constructed based on just two rounds of syndrome extraction.
	
	In the following, we will consistently work with this generalized check matrix in the form of a lower block bidiagonal Toeplitz matrix. Additionally, the Tanner graph induced by this generalized check matrix will be referred to as a space-time Tanner graph. This term is used because an error location affects at most two rounds of syndromes, meaning a check node connects to variable nodes from the same round and possibly the previous round of the syndrome extraction circuit. As a result, we can deploy  the variable and check nodes of the Tanner graph in a two-dimensional time-space domain.

\begin{figure}[t!]	   
		\[
		\includegraphics[width=1.025\linewidth]{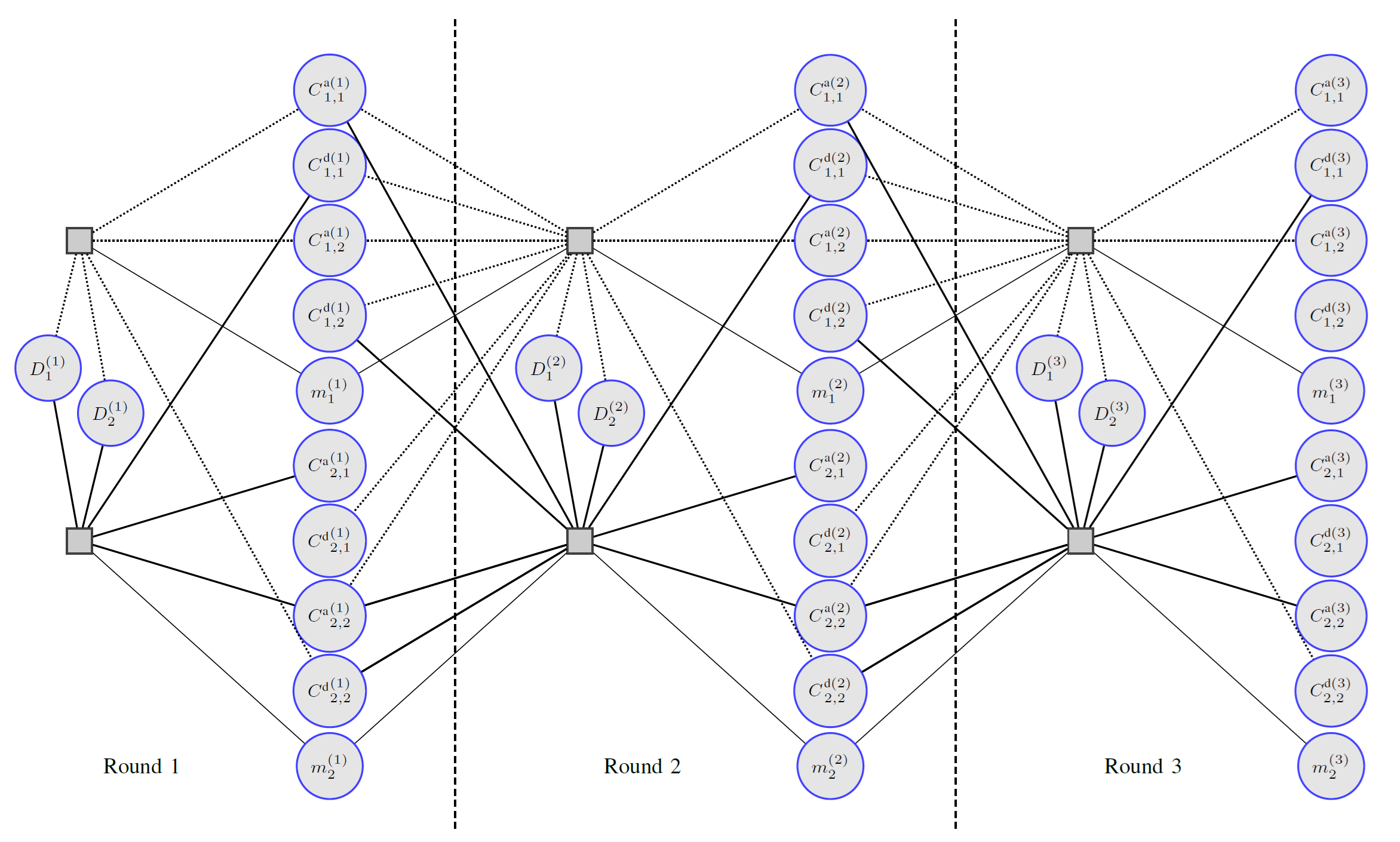}
		\]
		\caption{
			The space-time Tanner graph induced from Matrix~(c) in Fig.~\ref{fig:FT_mtx}. 
			\rmark{Here, each 16-ary variable is decomposed into two quaternary variables for clarity.}
		\bmark{Each square represents a check node, and each circle represents an error-variable node.}
			Time progresses from left to right.
			The style of each edge represents its type: a thin solid line connects to a binary variable node $m_i^{(\ell)}$, a thick solid line represents an edge of type~$X$, and a dotted line represents an edge of type~$Z$.		
		} \label{fig:space_time_Tanner} 
	\end{figure}

	Figure~\ref{fig:space_time_Tanner} illustrates the space-time Tanner graph of the example in Fig.~\ref{fig:FT_mtx} with three rounds of syndrome extraction, where \rmark{each CNOT error variable is decomposed into two single-qubit error variables for clarity}. The horizontal direction represents the timeline from left to right, while the vertical direction represents the space domain, with the variable nodes corresponding to a syndrome extraction round. Note that some variable nodes are omitted because they can be merged, as will be explained in the following subsection.

	\subsection{Merging  degenerate error locations of the same type}
	
	Degenerate errors, as defined in Def.~\ref{def:degeneracy}, represent errors that do not need to be distinguished during the decoding process. By identifying certain degenerate errors in advance, we can simplify the generalized check matrix.

	Two error locations of the same type are considered degenerate if all the errors they produce are degenerate.

	\begin{lemma} \label{lemma:merging_degenerate}
		
		 \rmark{Consider two degenerate error locations with initial error rates $\epsilon_1$ and $\epsilon_2$, respectively. If these variables belong to a $q$-ary alphabet, their merged location has an effective error rate given by: $$\epsilon' = \epsilon_1 + \epsilon_2 - \frac{q}{q-1} \epsilon_1 \epsilon_2$$
		 	for $q\in\{2,4,16\}$.
		 	
%
}
		 
%
%
		When $\epsilon$ is small, second-order terms can be neglected.
	\end{lemma}

	For example,  in Matrix (b) in Fig.~\ref{fig:FT_mtx},  an $X$ error at $D_1^{(1)}$ is equivalent to an $X$ error at $D_2^{(1)}$. Similarly, a $Z$ error at $D_1^{(1)}$ is equivalent to a $Z$ error at $D_2^{(1)}$. Therefore, these two locations are degenerate and can be merged into a single location.

	\rmark{
	\begin{corollary} \label{cor:merging_degenerate}
		In a raw stabilizer measurement circuit for a CSS code with purely $Z$-type or purely $X$-type stabilizers, the ancilla preparation error location and the measurement error location are degenerate.
	\end{corollary}
}

\bmark{

	\begin{proof}
		Consider a raw $Z$-type stabilizer measurement circuit. Fig.~\ref{fig:meas}(b) shows the case $w=4$ as an example. An ancilla qubit is prepared in the state $\ket{0}$, followed by a sequence of CNOT gates with the data qubits as controls and the ancilla as the target, and finally measured in the computational basis.
		
		The only nontrivial ancilla preparation error in this setting is an $X$ error. Such an $X$ error on the ancilla does not propagate through the CNOT gates to the data qubits. However, it flips the outcome of the final computational-basis measurement on the ancilla qubit.
		
		Therefore, an ancilla preparation $X$ error has no logical effect on the data qubits and produces the same syndrome as a measurement error. Consequently, the ancilla preparation error location is degenerate with the measurement error location under the circuit-level noise model.
		
		An analogous argument holds for a raw $X$-type stabilizer measurement circuit, completing the proof.
	\end{proof}
}

	According to Lemma~\ref{lemma:merging_degenerate} and Corollary~\ref{cor:merging_degenerate},  we derive  Matrix~(c)  by  deleting columns from  Matrix (b)	in Fig.~\ref{fig:FT_mtx}.

	\subsection{Probabilistic Error Consolidation}
	
		In the previous subsection, we discussed how two degenerate error locations can be merged into a single error location. However, in many cases, different error locations may induce degenerate errors, but cannot be directly merged at the level of circuit locations. For example, a CNOT gate may produce errors that are degenerate with idle qubit errors or measurement errors.

	\rmark{To address this, we propose a probabilistic error consolidation procedure that operates at the level of error variables rather than error locations. The key idea is to first identify groups of degenerate errors, and then perform probability consolidation within each group. 
		Such error variables are said to be \textit{probabilistically consolidated}, and BP can naturally accommodate the resulting modified input distributions.}

A conventional check matrix for an $n$-qubit quantum stabilizer code is a binary matrix that represents stabilizers by mapping single-qubit Pauli matrices $I,X,Z,Y$ to two binary bits $(0|0),(1|0),(0|1),(1|1)$, respectively. Similarly, a generalized check matrix can be represented in binary by mapping two-qubit Pauli operators to 4-bit strings, where each bit column corresponds to a Pauli $X$ or $Z$ variable. 

We propose a systematic procedure for consolidating error distributions based on this binary representation of the generalized check matrix, as summarized in Algorithm~\ref{alg:merge2}.

\rmark{
Our consolidation procedure operates in two phases: \textit{grouping} and \textit{distribution assignment}. 
In the grouping phase, we construct groups of error components that are potentially degenerate. 
This grouping is performed hierarchically, where larger-alphabet error variables are processed first, ensuring that probability correlations are largely preserved. 
In the distribution assignment phase, probabilities are consolidated within each group following a hierarchical decomposition strategy. 
The resulting modified distributions can be directly used as input for BP decoding.
}

\begin{proposition} \label{prop:C16}
	\rmark{Algorithm~\ref{alg:merge2} produces marginal distributions for the grouped binary, quaternary, and 16-ary error variables.} These marginals can then be combined to form a product-form approximation of the original quaternary or 16-ary error distributions.
\end{proposition}

As will be shown in Section~\ref{sec:sim}, this error consolidation procedure, which decouples correlations in the Pauli operators, not only reduces the excess columns in the matrix but also decreases the number of short cycles in the generalized check matrix. 

	\begin{algorithm}[H]
		\begin{flushleft}

			\caption{: Probabilistic error consolidation} \label{alg:merge2}
			\textbf{Input}: 
			A syndrome extraction circuit $\fM$, an $M\times N$ generalized check matrix $\cH$ over  a mixed alphabet
			of symbols from $\{I,X,Y,Z\}$, $\{I,X,Y,Z\}^2$, and $\{0,1\}$, 
			and initial error rates $\epsilon_{1},\dots,\epsilon_N$. \\
			
			{\bf Steps:}

			\begin{algorithmic}[1]
				
				\State Merge degenerate error locations of the same type following Lemma~\ref{lemma:merging_degenerate}. 
				\rmark{	At this stage, all 16-ary degenerate errors are merged.

				\State  Identify and group degenerate quaternary variables. 
				If a component quaternary variable is found to be degenerate with an existing group, decompose the corresponding 16-ary variable into two quaternary variables and assign them to the appropriate groups.
				
				\State Identify and group degenerate binary variables. In particular, decompose a 16-ary variable or a quaternary variable into binary variables if any of the component binary variables is degenerate with  
				any existing group of binary variables.
								
								\State	Within each group, consolidate the initial distributions by assigning the total probability mass to a representative variable, while setting the others to zero.
		}
				
				\State Return the updated (pseudo) error distributions for each column.
				
			\end{algorithmic}

		\end{flushleft}
	\end{algorithm}

	\bmark{To provide a concrete demonstration of Algorithm~\ref{alg:merge2}, we apply the consolidation procedure to a specific circuit-level scenario.}

	\begin{example}
		Consider  Matrix (c) in  Fig.~\ref{fig:202_circ}.
		\begin{enumerate}
			\item 	 (identical single-qubit columns)
			\rmark{
				The errors $IX$, $IZ$, and $IY$ at $C_{2,1}^{(1)}$ are degenerate with the single-qubit Pauli errors $X$, $Y$, and $Z$ at $D_1^{(2)}$, respectively. 
				Hence, they can be probabilistically consolidated.
			}
			Similarly, the errors $IX$, $IZ$, and $IY$ at $C_{1,2}^{(1)}$ are degenerate with the errors $X$, $Y$, and $Z$ at $D_1^{(2)}$ and can therefore be merged.
			The corresponding columns in Matrix~(c) are highlighted in red.

			\item  	The errors $XI$  at $C_{1,1}^{(1)}$ and $C_{1,2}^\text{(1)}$ can be probabilistically consolidated with the bit-flip error at $m_1^{(1)}$.
			The corresponding components in Matrix (c) are highlighted  in green.
			
			\item  	  
			
			The  error $ZI$ at $C^\text{(1)}_{1,1}$ and  the errors $IZ$ at $C^\text{(1)}_{1,2}$, $C^{(1)}_{2,1}$,   $C^{(1)}_{2,2}$, and $C^{(2)}_{1,1}$      can be probabilistically consolidated with the error  $Z$ at $D_1^{(2)}$.
			Note that  if errors $IZ$ at $C^\text{(1)}_{1,2}$ and $C^{(1)}_{2,1}$ have already been probabilistically consolidated with error  $Z$ at $D_1^{(2)}$ as in 1),
			they can be skipped in subsequent steps. 
			The corresponding components in Matrix (c) are highlighted  in blue.
			The components   highlighted  in magenta are interpreted as both blue and red. 
			
			\item 	 The error $ZI$ at $C_{1,2}^\text{(1)}$ can be ignored as it is  degenerate  to no error.

		\end{enumerate}
	\end{example}

	Additionally, we have another approximation using quaternary variables only.
	\begin{proposition} \label{prop:C4}
		A two-qubit gate error location $C_{i,j}$ can be approximated by two single-qubit error variables
		$C_{i,j}^{a}$ and $C_{i,j}^{d}$. 
		The distributions for the quaternary variables can then be approximated using the marginal  distributions for each single-qubit 
		$X$ and $Z$  error  generated by 	Algorithm~\ref{alg:merge2}.
	\end{proposition} 
	By generating product distributions for these quaternary variables instead of the original 16-ary variables, we can simplify the decoding complexity in SM-BP. 
	Figure~\ref{fig:space_time_Tanner} illustrates the Tanner graph induced from Matrix~(c) in~Fig.~\ref{fig:FT_mtx}, using this approximation.


	\section{ Lifetime of a quantum memory   } \label{sec:memory}
	\rmark{The \textit{lifetime simulation}} of a quantum memory begins with a noiseless encoded state and proceeds with continuous syndrome extraction. After collecting syndromes over a window of $r$ rounds, a decoding procedure is applied to determine a correction operator.
	\bmark{ The corrected state then undergoes subsequent rounds of syndrome extraction, followed by decoding. This process repeats until the quantum memory fails, i.e., when a logical error occurs.}

	\bmark{
		This lifetime simulation differs from conventional one-shot decoding, where each shot starts from a clean encoded state and thus neglects the fact that residual data errors due to miscorrections may persist and propagate to subsequent QEC cycles, as described in Def.~\ref{def:cldp}.
	}

	\begin{definition} \label{def:lifetime}
		The \textit{lifetime} of a quantum memory is the average number of rounds it survives before a logical error occurs, and the \textit{logical error rate} is the reciprocal of this lifetime.
	\end{definition}

To estimate the logical qubit lifetime, we use the strategy  in~\cite[Algorithm~1]{KL24},
which involves an \emph{actual} decoder and a \emph{virtual} decoder as we explain in the following.

\bmark{
	Rounds of noisy syndrome extraction are performed continuously. The actual decoder represents the realistic circuit-level decoding process operating on noisy syndromes. Every $r$ syndrome extraction rounds, the actual decoder operates only on the collected noisy syndrome data and applies an estimated correction to the code block, after which noisy syndrome extraction rounds continue to propagate errors stochastically.

		To monitor logical fidelity, we perform periodic \emph{virtual checks} every $r$ syndrome extraction rounds using a virtual decoder that serves as an ``oracle.'' This virtual decoder does not affect the physical evolution of the system. Instead, it uses an idealized terminal measurement (in a virtual round) solely to diagnose whether a logical error has occurred. These checks are purely diagnostic: they neither reset the state nor alter the underlying error propagation. The memory is declared to have failed at the first virtual check where a logical error is detected.

	We repeat this procedure until a logical failure occurs and record the total number of syndrome extraction rounds sustained. The memory lifetime is then defined as the average over many trials, i.e., the expected number of syndrome extraction cycles before logical failure.
	
}

	In the example shown in Fig.~\ref{fig:FT_mtx}, an additional round of perfect syndrome extraction is included to illustrate the process of virtual decoding.

	\begin{figure*}[t!]	  
		\includegraphics[width=1.0\linewidth]{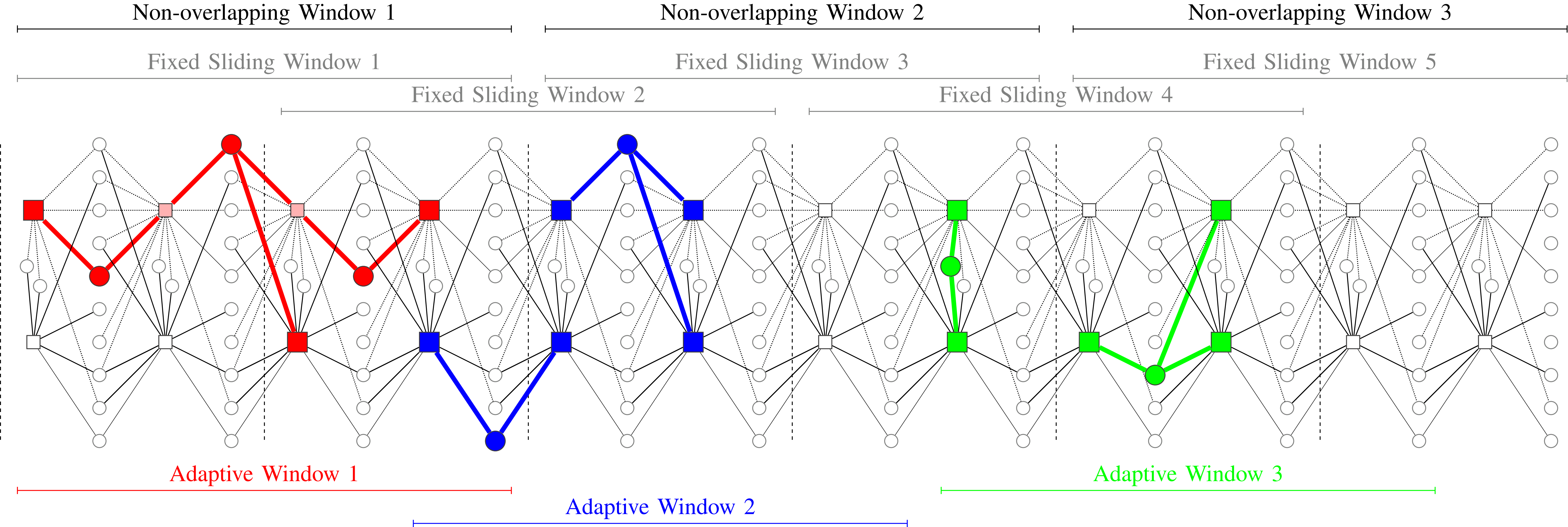}
		\caption{	
			A space-time Tanner graph corresponding to Matrix~(c) in Fig.~\ref{fig:FT_mtx}, with 11 rounds of syndrome extraction, extending Fig.~\ref{fig:space_time_Tanner}. 
			\bmark{Each square represents a check node, and each circle represents an error-variable node.}
		Variable nodes corresponding to nontrivial entries for a given error set are enlarged and highlighted in different colors. The check nodes connected to these variables are highlighted accordingly. The two pink check nodes are not enlarged because their syndrome bits are zero.
				\bmark{Top: A fixed-size window scheme boundary can artificially truncate connected error chains (e.g., the red errors spanning a fixed sliding window boundary; the blue errors spanning over a fixed non-overlapping window boundary), which can potentially lead to decoding failures. Bottom: Our adaptive scheme dynamically adjusts the window offsets to capture entire connected clusters, ensuring that error chains are not partially processed.}
	} \label{fig:FT_win}
\end{figure*}

	\begin{figure}
		\includegraphics[width=0.49\textwidth]{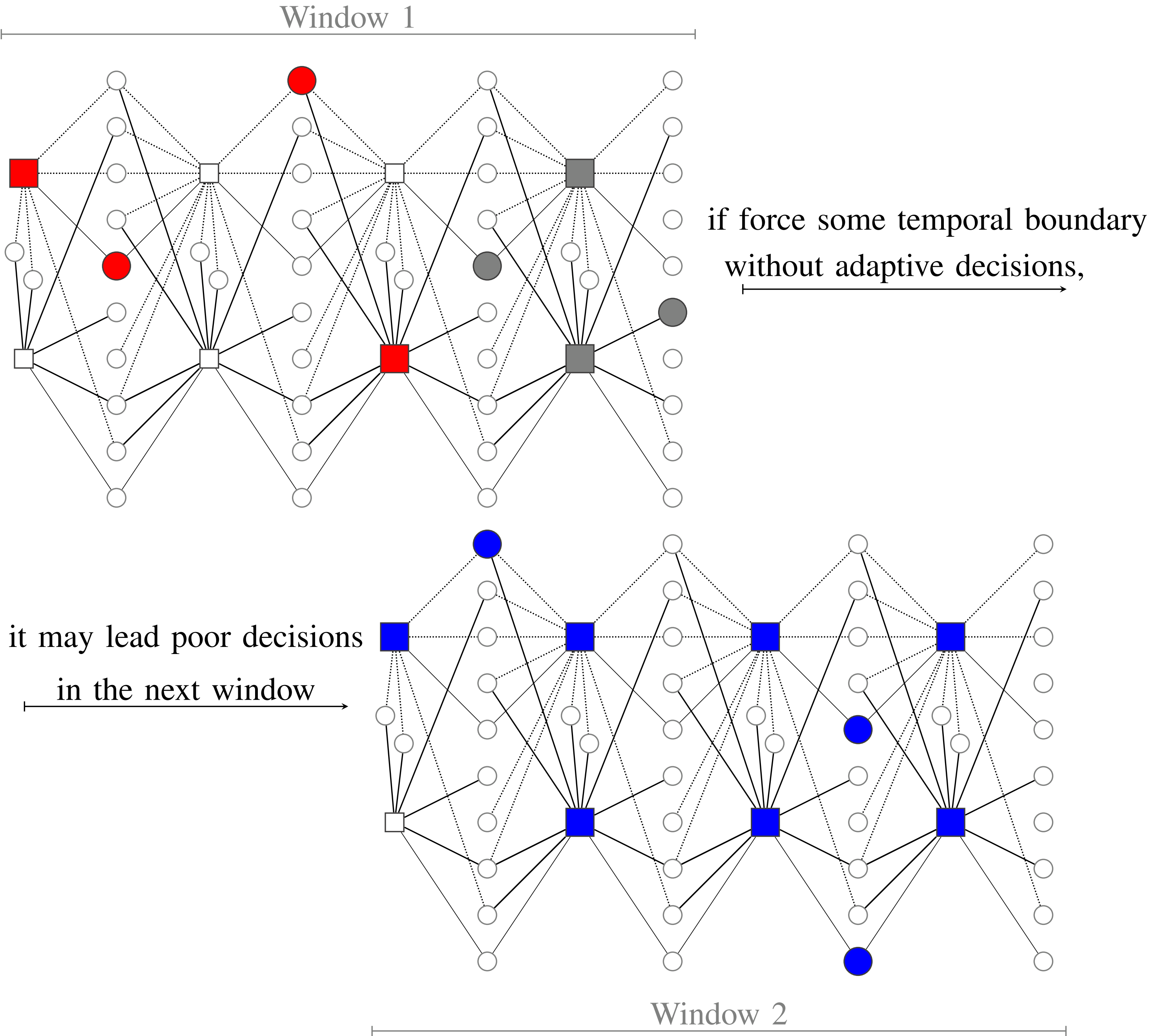}
		\caption{	
\bmark{
	Example of a non-adaptive scheme using a half-window offset decoding. 
	As shown in Window~1, the red variable nodes in rounds~1 and~2 are successfully decoded. However, the updated syndromes in rounds~3 to~6 may lead to incorrect outputs in Window~2, as boundary-crossing errors are split across windows and not jointly processed. 
	In contrast, our adaptive scheme resolves such boundary ambiguities directly during decoding.
}
		} \label{fig:FT_win_fix} 
	\end{figure}

		\subsection{Adaptive sliding window}

		Fixed-size decoding windows may lead to performance degradation in SM-BP decoding, as errors can propagate across multiple rounds of syndrome extraction. In particular, non-overlapping window schemes fail to capture such cross-boundary error correlations.

	To address this issue, a sliding window scheme is proposed in \cite{DKLP02}.
	For a window size of $r$ syndrome extraction rounds, a fixed window offset of $r/2$ is suggested. 
	In this method, $r$ rounds of syndrome extraction are performed first, followed by a decoding process. Error correction is applied only to the first $r/2$ rounds of error locations, while the remaining $r/2$ rounds of syndromes are carried forward to the next error correction cycle. Subsequently, another $r/2$ rounds of syndrome extraction are performed, and the remaining $r/2$ rounds from the previous cycle are decoded alongside the newly acquired $r/2$ rounds. This iterative procedure allows for continuous error correction while maintaining manageable decoding complexity.
	
	Note that in each window, errors occurring in the first $r/2$ rounds may be correlated with errors in the remaining $r/2$ rounds. This makes decoding with a fixed window offset suboptimal, as these correlations are not taken into account.

	In practice, additional information about the occurred errors can be leveraged to improve decoding performance.
	In the context of a space-time Tanner graph, we say that two error locations are \textit{connected} if their corresponding variable nodes are linked through a common check node.
	Given an estimated error from SM-BP decoding, if there is a series of consecutive error locations connected to some error locations in the first half of the window, these errors are likely correlated and should be handled together. In such cases, the window offset should be adjusted accordingly. We propose an adaptive sliding window procedure with the following modifications to the original sliding window method.

	\begin{proposition} \label{prop:adp_win} ({\bf Adaptive Sliding Window}) 
		Suppose that an error estimate  for a decoding window is generated by SM-BP.
		\begin{enumerate} \item Error correction  is always applied to locations in the first half of the window. 
			\item Additionally, error locations connected to the first half of the window will also be corrected (even if these connections are near the window boundary). 
			\item Update the corresponding syndromes after applying the error correction. 
			\item The first uncorrected error location in the second half of the window will mark the starting point for the next window. \end{enumerate} \end{proposition}

	For illustration, a space-time Tanner graph corresponding to Matrix~(c) in Fig.~\ref{fig:FT_mtx} is shown in Fig.~\ref{fig:FT_win}. 
	For a given set of errors highlighted in different colors, the check nodes with nontrivial syndromes are marked accordingly. 
	The sliding-window procedure with a fixed window offset is illustrated at the top of the Tanner graph, where each window covers four rounds of syndrome extraction with an offset of two rounds. 
	
	As observed, several nontrivial connected errors span across these decision boundaries, such as the red errors occurring in the second and third rounds, located at the boundary between Fixed Sliding Window~1 and Fixed Sliding Window~2. 
	\bmark{In this case, the fixed-window decoder processes these correlated errors in separate decoding steps, effectively breaking their dependency and leading to residual uncorrected  errors.}
	\bmark{Handling such boundary-crossing events typically requires additional post-processing to reconcile inconsistent decisions across windows, whereas our adaptive scheme eliminates the need for such post-processing, as described in the following.}

	Our adaptive windows are shown at the bottom of the Tanner graph in Fig.~\ref{fig:FT_win}, with adaptive window offsets. 
	In Adaptive Window 1, if the red errors are decoded, they will be fully corrected since they are connected and span the first half of the window. However, the blue check node at Round 4 might lead to a decoded error that connects to other error nodes outside the current window. If this decoded error is not connected to errors within the first half of the window, it will not be corrected and will be deferred to the next decoding window.

 \bmark{
 	For illustration, an example of non-adaptive sliding-window decoding is shown in Fig.~\ref{fig:FT_win_fix}. In this scheme, each window is processed using a half-window offset: the first half is decoded and applied, while the second half is deferred to the next window. This is necessary because the decoder does not have access to perfect measurement outcomes, and boundary-region decisions can be unreliable.
 	
 	As shown in Window~1, the red variables are successfully decoded, while the gray variables are not committed and are passed to the next decoding stage. However, due to boundary-crossing errors, the decoding outcome in Window~2 based on the updated syndromes may be  incorrect.
 	In contrast, using the adaptive sliding-window procedure from Proposition~\ref{prop:adp_win}, connected error events are grouped into a single decoding instance, avoiding artificial fragmentation across windows and thereby reducing decoding failures.
 	
 	As demonstrated in our simulations (with window size $r=O(d)$ syndrome extraction rounds), the adaptive scheme achieves improved performance, as will be shown in the next section.
 }

	\section{Simulation results} \label{sec:sim}

	In this section, we simulate the lifetime of a quantum memory protected by rotated toric codes, rotated 6.6.6 toric color codes, or twisted XZZX toric codes, all decoded using  SM-BP.

	The parameter $\alpha$ in SM-BP controls the step size in message passing. 
	\rmark{We consider an ensemble decoding strategy in which SM-BP is executed with a sequence of $\alpha$ values, and an optimal $\alpha^*$ is selected.} This procedure is referred to as SM-BP$(\alpha^*)$, as detailed in Appendix~\ref{app:AMBP} (see Algorithm~\ref{alg:FTAMBP}). 
	\bmark{This approach effectively runs multiple SM-BP instances and selects the most reliable outcome. Although it incurs higher complexity than a single SM-BP run, it can be optimized, and parallelization strategies are discussed in Appendix~\ref{sec:efficiency_paralellism}.}

	In the following simulations, we consider equal error rates $\epsilon$ for data Pauli errors, syndrome bit errors, ancilla initialization errors, and two-qubit gate errors. The simulations use a maximum number of iterations $T_{\max}=150$ \rmark{for a single SM-BP run with a serial message-update schedule, as described in \cite{KL20,KL22}}.
	
	\bmark{
	The BP update rules utilized in this work is a  serial schedule randomized at each iteration. This dynamic scheduling prevents the decoder from falling into persistent oscillations or bad convergence states that a fixed-order schedule might encounter for certain syndromes \cite{SLG07,Cre+23}.
}

	We simulate SM-BP$(\alpha^*)$ with 
	$\alpha^*\in\{1,0.99,\dots,0.4\}$ \bmark{(Algorithm~\ref{alg:FTAMBP})}.
	
	For clarity, we denote the decoder using the quaternary approximation from Proposition~\ref{prop:C4} as SM-BP$_{4}(\alpha^*)$, while the decoder utilizing Proposition~\ref{prop:C16} is referred to as SM-BP$_{16}(\alpha^*)$. 
	
	In the following simulation results, for a code with distance $d$, a bold dashed line representing a curve $a\epsilon^{t+1}$ is plotted, where $t = \lfloor \frac{d-1}{2} \rfloor$, and $a$ is a scalar used to align it with the code's performance curve. This dashed line characterizes the error floor performance of the code.

	The \textit{error threshold} refers to the critical physical error rate below which a family of quantum error correction codes can effectively correct errors by increasing the code length. 
	For a family of topological codes, the threshold is approximately located where the logical error rate performance curves for different code distances intersect. Since we are limited to simulating codes of finite sizes, we use the finite-size scaling (FFS) ansatz~\cite{WHP03,Har04} to estimate the error thresholds of the topological code families considered in this paper (see Appendix~\ref{app:threshold_ansatz}).
		\bmark{
		In this context, the FSS analysis serves as a heuristic tool to identify the effective threshold of a given decoder.}

	\bmark{We simulate the memory logical error rate, defined as the reciprocal of the memory lifetime in Def.~\ref{def:lifetime}. (Recall also the actual decoder and virtual decoder after Def.~\ref{def:lifetime}).
	For clarity, we explicitly 
	provide the operational definition of  the memory lifetime under a non-overlapping window decoding scheme in our simulations as follows (see also \cite[Algorithm~1]{KL24}).}

\bmark{
\begin{definition}[Memory lifetime under non-overlapping decoding] \label{def:T_by_J}
	 Suppose the memory is monitored over consecutive cycles, where each cycle consists of $r$ syndrome extraction rounds. Let $J$ denote the number of fully successful cycles of the actual decoder, before the first failure is detected by the virtual decoder. That is, cycles $1,\dots,J$ are decoded without indicating failure, while a failure is first detected in cycle $J+1$ by the virtual decoder.
	We define the lifetime $T$ as the index of the last successfully verified syndrome extraction round. Since each cycle contains $r$ rounds, this gives
	\[
	T = Jr + 1.
	\]
	The additional $+1$ term corresponds to the first syndrome extraction round in cycle $J+1$, which is used to detect the failure event under the virtual decoding procedure.
	
	The corresponding logical error rate is defined as $1/T$. In particular, when $J=0$ (i.e., failure is detected in the first cycle), we have $T=1$, yielding a logical error rate of $1$.
\end{definition}
}

	\bmark{
	Similarly we define the operational memory lifetime using the adaptive procedure in Proposition~\ref{prop:adp_win} as follows.}

\bmark{
	\begin{definition}[Memory lifetime under adaptive decoding]\label{def:T_by_adp}
		In the adaptive sliding-window scheme with dynamically adjusted offsets, we define the memory lifetime $T$ 
		as the total number of syndrome extraction rounds processed by the actual decoder, until the first failure is detected by the virtual decoder. Each decoding window contains up to $r$ syndrome extraction rounds, but the window offset is not fixed and may vary adaptively during the decoding process.
	\end{definition}
}

\bmark{Unless otherwise specified, the adaptive window scheme (Proposition~\ref{prop:adp_win}) is used. Error correction provides a gain when the memory logical error rate $1/T$ falls below the \emph{uncoded logical error rate}, defined as $1-(1-\epsilon)^k$ for a quantum memory with $k$ logical qubits at physical error rate $\epsilon$.}

	\subsection{Probabilistic error consolidation and adaptive sliding window  techniques}
	
\begin{figure}[t!]
		\centering 
		\includegraphics[width=0.5\textwidth]{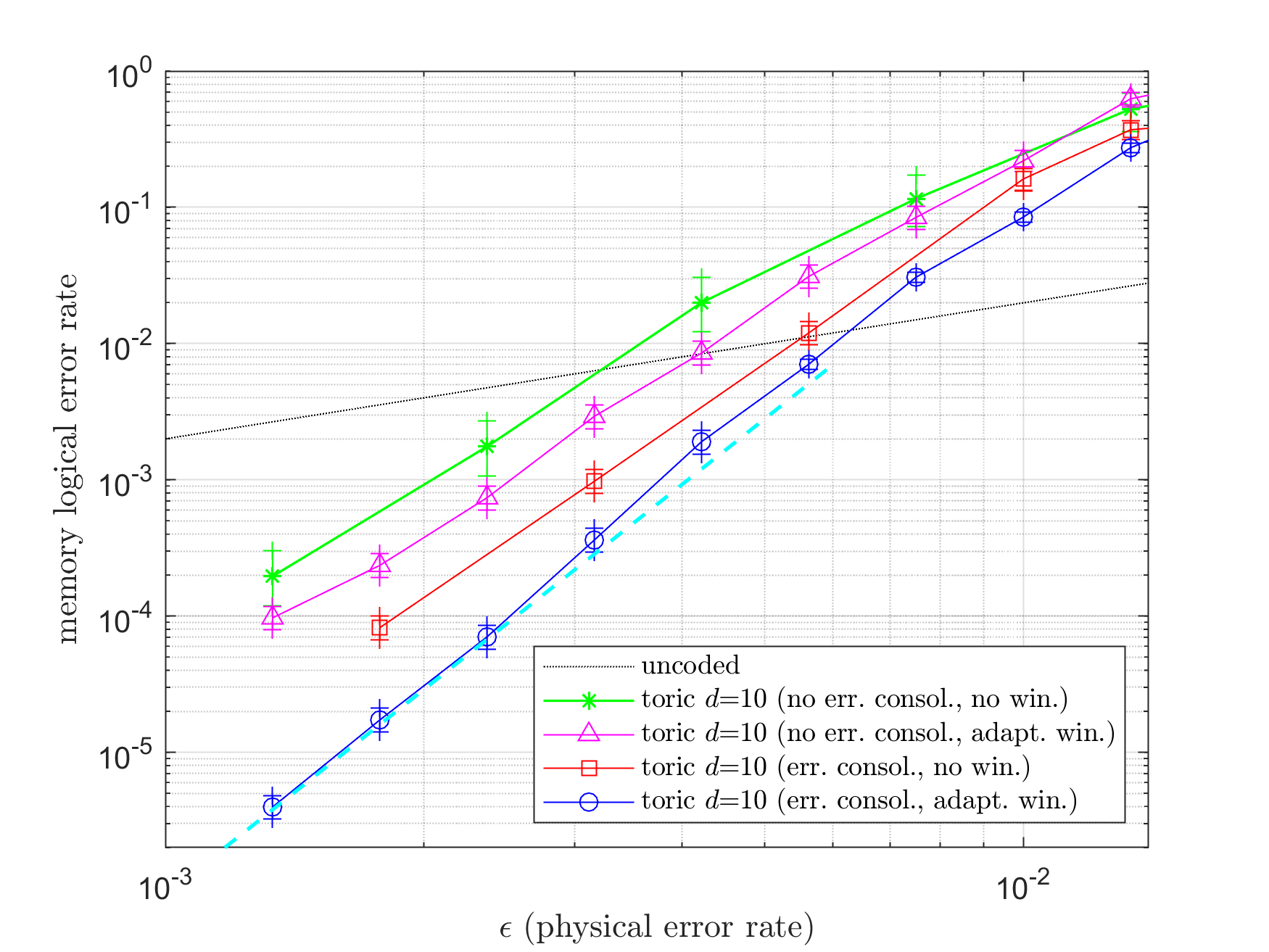}	
		\caption{
			\rmark{Performance of SM-BP$_{16}(\alpha^*)$ on the $[[100,2,10]]$ rotated toric code, comparing four configurations with and without adaptive sliding windows and probabilistic error consolidation. The window size is set to $r=10$.  The dashed line shows the expected scaling $a\epsilon^{t+1}$ with $t=\lfloor\frac{d-1}{2}\rfloor$ for an appropriate constant $a$.}
		}
	 \label{fig:FT_schemes} 
	\end{figure}

	First, we demonstrate the effects of probabilistic error consolidation (Lemma~\ref{lemma:merging_degenerate} and Algorithm~\ref{alg:merge2}) and adaptive sliding windows (Proposition~\ref{prop:adp_win}) on SM-BP decoding of the \rmark{$[[100,2,10]]$} rotated toric code. The performance results for various cases are shown in Fig.~\ref{fig:FT_schemes}.

	We construct a generalized check matrix for the $[[100,2,10]]$ code using Lemma~\ref{lemma:row_operation} with $r=d$.

	\bmark{
	Applying probabilistic error consolidation (Lemma~\ref{lemma:merging_degenerate}, Algorithm~\ref{alg:merge2}, and Proposition~\ref{prop:C16}) significantly improves the performance of SM-BP$_{16}$, as shown in Fig.~\ref{fig:FT_schemes}. Applying only the adaptive window scheme (Proposition~\ref{prop:adp_win}) also yields improvement, though to a lesser extent. When both techniques are combined, the performance further improves, particularly in the low logical error rate regime.  At $\epsilon=2\times 10^{-3}$, the logical error rate is reduced from around $10^{-3}$ to $3\times 10^{-5}$, corresponding to nearly two orders of magnitude improvement.
}

	In the following simulations, we will employ both  probabilistic error consolidation and adaptive sliding windows techniques.

	\subsection{Size of a decoding window} \label{sec:sim_win_size}

	In general, a generalized check matrix derived from a larger number of syndrome extraction rounds enhances circuit-level decoding, enabling the underlying stabilizer code to achieve its full error-correction potential by correcting up to $\lfloor (d-1)/2\rfloor$ errors, where $d$ is the minimum distance of the code.  
	However, increasing the number of syndrome extraction rounds also raises the  decoding complexity. 
	It is  believed that  $O(d)$ rounds of syndrome extraction are sufficient  to achieve good decoding performance~\cite{DKLP02}.

	\begin{figure}[t!]
		\centering \includegraphics[width=0.5\textwidth]{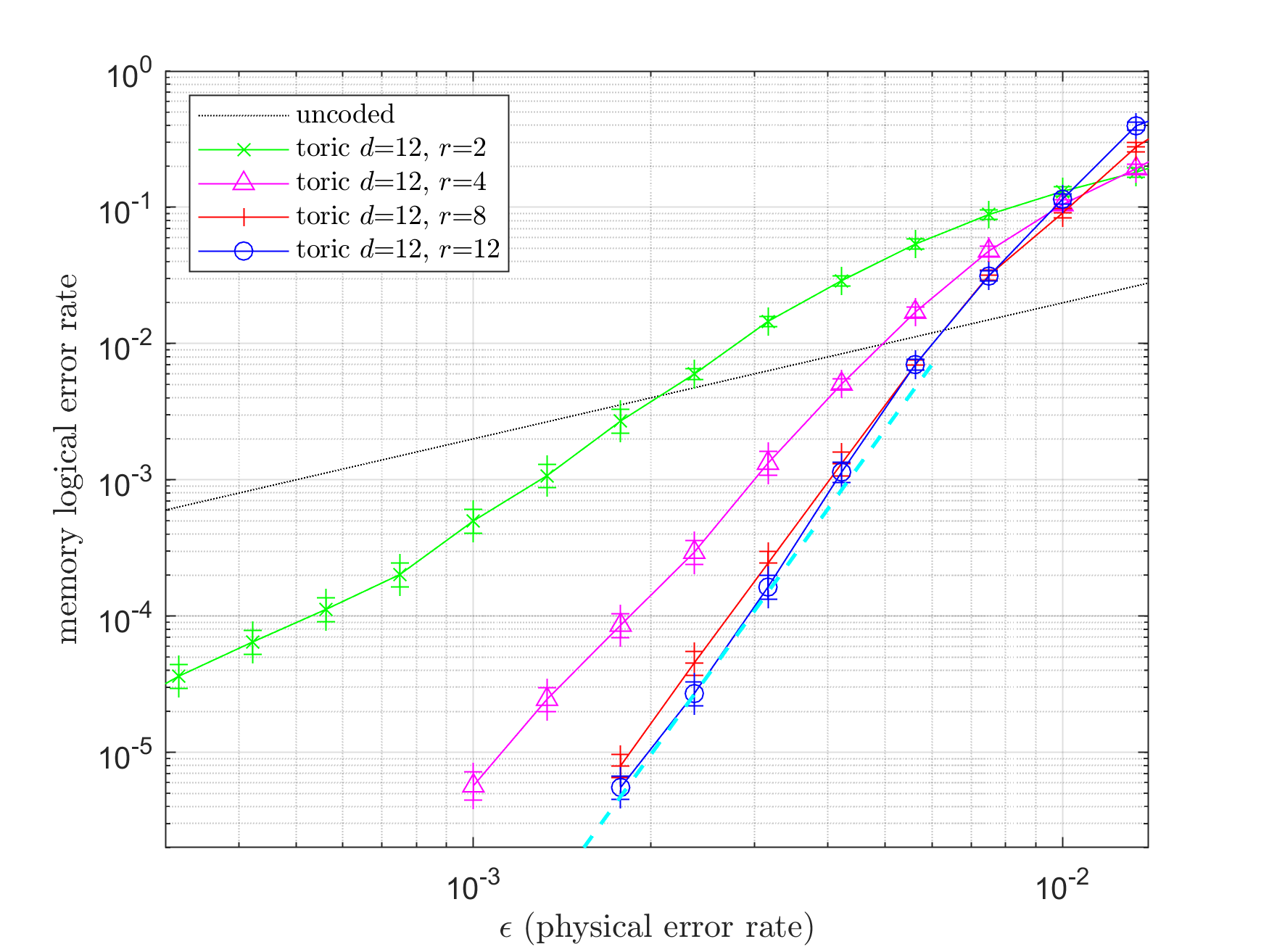}
		\caption{	
			\bmark{SM-BP$_{16}(\alpha^*)$ performance on the $[[144,2,12]]$ rotated toric code  for different window sizes.  The dashed line shows the expected scaling $a\epsilon^{t+1}$ with $t=\lfloor\frac{d-1}{2}\rfloor$ for an appropriate constant $a$.}
		} \label{fig:FT_toric_12} 
	\end{figure}

	\begin{figure}[t!]
		\centering \includegraphics[width=0.5\textwidth]{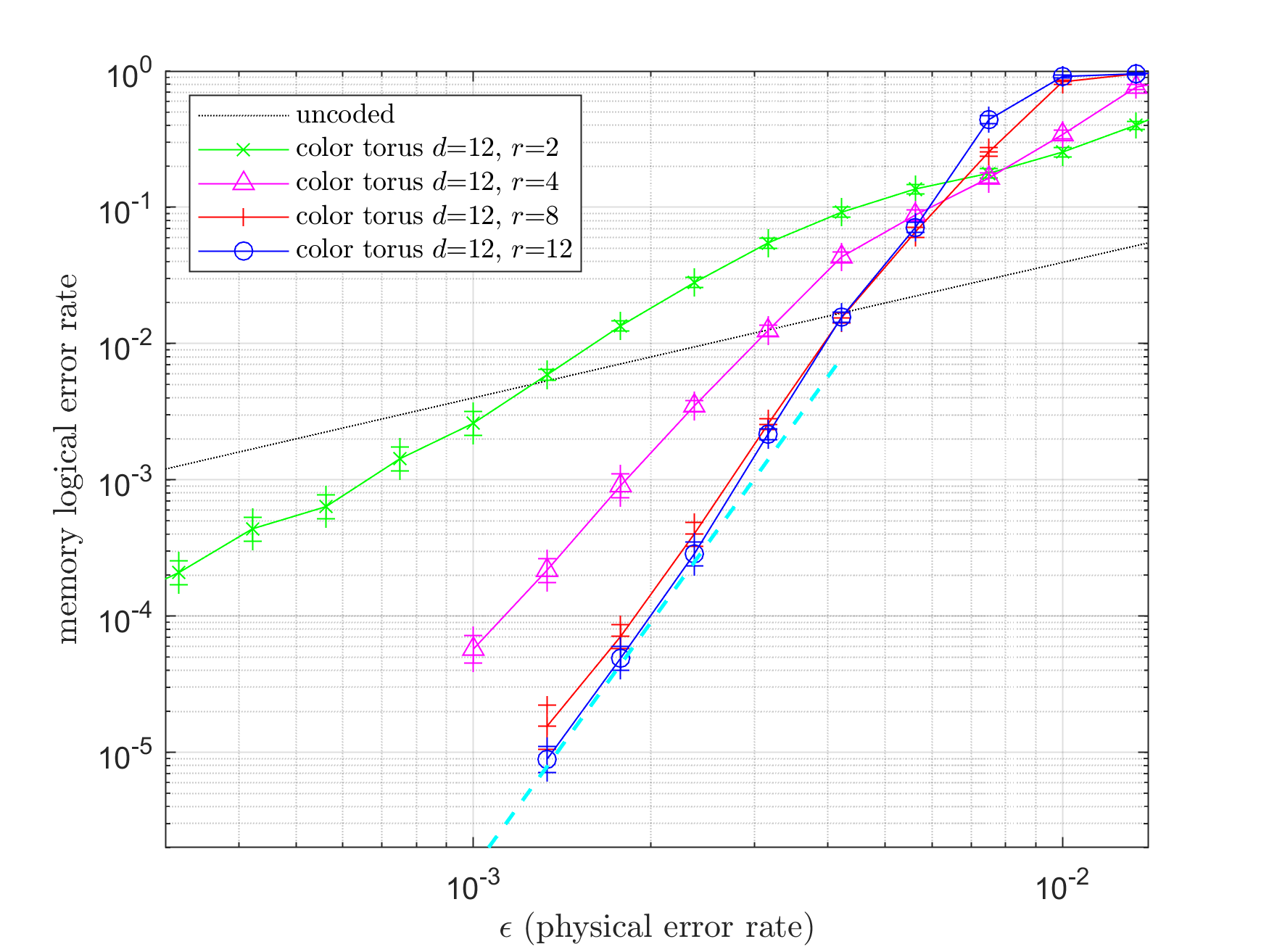}
		\caption{	
			\rmark{SM-BP$_{16}(\alpha^*)$ performance on the $[[162,4,12]]$ rotated color code  for different window sizes.  The dashed line shows the expected scaling $a\epsilon^{t+1}$ with $t=\lfloor\frac{d-1}{2}\rfloor$ for an appropriate constant $a$.}
		} \label{fig:FT_color_12} 
	\end{figure}

We examine the impact of the number of syndrome extraction rounds within a decoding window on the quantum memory lifetime. We present the SM-BP decoding performance for \bmark{the $[[144,2,12]]$ rotated toric code in Fig.~\ref{fig:FT_toric_12} and} the $[[162,4,12]]$ rotated 6.6.6 color code in Fig.~\ref{fig:FT_color_12}, comparing different window sizes $r$.
Generalized check matrices for various values of $r$ are constructed using Lemma~\ref{lemma:row_operation}.

At higher physical error rates, using a smaller window size $r$ generally yields better performance. This is because a larger $r$ leads to a more extensive space-time Tanner graph, increasing the number of possible error configurations, which complicates BP decoding. Moreover, at high error rates, a larger $r$ allows errors to accumulate over more rounds, increasing the likelihood of logical failure.

In contrast, in the low-error-rate regime, decoding performance improves as $r$ increases, \rmark{eventually saturating at around $r \approx d$ for the toric and color codes in our tests.}

	\begin{figure}[t!]
		\centering \includegraphics[width=0.5\textwidth]{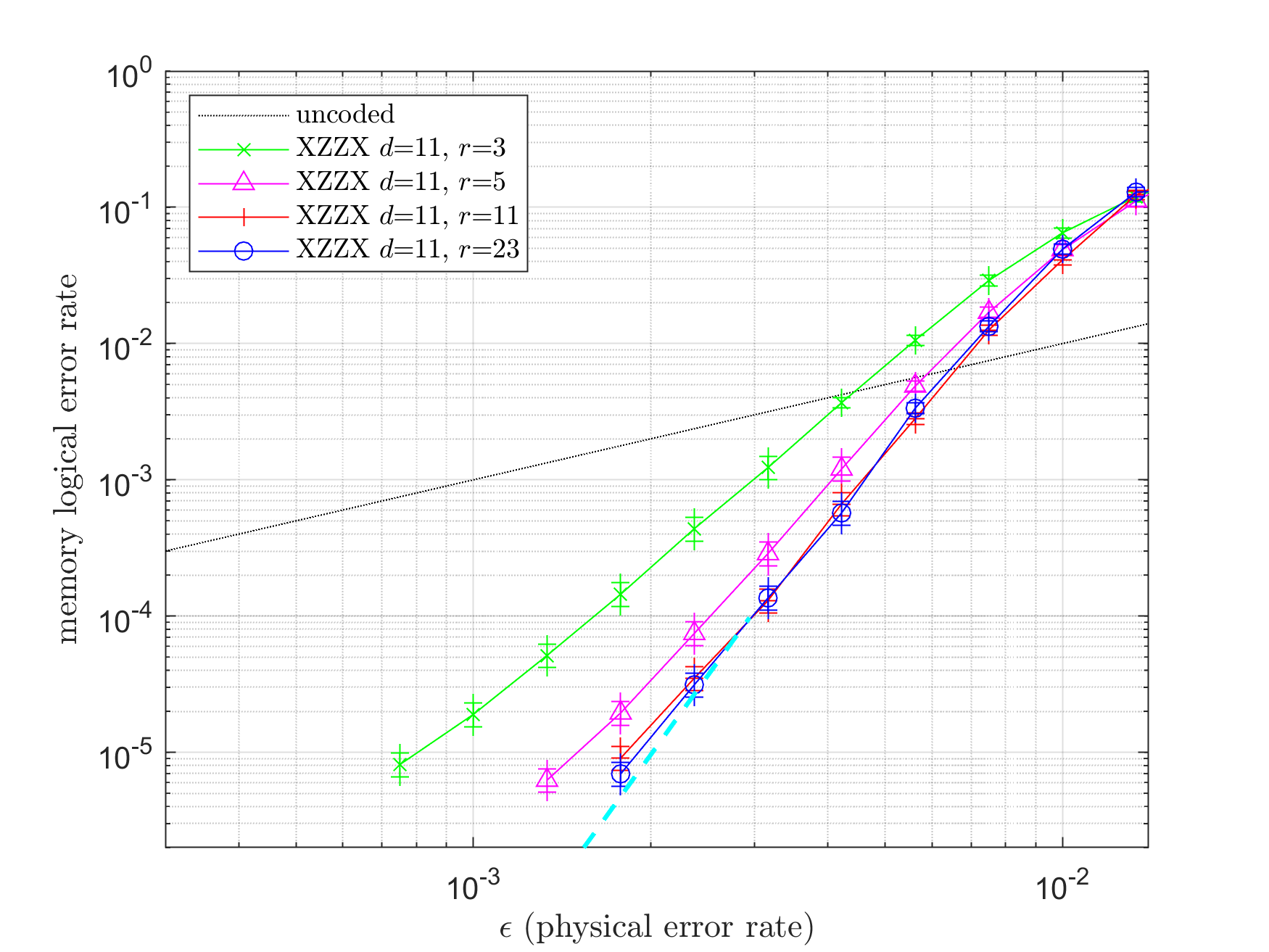}
		\caption{	
			\bmark{SM-BP$_{16}(\alpha^*)$ performance on the $[[61,1,11]]$ twisted XZZX code  for different window sizes.  The dashed line shows the expected scaling $a\epsilon^{t+1}$ with $t=\lfloor\frac{d-1}{2}\rfloor$ for an appropriate constant $a$.}
		} \label{fig:FT_XZZX_11} 
	\end{figure}


\bmark{For the twisted XZZX codes, the SM-BP performance for $d=11$ is shown in Fig.~\ref{fig:FT_XZZX_11}. Strong performance is already achieved with a small number of syndrome extraction rounds (e.g., $r=3$). However, in the low-error regime, the decoding performance deviates from the expected $a\epsilon^{t+1}$ scaling (dashed lines). Increasing $r$ does not significantly improve performance, suggesting that the limitation is not due to insufficient temporal redundancy, but is instead inherent to the decoding problem for this code family.
This observation indicates that careful design of syndrome extraction procedures is important for non-CSS codes.}

	\bmark{Further simulations indicate that twisted XZZX codes benefit from quaternary SM-BP (SM-BP$_4$, instead of SM-BP$_{16}$) in the low-error regime. Error consolidation can introduce bias in bit decisions; in SM-BP$_{16}$, this effect propagates across more components, whereas SM-BP$_4$ confines it more effectively. See Appendix~\ref{app:C4C16}.}

	\subsection{Quantum memory with rotated toric codes}

\begin{figure}[t!]
		\centering \includegraphics[width=0.5\textwidth]{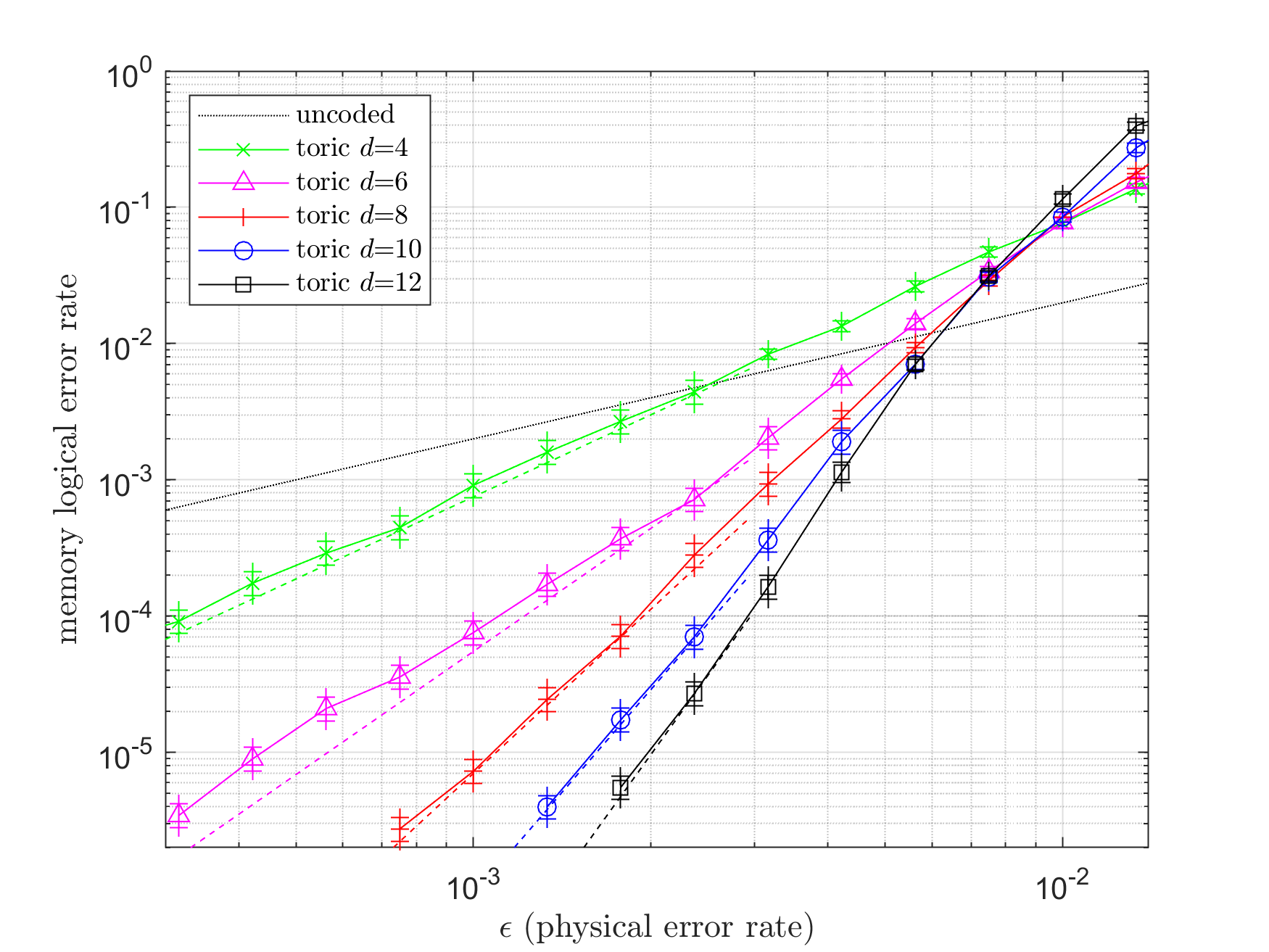}
		\caption{	
			SM-BP$_{16}(\alpha^*)$ decoding performance for the $[[d^2,2,d]]$ rotated toric codes with $d\le 12$.
			The window size is $r=d$.  The dashed lines show the expected scaling $a\epsilon^{t+1}$ with $t=\lfloor\frac{d-1}{2}\rfloor$ for appropriate constants $a$.
		} \label{fig:FT_toric} 
	\end{figure}

	We simulate quantum memory using the family of $[[d^2,2,d]]$ rotated toric codes for $d \leq 12$, with results shown in Fig.~\ref{fig:FT_toric}. 
	
	For $d=6,8,10,12$, the FFS ansatz estimates a threshold of $0.75\%$ for rotated toric codes (see Appendix~\ref{app:threshold_ansatz}), which is consistent with the intersection of the performance curves in Fig.~\ref{fig:FT_toric}. This threshold is comparable to those obtained with MWPM in~\cite{RH07,WFSH10,Ste14s,CKYZ20},
	 while using SM-BP with adaptive sliding-window decoding in lifetime simulations.
	 
	
	For reference, a threshold of $0.68\%$ for 2D surface codes was reported in~\cite{TZC+23} using MWPM with parallel sliding-window decoding, where the decoding window spans $\frac{3d}{2}$ syndrome extraction rounds.
	
	\bmark{A comprehensive comparison with other decoders, including BP+OSD, standard and windowed MWPM, and weighted UF, is provided in Appendix~\ref{app:versus_others}.}

	\subsection{Quantum memory with rotated 6.6.6 toric color codes}

	\begin{figure}[t!]
		\centering \includegraphics[width=0.5\textwidth]{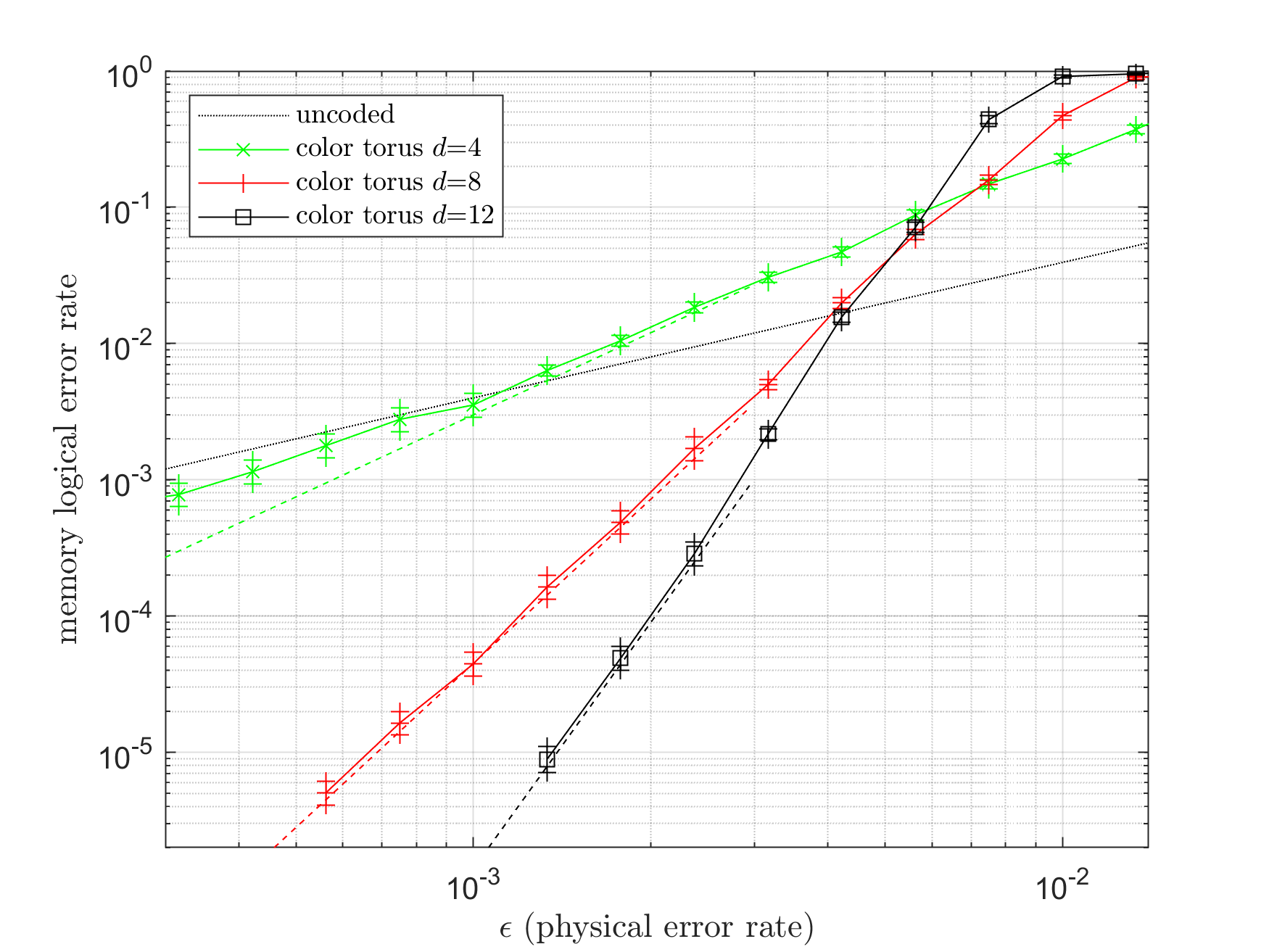}
		\caption[Roots associated to the Cartan matrix]{	
			SM-BP$_{16}(\alpha^*)$ decoding performance for the $[[\frac{9}{8}d^2,4,d]]$ rotated 6.6.6 color codes with $d=4,8,12$.
			The window size is $r=d$.  The dashed lines show the expected scaling $a\epsilon^{t+1}$ with $t=\lfloor\frac{d-1}{2}\rfloor$ for appropriate constants $a$.
		} \label{fig:FT_color} 
	\end{figure}

	We consider the family of $[[\frac{9}{8}d^2, 4, d]]$ rotated 6.6.6 toric color codes for $d=4,8,12$, with results shown in Fig.~\ref{fig:FT_color}.
	
		\rmark{
	The performance curves for $d=4,8$ intersect around $\epsilon=0.6\%$, and those for $d=8,12$ intersect around $0.5\%$, as shown in Fig.~\ref{fig:FT_color}. Using $d=4,8,12$, the scaling ansatz estimates a threshold of approximately $0.5\%$ (see Appendix~\ref{app:threshold_ansatz}). Therefore, we expect the SM-BP decoding threshold for the rotated 6.6.6 toric color codes to lie between $0.5\%$ and $0.6\%$.
}

		\rmark{
	For $d=4$, the code exhibits a limited coding gain at small physical error rates. This is attributed to the higher stabilizer weight (6) of color codes, which increases the complexity of the syndrome extraction circuit compared to toric codes (weight 4). For larger distances ($d \ge 8$), the performance approaches the expected scaling $a\epsilon^{t+1}$ (dashed lines).
}

	\rmark{
	For reference, thresholds of $0.1\%$--$0.143\%$ have been reported for $[[\frac{(d-1)^2}{2}+d,1,d]]$ 4.8.8 color codes~\cite{BM06,WFHH10,Ste14c}, and approximately $0.2\%$ for $[[\frac{3d^2+1}{4},1,d]]$ triangular 6.6.6 color codes under modified MWPM decoding~\cite{CKYZ20}.
}

	\subsection{Quantum memory with twisted XZZX toric codes}
	
	\begin{figure}[t!]
		\centering \includegraphics[width=0.5\textwidth]{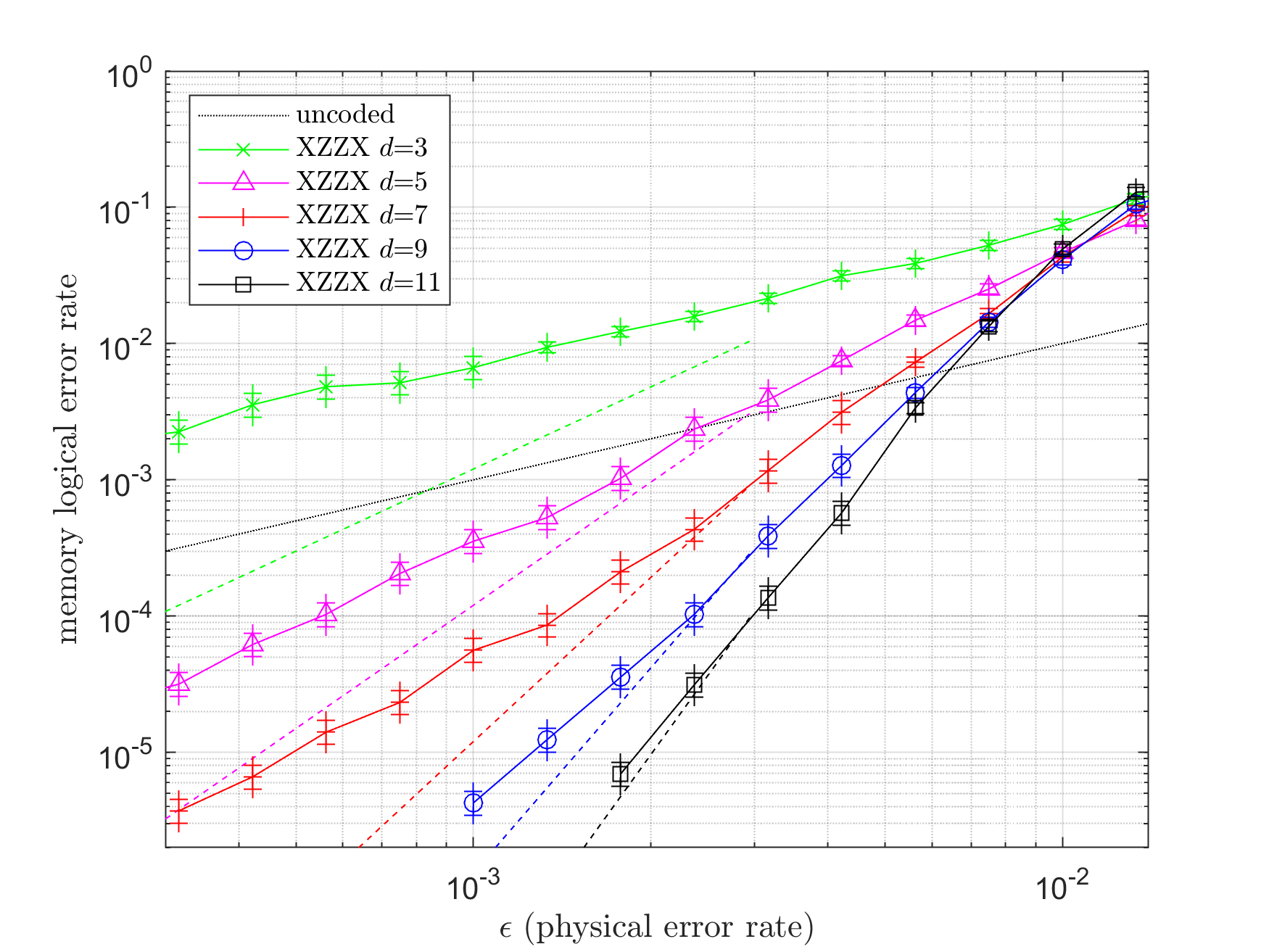}
		\caption{	
			SM-BP$_{16}(\alpha^*)$ decoding performance for the $[[(d^2+1)/2,\, 1,\, d]]$ twisted XZZX toric codes with ${d\le 12}$. The window size is $r=2d+1$.  The dashed lines show the expected scaling $a\epsilon^{t+1}$ with $t=\lfloor\frac{d-1}{2}\rfloor$ for appropriate constants $a$.
		} \label{fig:FT_XZZXtoric}
	\end{figure}
	
	We apply SM-BP decoding to the $[[\frac{d^2+1}{2},\, 1,\, d]]$ twisted XZZX toric codes. These codes have weight-4 stabilizers, similar to toric codes. The simulation results are shown in Fig.~\ref{fig:FT_XZZXtoric}.

	In contrast to toric codes, the $XZZX$ measurement circuit can induce more involved error propagation during syndrome extraction. For small code distances ($d \leq 7$), the raw stabilizer measurements are insufficient to achieve effective decoding performance.
	
	For larger distances ($d \geq 9$), the performance curves become closer to the expected scaling $a\epsilon^{t+1}$ (dashed lines), indicating improved decoding behavior with increasing $d$.

		Using $d=7,9,11$, the FFS ansatz estimates a threshold of $0.88\%$ for twisted XZZX toric codes (see Appendix~\ref{app:threshold_ansatz}). However, the performance curves appear to intersect around $0.75\%$--$0.8\%$ in Fig.~\ref{fig:FT_XZZXtoric}. Thus we expect the SM-BP$_{16}(\alpha^*)$ decoding threshold  for twisted XZZX toric codes
	lies between  $0.75\%$ and $0.88\%$.

	\section{Conclusion} \label{sec:con}
	
	In conclusion, we have proposed methods for circuit-level decoding of general quantum codes using BP, extending the quantum data-syndrome code framework to operate under circuit-level noise. Several techniques were introduced to improve decoding performance in quantum memory lifetime simulations, including the use of sparse generalized check matrices,  probabilistic error consolidation, and adaptive sliding windows. Our approach can be applied to general topological codes, including non-CSS codes. This discussion can be easily extended to accommodate other syndrome extraction procedures.

	Note that the generalized check matrix constructed using Lemma~\ref{lemma:row_operation} may resemble the decoding graph used in MWPM for toric codes \cite{DKLP02}. However, our approach involves mixed variable nodes, and we do not utilize separate primal and dual lattices for decoding $X$ and $Z$ errors. Additionally, our derivation employs the method of syndrome assignment from coding theory, emphasizing a preference for sparser matrices.

	For SM-BP, two-qubit gate errors are treated as 16-ary errors. However, in our approach, some correlations are disregarded to reduce degeneracy and minimize short cycles in the Tanner graph. Further research is needed to effectively incorporate these correlations while maintaining efficient SM-BP decoding to enhance overall decoding performance.

	The simulation results could be further optimized by incorporating additional techniques, such as more syndrome extraction rounds in a decoding window, fixed initialization in BP \cite{HFI12,KL22,KL24} or expanding the range of the parameter $\alpha$ in SM-BP.

	In our simulations, a serial schedule for BP was employed. Future work could explore the effects of different  probabilistic error consolidation sequences and BP scheduling strategies.

	    \bmark{
		Our probabilistic error consolidation can be viewed as a refinement of the standard independent binary error model, partially preserving correlations induced by circuit-level noise while maintaining manageable decoding complexity. We emphasize that the primary limitations observed in decoding performance  stem from the structure of the induced decoding problem, in particular the presence of short cycles and trapping sets in the generalized Tanner graph. Such structural challenges persist even under fully binary representations and are intrinsic to circuit-level decoding, especially for irregular QLDPC codes. For topological codes with more regular structure and localized correlations, the proposed approach is highly effective in practice. These observations suggest that further improvements for general code families will likely rely on complementary techniques, such as reliable-subset reduction (RSR) combined with OSD~\cite{KKL24}.
	}

	
	\bmark{
	The source files of our computer programs can be found
	in \cite{SM-BP26}.
}

\bmark{\section*{Author contribution statements}
C.-Y.L. and K.-Y.K. formulated the initial idea and developed the theory. K.-Y.K. performed
the simulations. All co-authors contributed to the preparation of the manuscript.}

\bmark{\section*{Acknowledgment}
We thank Mingyu Kang for helpful discussions on the ordering of measurement gates.}

%

\bibliographystyle{quantum}

\onecolumn
\appendix

	\section{Adaptive version of SM-BP}\label{app:AMBP}

	\begin{algorithm}[H]
				\caption{:  {SM-BP$(\alpha^*)$}} \label{alg:FTAMBP}
				
				{\bf Input}: An $M\times N$ generalized check matrix $\cH$ over  a mixed alphabet
				of symbols from $\{I,X,Y,Z\}$, $\{I,X,Y,Z\}^2$, and $\{0,1\}$, an error syndrome vector $ s \in\{0,1\}^{M}$, 
				an~integer $T_{\max}>0$, 
				a~sequence of $\ell$ parameters $\alpha_1 > \alpha_2 >\dots > \alpha_\ell > 0$,
				and initial LLRs $\Lambda_{1},\dots,\Lambda_{N}$.
				
				{\bf Initialization}:  Let $i=1$. 
				
				\begin{itemize}
					\item {\bf BP Step}: Run SM-BP$( \cH,\, s,\, T_{\max},\, \alpha_i,\, \text{initial LLRs})$,
					\begin{itemize}
						\item[] which returns  ``CONVERGE'' or ``FAIL'' with an error vector $\hat{\cE}$ over $\{I,X,Y,Z\},\{I,X,Y,Z\}^2, \{0,1\}$.
					\end{itemize}
					\item  {\bf Adaptive Check}: 
					\begin{itemize}
						\item If ``CONVERGE'', output  $\hat \cE $ and $\alpha^* = \alpha_i$ and return ``SUCCESS'';
						\item Else, if $i<\ell$, update $i\leftarrow i+1$ and repeat from the BP Step;
						\item Else, return ``FAIL''.
					\end{itemize}
				\end{itemize}
				
	\end{algorithm}

	%
	For enhanced performance, the optimal value $\alpha^*$ can be adaptively selected \cite{KL22}. 
	This adaptive algorithm, SM-BP$(\alpha^*)$, builds upon the AMBP algorithms~\cite{KL22} and extends SM-BP. 
	The details of this algorithm are presented in   Algorithm~\ref{alg:FTAMBP}.
	\bmark{It executes multiple SM-BP instances sequentially and selects the first solution that converges.}
	Since BP typically converges in about $\log N$ iterations, selecting $T_{\max} = O(\log N)$ is generally sufficient. For the decrement sweep of $\alpha_1 > \alpha_2 > \dots > \alpha_\ell > 0$, a step size of 0.01 is sufficient because $\alpha$ controls the strength of message passing, and BP tends to converge within $O(\log N)$ iterations, making it relatively insensitive to changes in $\alpha$.	Additionally, multiple SM-BP instances with different $\alpha$ values can be executed in parallel.

\section{SM-BP efficiency} \label{sec:efficiency_paralellism}

\begin{figure}[t!]
	\centering \includegraphics[width=0.48\textwidth]{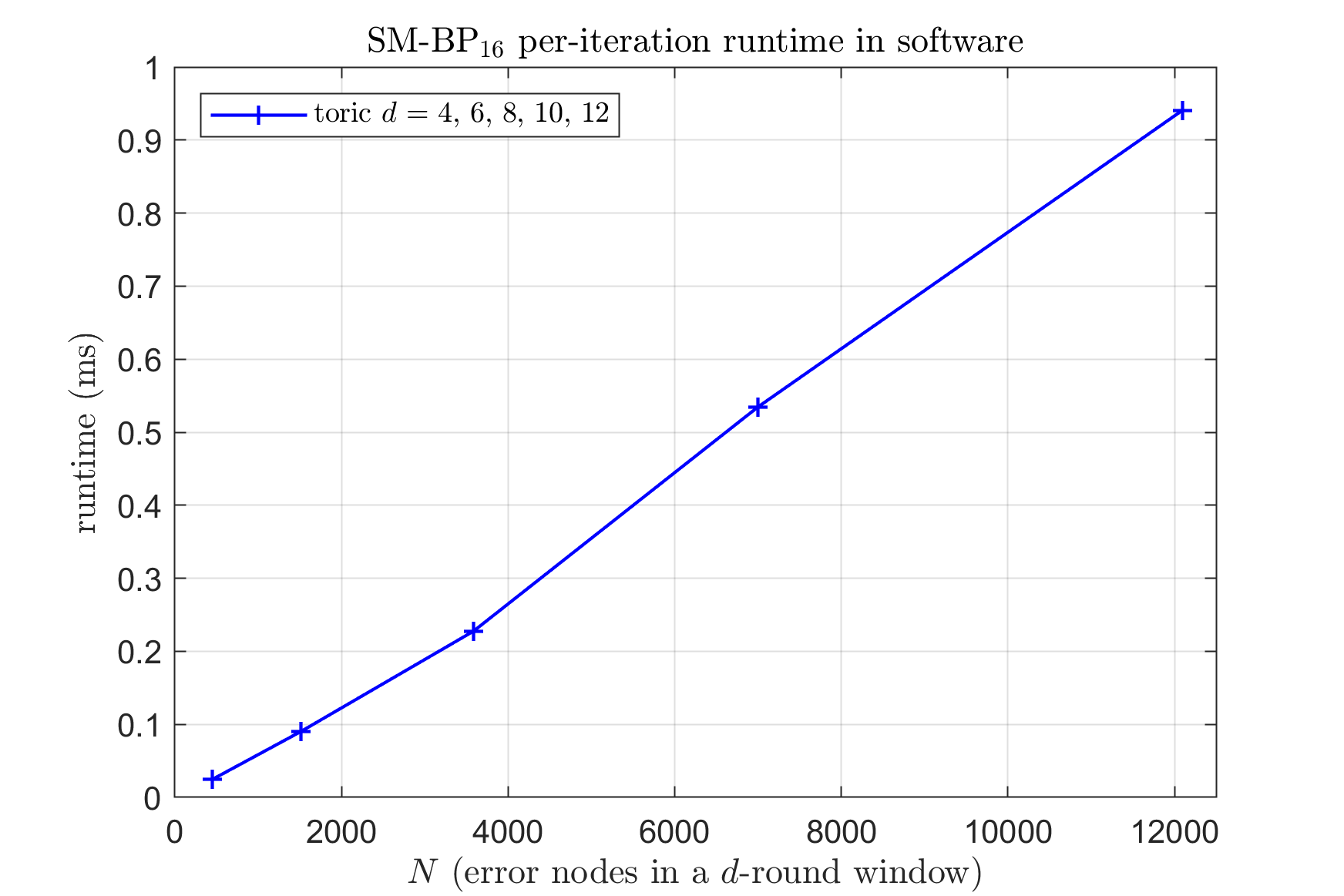}
	\caption{	
		\bmark{
			Per-iteration runtime of a single-instance SM-BP$_{16}$ versus the number of error locations $N={(n+wm+2m)r}=7d^3$ for rotated toric codes ($n=d^2$, $w=4$, $m=n$, $r=d$).
			The runtime scales approximately linearly with $N$.
			%
		}
	} \label{fig:SM-BP_runtime} 
\end{figure}

\bmark{  
	We measure the per-iteration runtime of a single-instance SM-BP$_{16}$ using rotated toric codes. Recall that SM-BP has per-iteration complexity $O(N)$ and is therefore expected to exhibit approximately linear runtime in $N$ in software implementations. 
	We select the setting physical error rate $\epsilon=0.001$ and $\alpha=1$, as these parameters have negligible impact on runtime. Experiments are conducted on a single core of a 4.7\,GHz AMD 9800X3D CPU.
	
	The decoder runtime is evaluated as a function of $N$, the number of error locations (columns) in the generalized check matrix, where $N=(n+wm+2m)r$, with $n$ the number of data qubits, $w$ the average stabilizer weight, $m$ the number of stabilizers, and $r$ the window size (in syndrome extraction rounds).  
	For rotated toric codes,  $N=(d^2+4d^2+2d^2)d=7d^3$ and the results for $d=4$ to $12$ are shown in Fig.~\ref{fig:SM-BP_runtime}. 
	The measured runtime scales approximately linearly with $N$.
}

%

\begin{figure}
	\centering \includegraphics[width=0.5\textwidth]{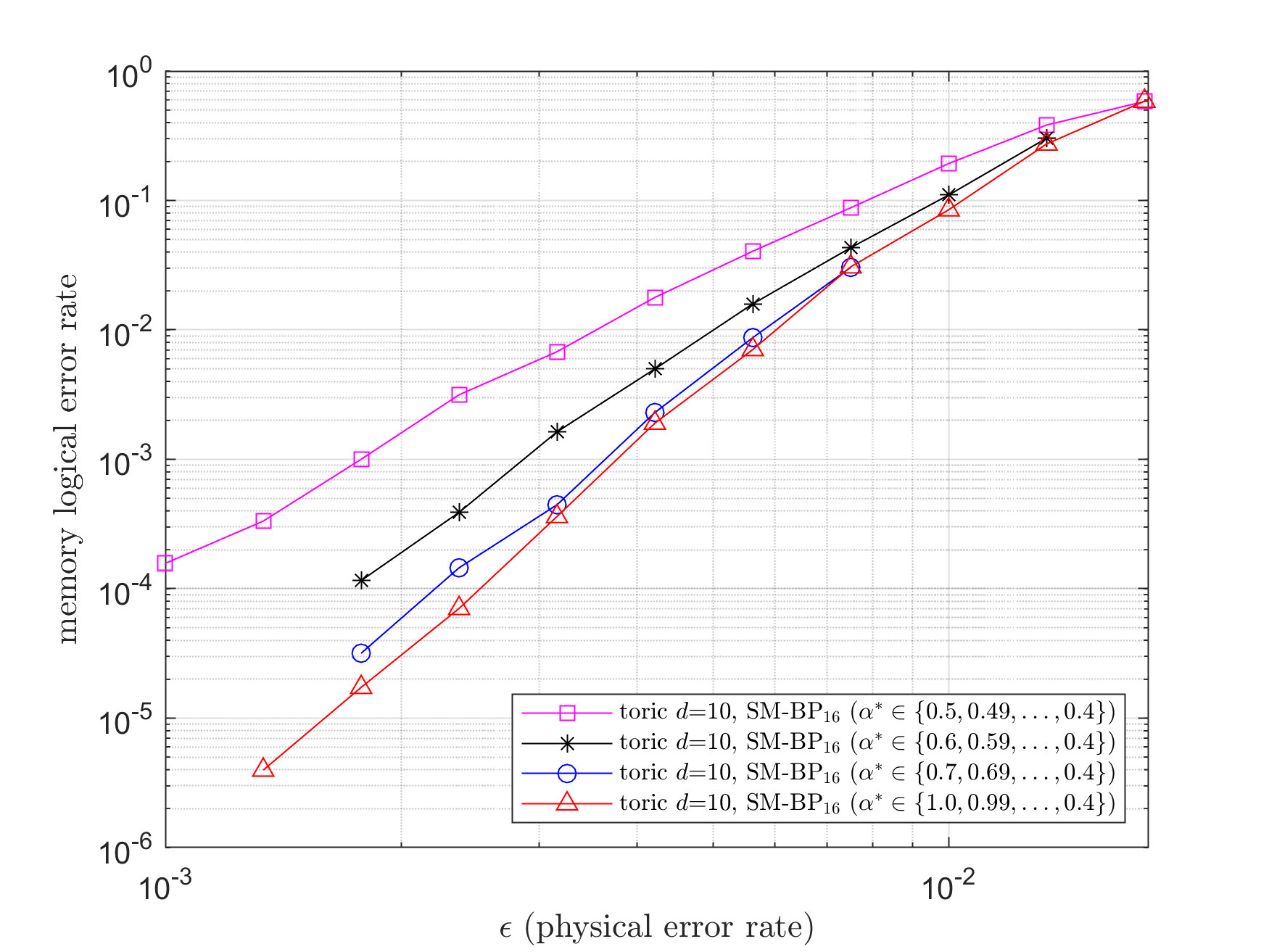}
	\caption{	
		\bmark{Comparison of SM-BP$_{16}(\alpha^*\in\{\alpha_1,\dots,0.4\})$ with starting values $\alpha_1=0.5,\,0.6,\,0.7$, and $1.0$. 
			%
		}
	} \label{fig:diff_alpha_1} 
\end{figure}

\bmark{  
	Next, we investigate the efficiency of the multiple-instance version, SM-BP$_{16}(\alpha^*)$, where $\alpha^*$ is selected from a decreasing sequence $\alpha_1 > \alpha_2 > \cdots > \alpha_\ell > 0$. In the main text, we use $\alpha_1 = 1.0$, $\alpha_2 = 0.99$, $\dots$, and $\alpha_\ell = 0.4$. The decoder runs multiple BP instances sequentially, each with a different $\alpha$, until one converges. The efficiency therefore depends on how quickly a suitable $\alpha$ is reached.

	To reduce total BP iterations, it is desirable to choose a good initial value $\alpha_1$. Since the BP step size scales as $1/\alpha$, smaller physical error rates  require finer exploration of the error space, thus larger $\alpha$. Empirically, $\alpha$ increases approximately linearly with $-\log_{10}(\epsilon)$, with  slope depending on the mean row weight of the check matrix as discussed in \cite[arXiv version Sec.~3]{KL22}.

	Then we perform preliminary simulations to estimate this relation, using the $d=10$ rotated toric code with error consolidation and adaptive windows, and window size $r=d$. We test SM-BP$_{16}(\alpha^*)$ (Algorithm~\ref{alg:FTAMBP}) with different initial values $\alpha_1 = 0.5, 0.6, 0.7$, and $1.0$. The results are shown in Fig.~\ref{fig:diff_alpha_1}. As expected, smaller $\epsilon$ requires larger $\alpha_1$.
Using representative operating points,
	\begin{itemize}
		\item $\alpha_1=0.6$ at $\epsilon\approx 0.0133$ \, ($\log_{10}(\epsilon)= -1.875$);
		\item $\alpha_1=0.7$ at $\epsilon\approx 0.0042$ \, ($\log_{10}(\epsilon)= -2.375$),
	\end{itemize} 
we obtain an approximate linear relation
	\begin{equation}\label{eq:alpha1_func}
		\alpha_1= -0.2\log_{10}(\epsilon) + 0.225,
	\end{equation}
	with $\alpha_1$ clipped to the range $[0.4, 1.0]$ in practice.
}

\begin{figure}
	\centering \includegraphics[width=0.5\textwidth]{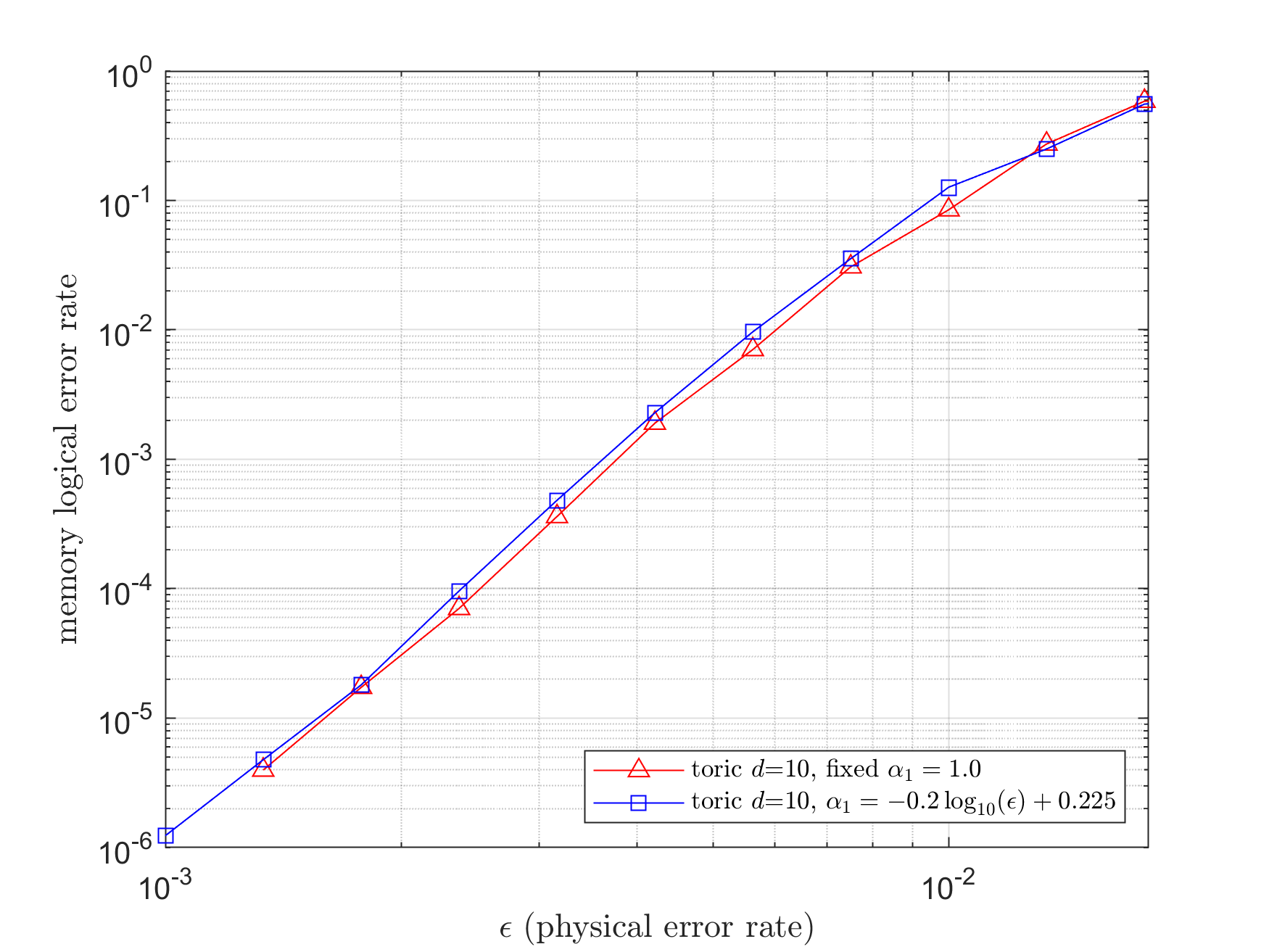}
	\caption{	
		\bmark{
			Comparison of SM-BP$_{16}(\alpha^*\in\{\alpha_1,\dots,0.4\})$ with $\alpha_1$ determined by \eqref{eq:alpha1_func}, versus the baseline choice $\alpha_1=1.0$. 
			Using \eqref{eq:alpha1_func} achieves comparable accuracy with significantly fewer iterations (Table~\ref{tb:avg_iter}). 
		}
	} \label{fig:auto_alpha1} 
\end{figure}

\bmark{
	We repeat the simulation using the relation in \eqref{eq:alpha1_func}. The performance closely matches that of the fixed $\alpha_1=1.0$ setting, as shown in Fig.~\ref{fig:auto_alpha1}.
	However, the average number of BP iterations differs significantly, as detailed in Table~\ref{tb:avg_iter}. These counts are obtained by summing the iterations across all SM-BP$_{16}(\alpha^*)$ instances used for decoding a quantum memory, and then normalizing by the number of syndrome extraction rounds before failure.
		This indicates that the performance reported in the main simulations can be achieved at significantly reduced computational cost.
		

	For the same topological-code family, the mean row weight of the check matrix is approximately constant, which, as noted, determines the slope in relations of the form \eqref{eq:alpha1_func}. Therefore, we expect \eqref{eq:alpha1_func} to be applicable to other codes within the same toric family, as supported by additional tests for $d=12$ (not shown).

}

\begin{table}[h!] 
	\caption{
		\bmark{
			SM-BP$_{16}(\alpha^*)$ iteration counts for different $\alpha_1$ schemes on the $d=10$ rotated toric code ({using} \eqref{eq:alpha1_func} vs fixed $\alpha_1=1.0$).
			The corresponding performance is shown in Fig.~\ref{fig:auto_alpha1}.
		}
	} 
	\centering      
	\begin{tabular}{lrrrr}
		\hline
		\hline
		average iteration counts  & $\epsilon=0.1\%$  & $0.13\%$  & $0.32\%$  & $0.75\%$ \\
		\hline
		{fixed} $\alpha_1=1.0$  & 33.4  & 64.4  & 296.1  & 881.5 \\ 
		{damping $\alpha_1$ by} \eqref{eq:alpha1_func}  & 10.1  & 18.1  & 27.4  & 50.1 \\  
		\hline
		\hline
	\end{tabular} \label{tb:avg_iter} 
\end{table}

	\section{Error threshold estimates through finite-size scaling ansatz}
	\label{app:threshold_ansatz}

	We use the finite-size scaling ansatz~\cite{WHP03,Har04} to estimate the error thresholds of the topological code families, as we are limited to simulating codes of finite sizes.
	\bmark{
	Note that the threshold reported here is decoder-dependent. Although it may be lower than the theoretical maximum-likelihood threshold of the code family, it provides a realistic performance benchmark for practical decoding in large-scale QLDPC implementations.}
	
	The ansatz assumes that  the logical error rate $P_L(\epsilon,d)$ for a code of distance $d$ at depolarizing rate $\epsilon$   can be modeled as:
	\begin{align*}
		P_L(\epsilon,d)=f( d^{-\nu}(\epsilon-\tau)),
	\end{align*}
	where $\tau$ is the error threshold and  $\nu$  is critical exponent that characterizes the scaling behavior near the threshold.	
	The function $f(x)$ is a universal scaling function, and the simulation data are fitted using a degree-3 polynomial in~$x$.
	
\begin{figure}
		\centering \includegraphics[width=0.5\textwidth]{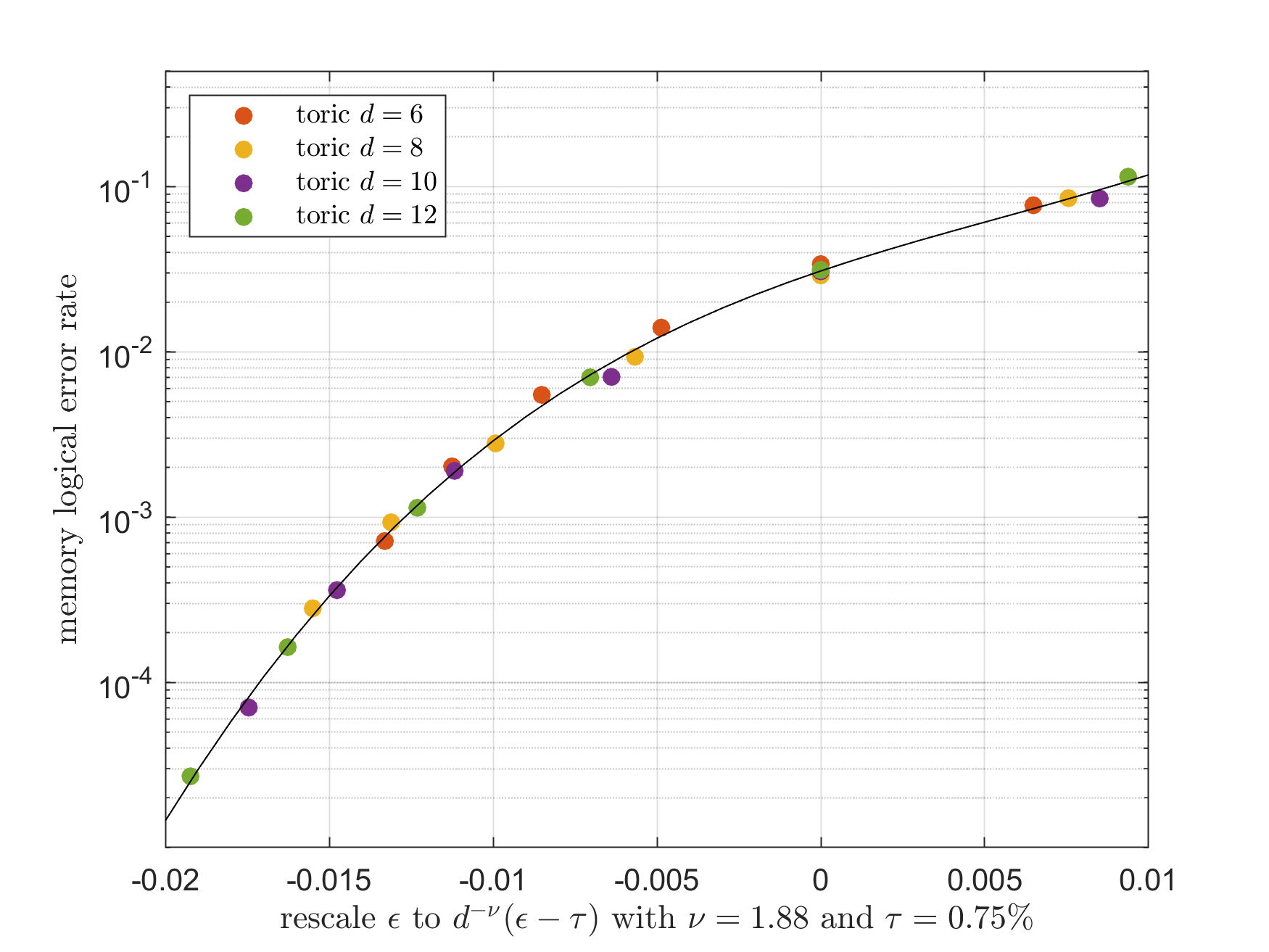}
		\caption
		{
			\rmark{Rescaled SM-BP$_{16}(\alpha^*)$ decoding threshold performance for rotated toric codes with $d=6,8,10,12$.
			The original data points are from Fig.~\ref{fig:FT_toric}.}  
		} \label{fig:TH_toric} 
	\end{figure}

	\rmark{The rescaled SM-BP$_{16}(\alpha^*)$ decoding threshold performance for rotated toric codes is shown in Fig.~\ref{fig:TH_toric} for $d=6$ to $12$. The ansatz estimates a threshold of $0.75\%$. We exclude $d=4$, as its inclusion overestimates the threshold as $0.91\%$.}
	
	\begin{figure}
		\centering \includegraphics[width=0.5\textwidth]{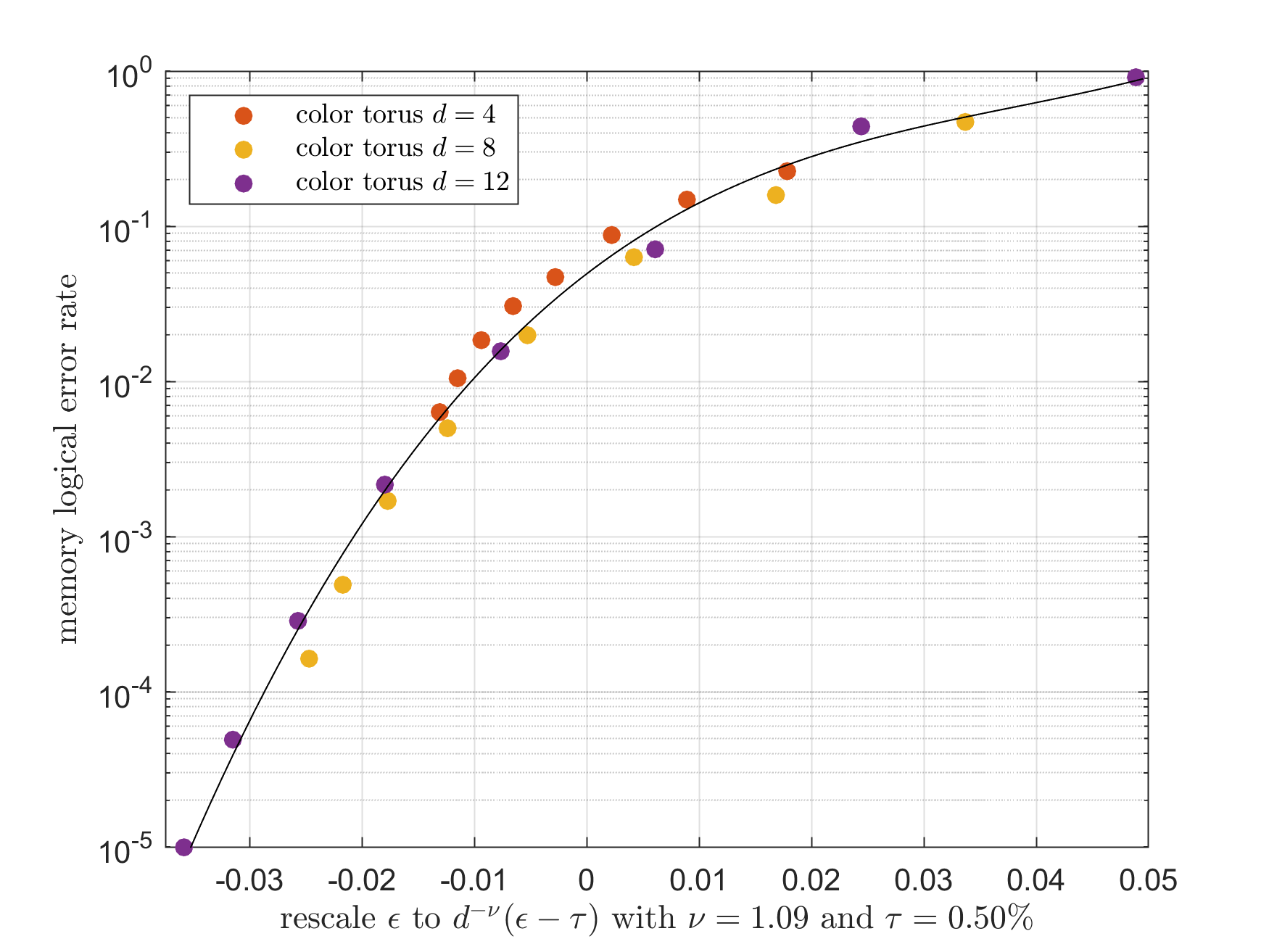}
		\caption
		{
			\rmark{Rescaled SM-BP$(\alpha^*)$ decoding threshold performance  for  rotated 6.6.6 color codes with $d=4,8,12$. The original data points are from Fig.~\ref{fig:FT_color}.}
		} \label{fig:TH_color} 
	\end{figure}

\rmark{For SM-BP$(\alpha^*)$ decoding of the rotated 6.6.6 color codes, the ansatz estimates a threshold of $0.50\%$ using $d=4,8,12$ (see Fig.~\ref{fig:TH_color}); excluding $d=4$ gives a similar estimate.}


	\begin{figure}
		\centering \includegraphics[width=0.5\textwidth]{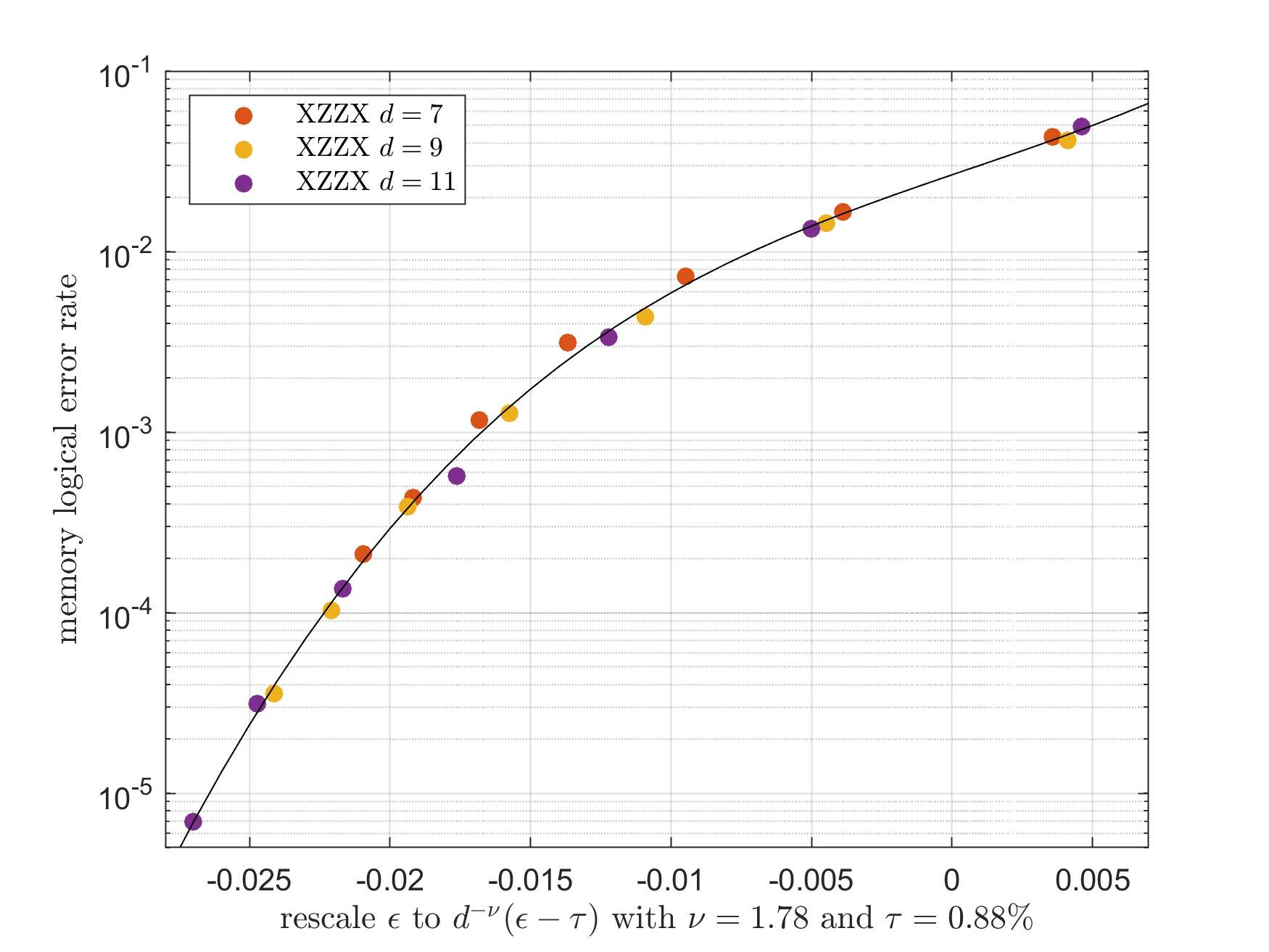}
		\caption
		{
			\rmark{Rescaled SM-BP$_{16}(\alpha^*)$ decoding threshold performance for twisted XZZX toric codes with $d=7$ to $11$. The original data points are from Fig.~\ref{fig:FT_XZZXtoric}.} 
		} \label{fig:TH_XZZX} 
	\end{figure}

	
	\rmark{Similarly, for the twisted XZZX toric codes, the ansatz method estimates a threshold of $0.88\%$ using $d=7,9,11$, as shown in Fig.~\ref{fig:TH_XZZX}. We exclude $d=3$ and $d=5$, as their inclusion leads to a systematic overestimation of the threshold.}

	\section{ SM-BP$_4$ v.s.  SM-BP$_{16}$}\label{app:C4C16}

\rmark{We compare SM-BP$_{16}$ and SM-BP$_{4}$ decoders, which employ probabilistic error consolidation based on Proposition~\ref{prop:C16} and Proposition~\ref{prop:C4}, respectively.}

\begin{figure}[t!]
		\centering \includegraphics[width=0.5\textwidth]{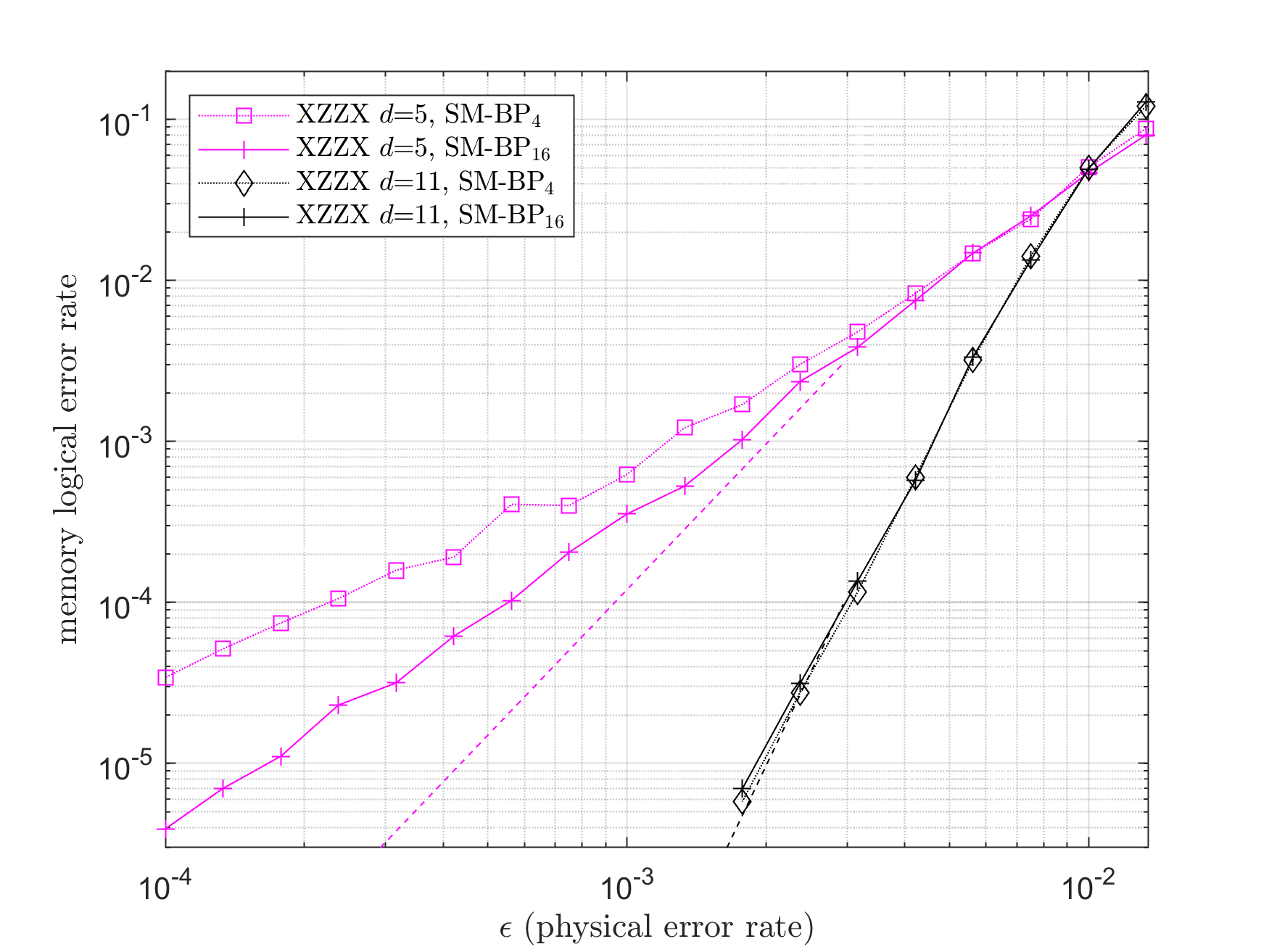}
		\caption
		{	
			\bmark{
			Comparison of SM-BP$_4(\alpha^*)$ and SM-BP$_{16}(\alpha^*)$ on twisted XZZX codes with $d=5$ and~$11$. 
			Data points for SM-BP$_{16}(\alpha^*)$ are from Fig.~\ref{fig:FT_XZZXtoric}, with the error consolidation following Proposition~\ref{prop:C16}. SM-BP$_{4}(\alpha^*)$ uses the same settings, except that the error consolidation follows Proposition~\ref{prop:C4}.  The dashed lines show the expected scaling $a\epsilon^{t+1}$ with $t=\lfloor\frac{d-1}{2}\rfloor$ for appropriate constants $a$.
			}
		} \label{fig:XZZX_C4C16} 
	\end{figure}

\bmark{The comparison is performed using adaptively chosen $\alpha^*$ following Algorithm~\ref{alg:FTAMBP}. A representative case is shown for twisted XZZX codes with $d=5$ and $d=11$ in Fig.~\ref{fig:XZZX_C4C16}. 
	As expected, SM-BP$_{16}$ performs better at $d=5$, since it captures correlations within 16-ary error structures that are not exploited by SM-BP$_{4}$. 
	As the code distance increases, this advantage diminishes, and the two decoders exhibit comparable performance at $d=11$.}
\bmark{
This behavior can be attributed to several factors: increased degeneracy at larger distances allows SM-BP$_4$ to converge to valid degenerate solutions; its underlying generalized check matrix contains fewer short cycles; and error consolidation in SM-BP$_{16}$ can bias the decoder toward certain bit decisions. 
Improving error consolidation remains an open research direction.}

\section{Comparisons with other decoders} \label{app:versus_others}

In this section, we compare our SM-BP decoder with other state-of-the-art decoding methods for quantum error correction. 
\subsection{Virtual Decoder and Threshold Estimation}

\begin{figure}
	\centering
	\includegraphics[width=0.325\textwidth]{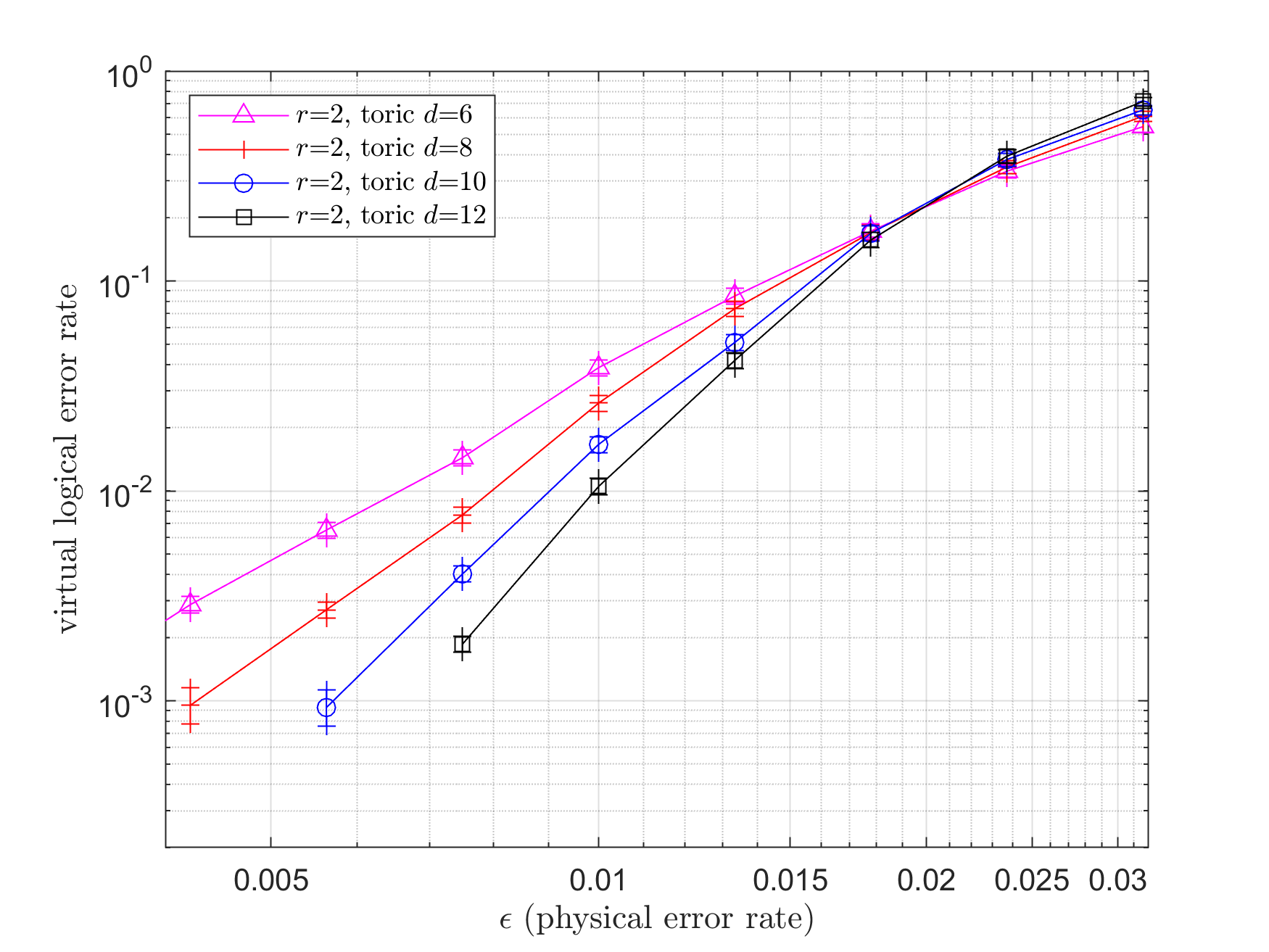}
	\includegraphics[width=0.325\textwidth]{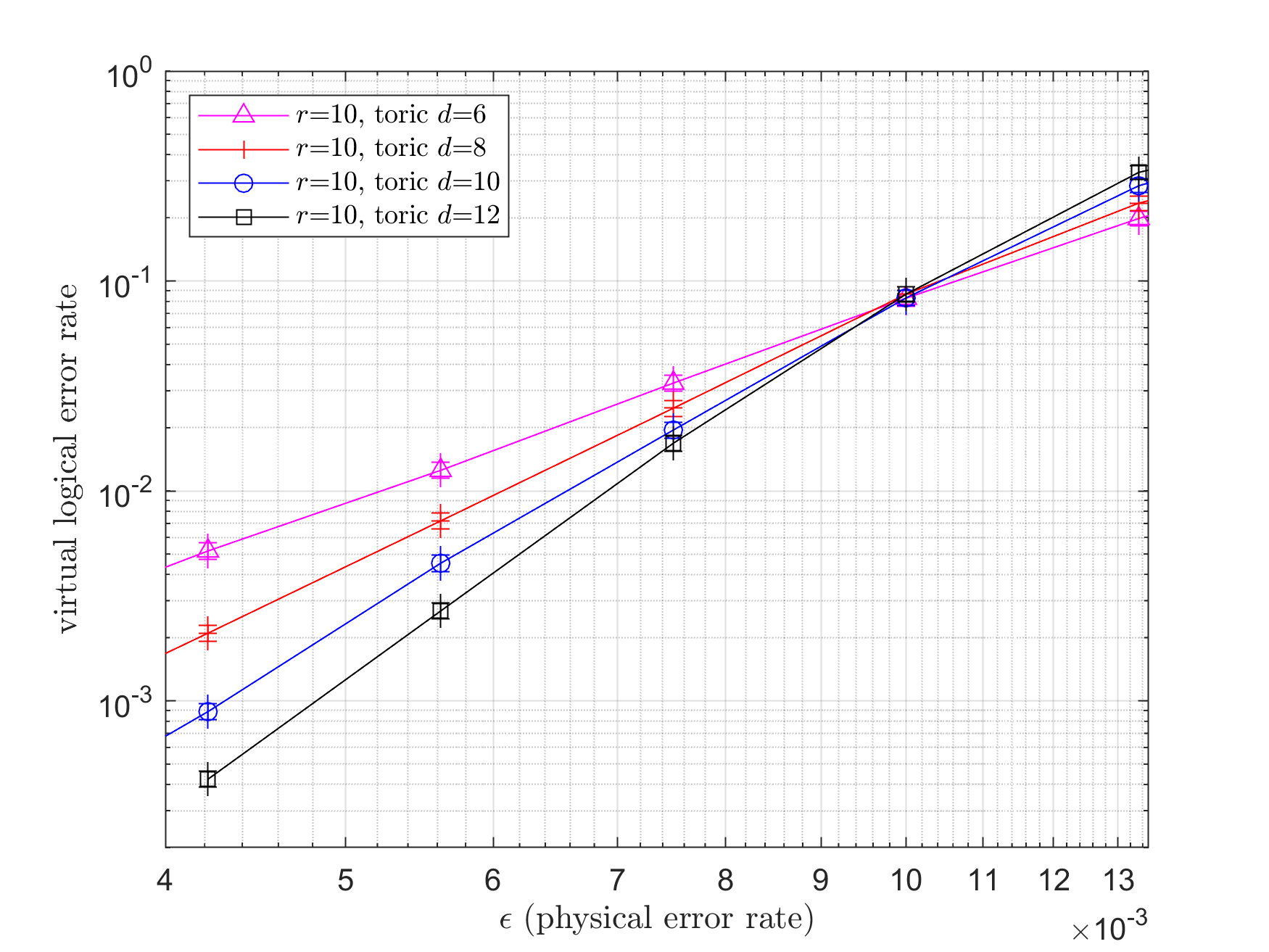}
	\includegraphics[width=0.325\textwidth]{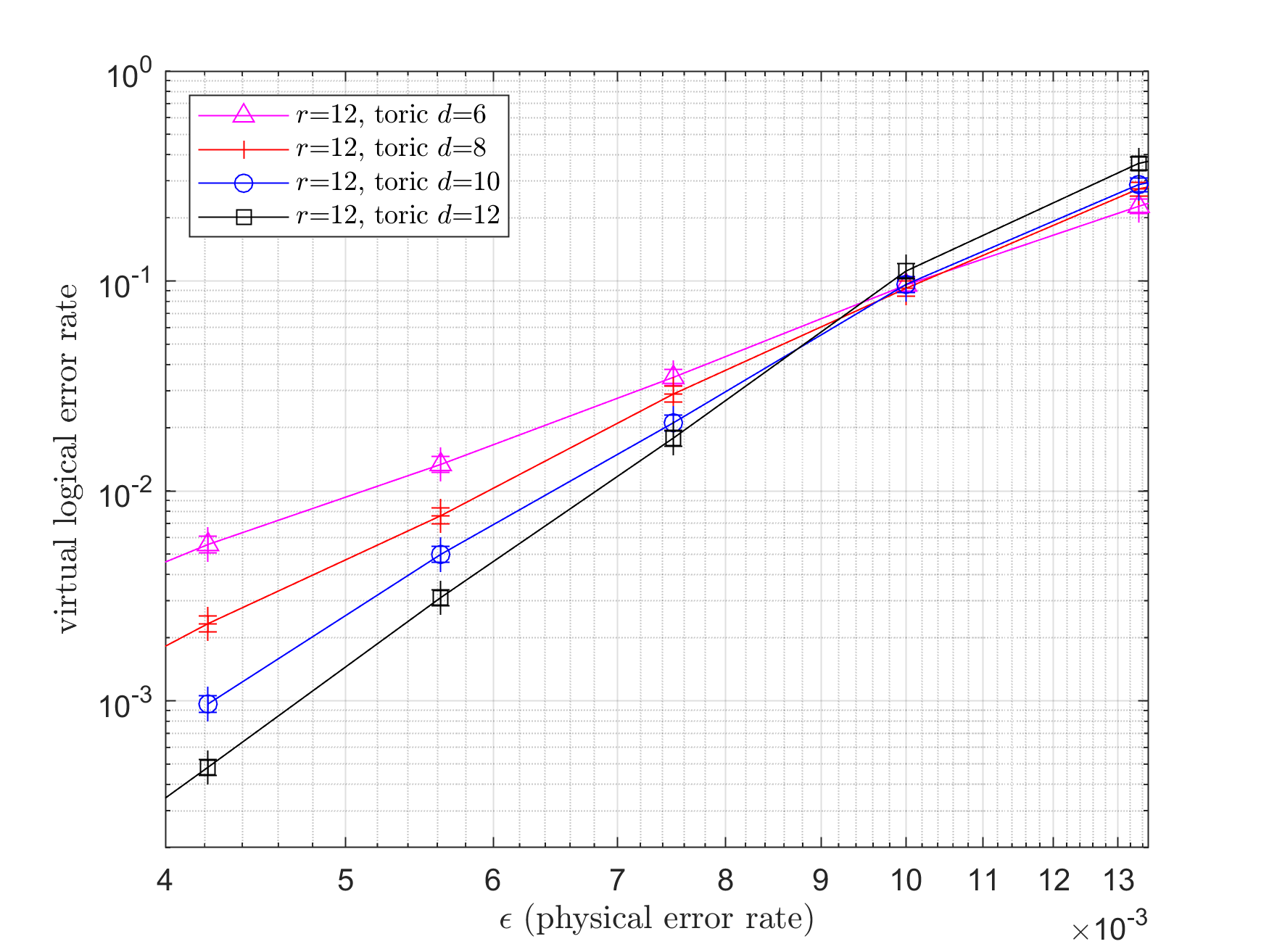}
	\caption{	
		\bmark{
			Estimating the virtual decoder’s threshold. Increasing $r$ reduces the degradation of the threshold to a negligible level. Here, we show results for $r=2$, $10$, and $12$. For $r \ge 10$, a similar threshold close to $0.9\%$ is observed for $d=6$ to $12$.
		}
	} \label{fig:vir_thd_in_r} 
	\vspace*{\floatsep}	
	\centering
	\includegraphics[width=0.45\textwidth]{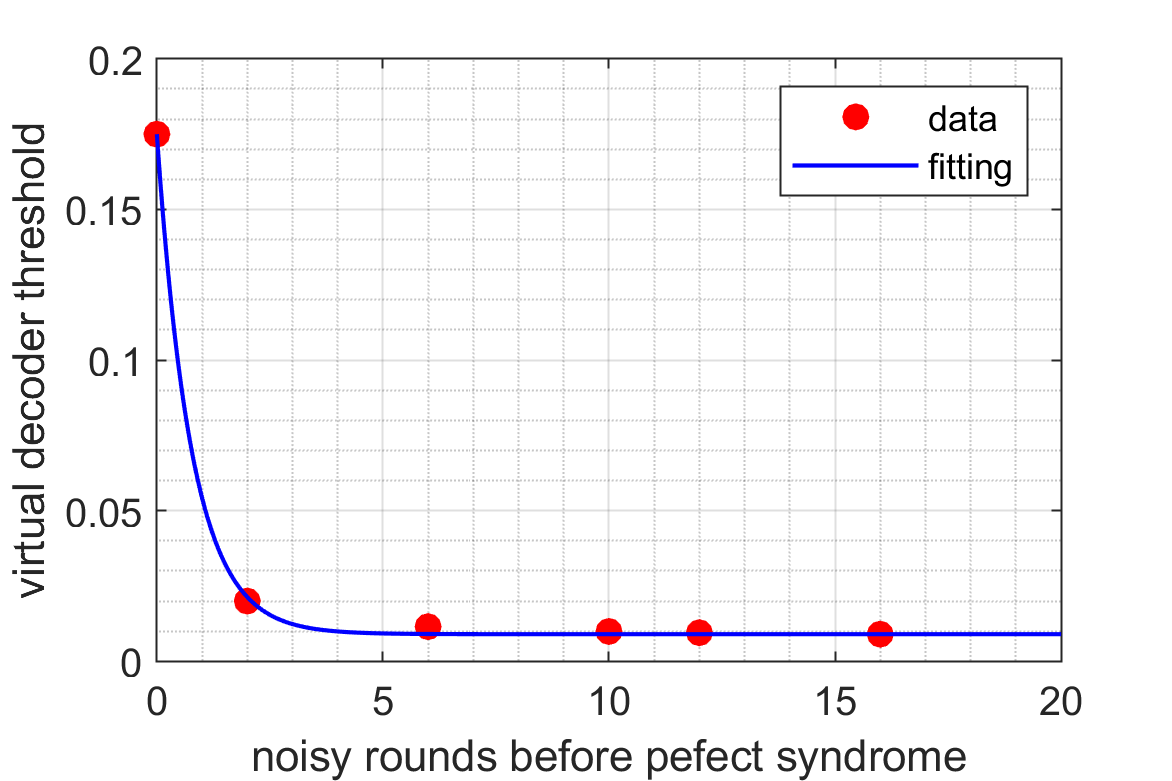}
	\quad
	\includegraphics[width=0.52\textwidth]{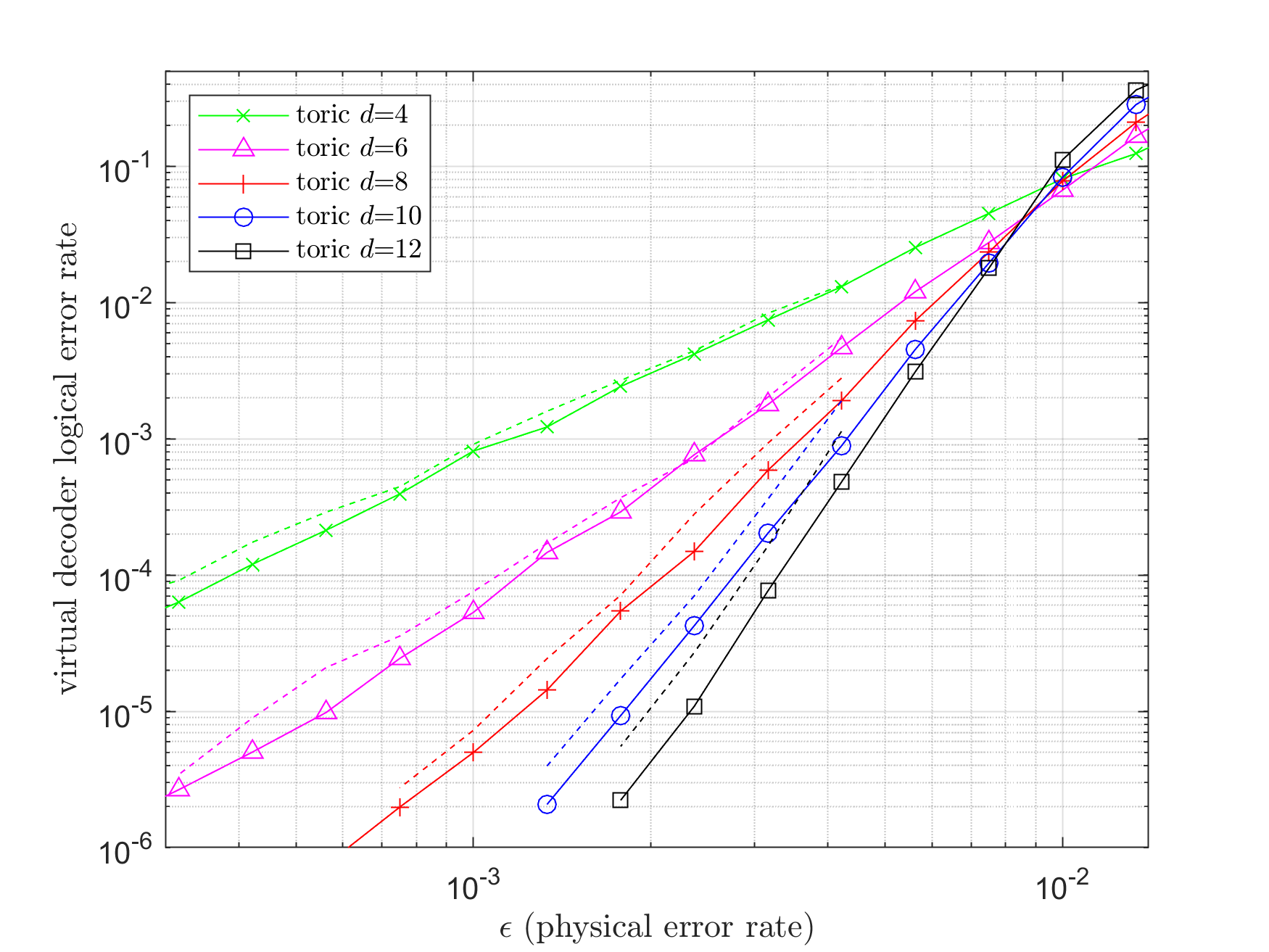}
	\caption{	
		\bmark{
		(left) Virtual threshold fitting curve obtained by repeating the cases in Fig.~\ref{fig:vir_thd_in_r} for various $r$. The fit follows \eqref{eq:vir_th_fitting} with $\gamma=1.3$ and $p_\infty=0.9\%$.
		(right) Virtual threshold performance of toric codes with $r=d$, using SM-BP$_{16}(\alpha^*)$ with \eqref{eq:alpha1_func} (solid lines).
		This provides a best-case performance bound for the actual decoder in Fig.~\ref{fig:FT_toric}, whose data points are shown here as dashed lines. 
		For virtual logical error rate, it is computed as in Definition~\ref{def:T_by_J}. The intersection (virtual threshold) is roughly at $0.9\%$, close to $p_\infty$.  
		}		
	} \label{fig:toric_vir_thd} 
	\end{figure}

\bmark{We consider the virtual decoder scenario that assumes a perfect syndrome measurement is available after every $r$ noisy syndrome extraction rounds.

We define $p_\text{th}(r)$ as the threshold of the virtual decoder given $r$. It can be estimated by the ansatz method or simply by the intersection, as noted earlier.
	The asymptotic value $p_\infty = \lim_{r\to\infty} p_\text{th}(r)$ is called the sustained threshold.

	In the virtual decoder scenario, $r=0$ (data noise only \cite{KL22}) yields the best performance, which degrades as $r$ increases due to accumulated syndrome errors; see Fig.~\ref{fig:vir_thd_in_r}. 
	The degradation becomes insignificant when $r \ge d$. 
	In the following, we focus on the case $r=d$, which serves as a practical upper bound for memory performance, as global decoding over $T \gg d$ rounds is computationally impractical \cite{DKLP02}.
}

	\bmark{
	By varying $r$ and estimating $p_\text{th}(r)$ for each value (Fig.~\ref{fig:vir_thd_in_r}), we obtain the fitting curve shown in Fig.~\ref{fig:toric_vir_thd}\,(left) following \cite{BNB16,QVRC21}:
	\begin{equation} \label{eq:vir_th_fitting}
		p_\text{th}(r) = p_\infty\left(1 - (1 - p_\text{th}(0)/p_\infty)e^{-\gamma r}\right),
	\end{equation}
	where $p_\text{th}(0)$ is the threshold for data noise only (17.5\% in our case \cite{KL22}), and $p_\infty$ is a fitting parameter. The fit yields $\gamma=1.3$ and $p_\infty=0.9\%$, which closely matches the intersection-based threshold estimate at $r=d$ shown in Fig.~\ref{fig:toric_vir_thd}\,(right).
	
	All virtual decoder results presented here use SM-BP$_{16}(\alpha^*)$ with the damping $\alpha$ schedule in \eqref{eq:alpha1_func}. 
	The logical error rate is evaluated as described in Definition~\ref{def:T_by_J}.}

	\subsection{Comparing to BP-OSD}
\begin{figure}
		\centering \includegraphics[width=0.5\textwidth]{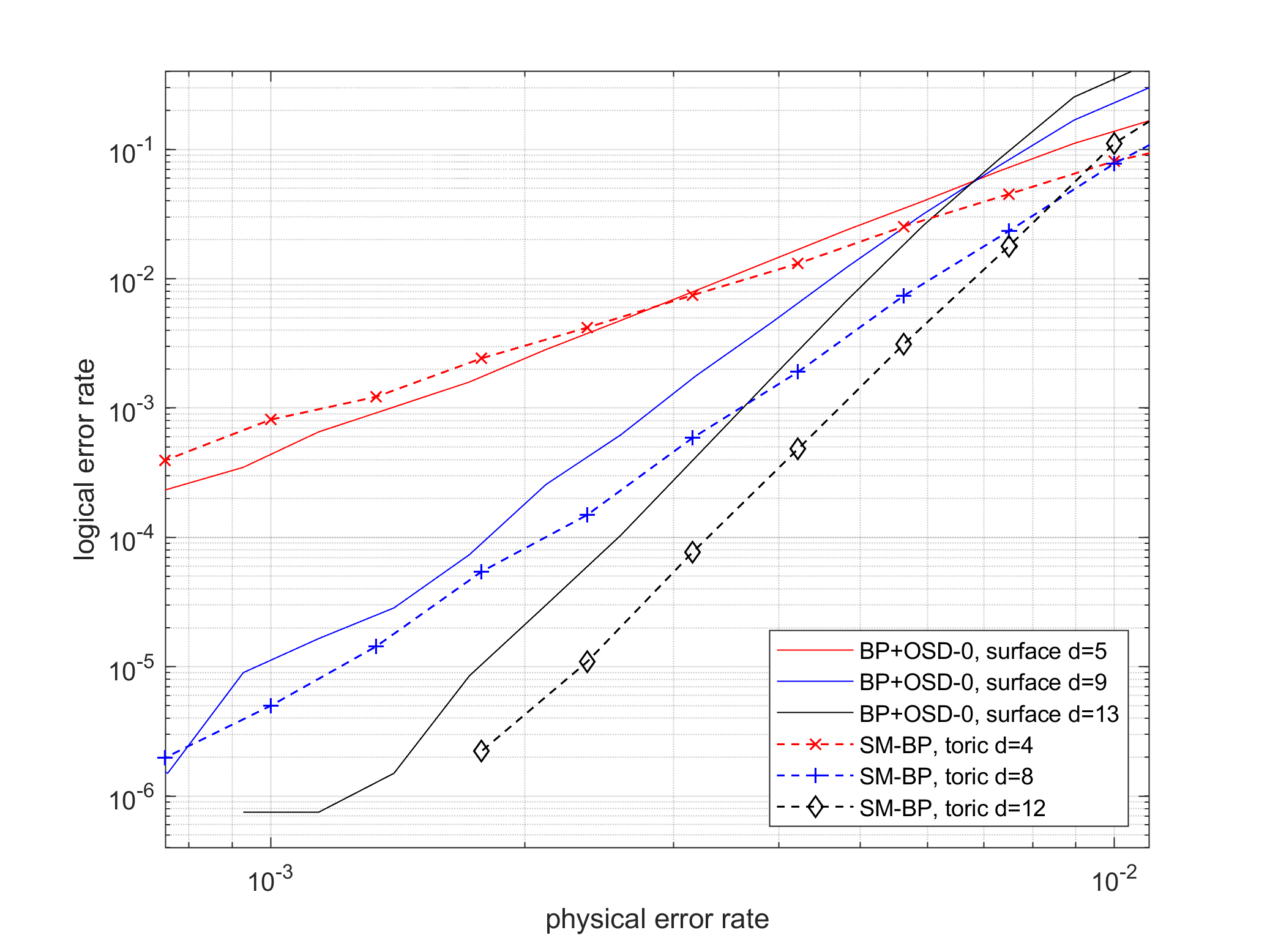}
		\caption{	
			\bmark{Comparison of BP+OSD-0 from \cite{Hil+25} (surface codes) with SM-BP$_{16}(\alpha^*)$ (toric codes). Curves are matched by setting $d_\text{toric} = d_\text{surf}-1$. 
			}
		} \label{fig:vs_BPOSD} 
	\end{figure}

\bmark{

	BP+OSD-0 on surface codes under circuit-level noise is analyzed in Ref.~\cite{Hil+25}, where three approaches are compared: BP+OSD-0, BeliefFind (BP+UF), and BP+LSD-0. Among these, BP+OSD-0 achieves the best accuracy at the highest computational cost, while BP+LSD-0 provides a lower-complexity alternative with slightly reduced accuracy for large $d \ge 17$.

	Their work studies surface codes with $d_{\text{surf}}=5,9,13,17$ in a virtual-decoder setting, while we consider toric codes with $d_{\text{toric}}=4,6,8,10,12$ in Fig.~\ref{fig:toric_vir_thd}\,(right). Although the code distances are not identical, the results provide a useful reference for comparison. Figure~\ref{fig:vs_BPOSD} shows the comparison for comparable distances.

	For direct comparison with matching curve colors, we set $d_\text{toric} = d_\text{surf}-1$ (note that this implies the number of correctable errors $t_\text{toric} = t_\text{surf}-1$). The logical error rates reported in \cite{Hil+25} are for a single check type ($Z$-checks only), so we scale them by $3/2$ to account for both check types, slightly raising the curves but leaving the slope and threshold unchanged.
	
	As shown, SM-BP achieves a comparable threshold while maintaining competitive accuracy across different code distances.}

	\subsection{Comparing to MWPM}

\begin{figure}
		\centering \includegraphics[width=0.5\textwidth]{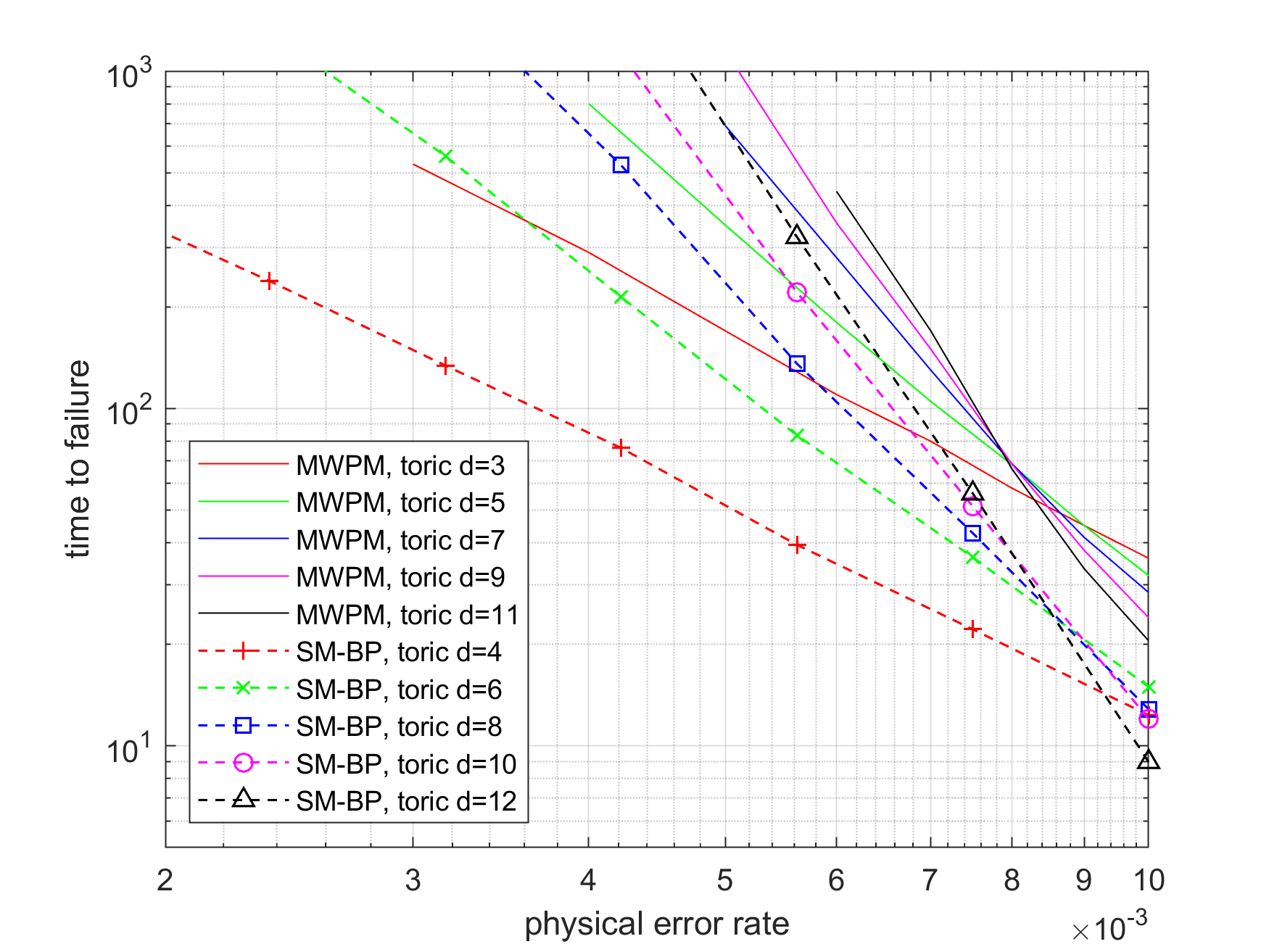}
		\caption{	
			\bmark{
				Comparison of MWPM from \cite{WFSH10} with SM-BP$_{16}(\alpha^*)$ (virtual decoder scenario). 
				We compare to \cite[Fig.~11\,(top)]{WFSH10}, while our data is from Fig.~\ref{fig:toric_vir_thd}\,(right).
				Both decoders show similar slopes at comparable $d$.
				 MWPM achieves a longer lifetime but the gap vanishes as $d$ increases.
			}
		} \label{fig:vs_MWPM} 
	\end{figure}

\begin{figure}
		\centering \includegraphics[width=0.5\textwidth]{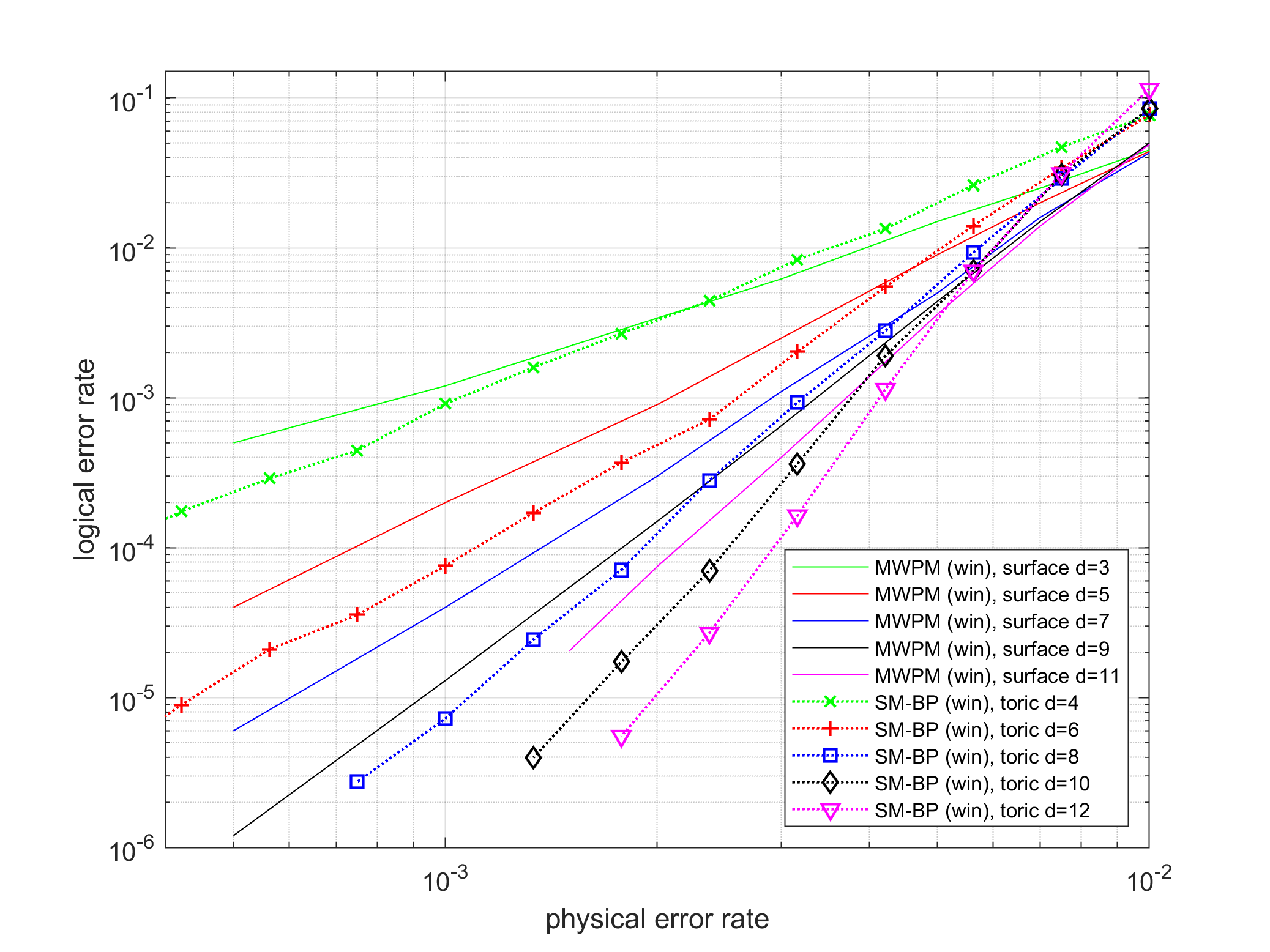}
		\caption{	
			\bmark{
				Comparison of windowed MWPM from \cite{Zha+23} {(surface codes)} with SM-BP$_{16}(\alpha^*)$ using the adaptive window scheme {(toric codes, Fig.~\ref{fig:FT_toric})}. 
				{Our results show better slope in the mid-error regime, achieving better accuracy in the mid- to low-error rate region.}
			}
		} \label{fig:vs_MWPM_win} 
	\end{figure}

\bmark{
Figure~\ref{fig:vs_MWPM} compares MWPM performance from \cite[Fig.~11 (top)]{WFSH10} with our SM-BP decoder. For this comparison, we convert our logical error rates (per round) back to time-to-failure $T$ to match their presentation format.

The comparison shows that both decoders exhibit similar slopes at comparable code distances. 
MWPM achieves a longer lifetime but the gap vanishes as $d$ increases.

We also compare with the windowed MWPM decoder from \cite{Zha+23}, which processes surface codes using only noisy syndromes with a sliding window approach. Figure~\ref{fig:vs_MWPM_win} shows this comparison, where our results use the adaptive window scheme {(Fig.~\ref{fig:FT_toric})}.

Our adaptive window scheme improves the slope in the mid-error regime (recall Fig.~\ref{fig:FT_schemes}), 
{resulting in better accuracy in the mid- to low-error rate region.}
}

	\subsection{Comparing to UF}

\begin{figure}
		\centering \includegraphics[width=0.5\textwidth]{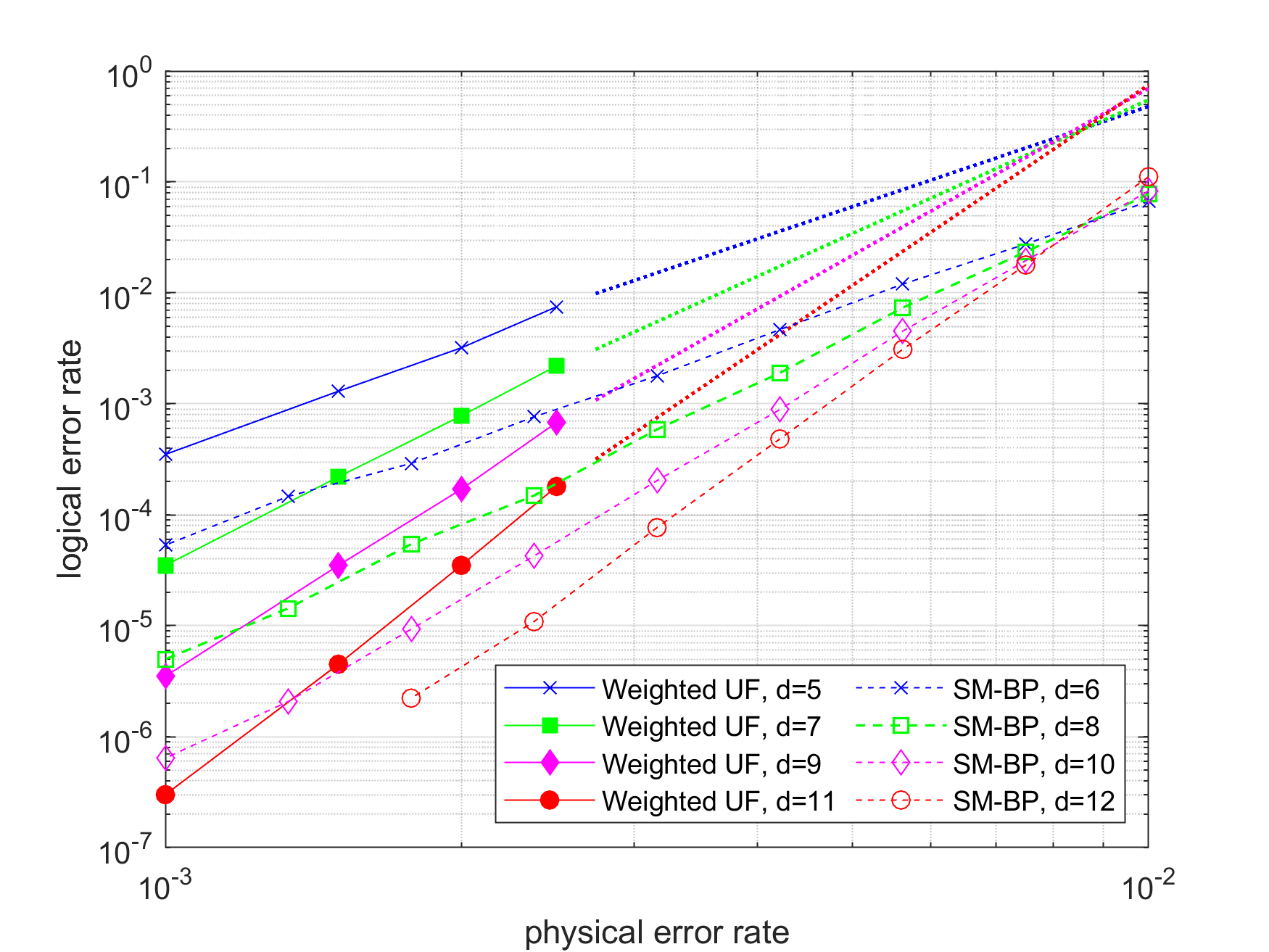}
		\caption{	
			\bmark{
				Comparison of weighted UF from \cite{HNB20} with SM-BP$_{16}(\alpha^*)$ (virtual decoder). 
				Dotted lines indicate the expected exponential scaling $a\epsilon^{(t+1)}$ for each distance.
				Ref.~\cite{HNB20} assumes preparation and measurement error rates of $2\epsilon/3$, leading to steeper slopes than our uniform-$\epsilon$ noise model.
			}
		} \label{fig:vs_UF} 
	\end{figure}

\bmark{
	Ref.~\cite{HNB20} evaluates weighted UF on unrotated toric codes under circuit noise in the virtual decoder setting, which is comparable to our decoder in Fig.~\ref{fig:toric_vir_thd}\,(right). 
	Figure~\ref{fig:vs_UF} presents this comparison, reproduced from \cite[Fig.~4 (top)]{HNB20}. 
	
	SM-BP exhibits better logical error rates across the compared region. 
	The UF results show steeper slopes, which is expected since \cite{HNB20} assumes a reduced preparation and measurement error rate of $2\epsilon/3$ (compared to $\epsilon$ for two-qubit gates). 
}

\end{document}